\documentclass[aps,prd,superscriptaddress,longbibliography,amsmath,amssymb,twocolumn,fleqn,epsf,epsfig,showpacs,nofootinbib]{revtex4-1} 
\usepackage{dcolumn}
\usepackage{bm}
\usepackage{amsfonts}
\usepackage{color}
\usepackage{graphicx}
\usepackage{colordvi}
\usepackage{mathtools}
\usepackage{empheq}
\usepackage[utf8]{inputenc}
\usepackage{hyperref}

\def\be{\begin{equation}}
\def\ee{\end{equation}}
\def\ba{\begin{eqnarray}}
\def\ea{\end{eqnarray}}

\newcommand{\dd}{\mathrm{d}}

\begin{document}

\title{Cosmological constraints on post-Newtonian parameters in effectively massless scalar-tensor theories of gravity}

\author{Massimo Rossi}\email{rossi.massim@gmail.com}
\affiliation{INAF/OAS Bologna, via Gobetti 101, I-40129 Bologna, Italy}
\author{Mario Ballardini}\email{mario.ballardini@inaf.it}
\affiliation{Dipartimento di Fisica e Astronomia, Alma Mater Studiorum Universit\'a di Bologna,
Via Gobetti, 93/2, I-40129 Bologna, Italy}
\affiliation{Department of Physics and Astronomy, University of the Western Cape, Cape Town 7535, South Africa}
\affiliation{INAF/OAS Bologna, via Gobetti 101, I-40129 Bologna, Italy}
\author{Matteo Braglia}\email{matteo.braglia2@unibo.it}
\affiliation{Dipartimento di Fisica e Astronomia, Alma Mater Studiorum, Universit\`a degli Studi di Bologna, via Gobetti 101, I-40129 Bologna, Italy}
\affiliation{INAF/OAS Bologna, via Gobetti 101, I-40129 Bologna, Italy}
\affiliation{INFN, Sezione di Bologna, Via Irnerio 46, I-40127 Bologna, Italy}
\author{Fabio Finelli}\email{fabio.finelli@inaf.it} 
\affiliation{INAF/OAS Bologna, via Gobetti 101, I-40129 Bologna, Italy}
\affiliation{INFN, Sezione di Bologna, Via Irnerio 46, I-40127 Bologna, Italy} 
\author{Daniela Paoletti}\email{daniela.paoletti@inaf.it}
\affiliation{INAF/OAS Bologna, via Gobetti 101, I-40129 Bologna, Italy}
\affiliation{INFN, Sezione di Bologna, Via Irnerio 46, I-40127 Bologna, Italy}
\author{Alexei A. Starobinsky}\email{alstar@landau.ac.ru}
\affiliation{Landau Institute for Theoretical Physics, 119334 Moscow, Russia}
\affiliation{Bogolyubov Laboratory of Theoretical Physics, Joint Institute for Nuclear
Research, Dubna 141980, Russia}
\author{Caterina Umilt\`a}\email{umiltca@ucmail.uc.edu}
\affiliation{Department of Physics, University of Cincinnati,
345 Clifton Ct, Cincinnati, OH 45221, U.S.A.}

\date{\today}

\begin{abstract}
We study the cosmological constraints on the variation of the Newton's constant and on 
post-Newtonian parameters for simple models of scalar-tensor theory of gravity beyond 
the extended Jordan-Brans-Dicke theory.
We restrict ourselves to an effectively massless scalar field with a potential $V \propto F^2$, 
where $F(\sigma)=N_{pl}^2+\xi\sigma^2$ is the coupling to the Ricci scalar considered. 
We derive the theoretical predictions for cosmic microwave background (CMB) anisotropies 
and matter power spectra by requiring that the effective gravitational strength at present 
is compatible with the one measured in a Cavendish-like experiment and by assuming adiabatic 
initial condition for scalar fluctuations.
When comparing these models with $Planck$ 2015 and a compilation of baryonic acoustic oscillations 
data, all these models accomodate a marginalized value for $H_0$ higher than in $\Lambda$CDM. 
We find no evidence for a statistically significant deviation from Einstein's general relativity. 
We find $\xi < 0.064$ ($|\xi| < 0.011$) at 95\% CL for $\xi > 0$ (for $\xi < 0$, $\xi \ne -1/6$). 
In terms of post-Newtonian parameters, we find $0.995 < \gamma_{\rm PN} < 1$ 
and $0.99987 < \beta_{\rm PN} < 1$ ($0.997 < \gamma_{\rm PN} < 1$ and 
$1 < \beta_{\rm PN} < 1.000011$) for $\xi >0$ (for $\xi < 0$).
For the particular case of the conformal coupling, i.e. $\xi=-1/6$, we find constraints on the 
post-Newtonian parameters of similar precision to those within the Solar System.
\end{abstract}



\maketitle

\section{Introduction}
\label{sec:intro}
The astrophysical and cosmic tests for the change of the fundamental physical constants are 
improving thanks to the increasing precision of observations \cite{Uzan:2010pm,Ade:2013zuv}. 
In most of the cases these tests cannot compete with the precision which can be achieved in 
laboratories, but can probe lengths and/or timescales otherwise unaccessible on ground. 
There are however exceptions: for instance, current cosmological data can constrain the time 
variation of the Newtonian constant at the same level of experiments within the Solar System 
such as the Lunar Laser ranging \cite{Umilta:2015cta,Ballardini:2016cvy}. 

As far as cosmological tests are concerned, one of workhorse model to test deviations from 
general relativity (GR) is the extended Jordan-Brans-Dicke (eJBD) \cite{Jordan:1949zz,Brans:1961sx} 
theory, which has been extensively studied 
\cite{Chen:1999qh,Nagata:2003qn,Acquaviva:2004ti,Avilez:2013dxa,Li:2013nwa,Ooba:2016slp,Umilta:2015cta,Ballardini:2016cvy}.
eJBD is perhaps the simplest extension of GR within the more general Horndeski theory 
\cite{Horndeski:1974wa}:
\begin{align}
\label{eqn:action}
S = &\int \dd^{4}x \sqrt{-g} \Bigl[ G_2(\sigma,\chi) + G_3(\sigma,\chi)\square \sigma \notag\\
&+G_4(\sigma,\chi) \, R 
-2 G_{4,\chi}(\sigma,\chi)  \left(\square\sigma^2-\sigma^{;\mu\nu}\sigma_{;\mu\nu}\right) \notag\\
&+G_5(\sigma,\chi) G_{\mu\nu}\sigma^{;\mu\nu} 
+\frac{1}{3}G_{5,\sigma}(\sigma,\chi)
\left(\square\sigma^3 \right.\notag\\
&\left.-3 \sigma_{;\mu\nu}\sigma^{;\mu\nu} \square\sigma
+2\sigma_{;\mu\nu}\sigma^{;\nu \rho}{\sigma^{;\mu}}_{;\rho}\right) + {\cal L}_m \Bigr] \,,
\end{align}
where $\chi = -g^{\mu\nu} \partial_\mu \sigma \partial_\nu \sigma$, "$_{;}$" denotes the covariant derivative, $R$ is the Ricci scalar,  
$G_{\mu\nu}=R_{\mu\nu}-g_{\mu\nu} R/2$, and ${\cal L}_m$ is the density Lagrangian for the rest 
of matter. The eJBD theory corresponds to $G_3=G_5=0$, $G_2 = \omega_{\rm BD} \chi/\sigma - V(\sigma)$, 
$G_4 = \sigma$ (in the equivalent induced gravity (IG) formulation with a standard kinetic term 
the two last conditions become $G_2 = \chi/2 - V(\sigma)$, 
$G_4 = \xi \sigma^2/2$ with $\xi = 1/(4 \omega_{\rm BD})$). 

Cosmology puts severe test on eJBD theories.
The constraints from $Planck$ 2015 and a compilation of baryon acoustic oscillations (BAO) data 
lead to a 95\% CL upper bound $\xi < 0.00075$, weakly dependent on the index for a power-law 
potential \cite{Ballardini:2016cvy} (see \cite{Umilta:2015cta} for the $Planck$ 2013 
constraints obtained with the same methodology). In terms of the first post-Newtonian parameter 
$\gamma_{\rm PN} = (1+\omega_{\rm BD})/(2+\omega_{\rm BD}) = (1+4 \xi)/(1+8\xi)$, the above 95\% CL 
constraint read as $|\gamma_{\rm PN}-1| < 0.003$ \cite{Ballardini:2016cvy}.
With the same data, a 95\% CL bound is obtained on the relative time variation of the effective 
Newton's constant $10^{13}|\dot G_\textup{eff}/G_\textup{eff}| \lesssim 6\times 10^{-3}\ H_0$ 
at 95\% CL with an index for a power-law potential in the range $[0,8]$. 
The combination of future measurement of CMB anisotropies in temperature, polarization and lensing 
with Euclid-like (galaxy clustering and weak lensing) data can lead to constraints on 
$\gamma_{\rm PN}$ at a slightly larger level than the current Solar System constraints 
\cite{Ballardini:2019tho} (see also \cite{Alonso:2016suf} for forecasts 
for different experiments with different assumptions).

However, theoretical priors can play an important role in the derivation of the cosmological 
constraints and need to be taken into account in the comparison with other astrophysical or 
laboratory tests. Indeed, for eJBD theories only the first post-Newtonian parameter $\gamma_{\rm PN}$ 
is nonzero and fully encodes the deviations from GR, being the second post-Newtonian parameter 
$\beta_{\rm PN} \propto \dd \gamma_{\rm PN}/\dd \sigma$. In this paper we wish to go beyond the 
working assumption of $\beta_{\rm PN}=0$ implicit within in eJBD theories. For this purpose we 
therefore consider nonminimally coupled (NMC) scalar fields with $2G_4  = N_{\rm pl}^2 + \xi \sigma^2$ 
as a minimal generalization of the eJBD theories. 
NMC with this type of coupling are also known as extended quintessence models in the context of 
dark energy \cite{Uzan:1999ch,Perrotta:1999am,Bartolo:1999sq,Amendola:1999qq,Chiba:1999wt}. 
As for eJBD, NMC are also within the class of Horndeski theories consistent with the constraints 
on the velocity of propagation of gravitational waves \cite{Baker:2017hug,Creminelli:2017sry,Ezquiaga:2017ekz} which followed the 
observation of GW170817 and its electromagnetic counterpart \cite{ligobinary} (see also \cite{Lombriser:2015sxa,Lombriser:2016yzn}).

The outline of this paper is as follows. In Section~\ref{sec:two} we discuss the background 
dynamics and the post-Newtonian parameters $\gamma_{\rm PN}$ and $\beta_{\rm PN}$ for this 
class of scalar-tensor theories. We study the evolution of linear fluctuations in 
Section~\ref{sec:three}. We show the dependence on $\xi$ of the CMB anisotropies power spectra 
in temperature and polarization in Section~\ref{sec:four}.
We present the $Planck$ and BAO constraints on these models in Section~\ref{sec:five}.
We conclude in Section~\ref{sec:conclusion}. The initial conditions for background and 
cosmological fluctuations are collected in Appendix~\ref{sec:appendix_A}.

\section{Dark Energy as an effectively massless scalar field non-minimally coupled to gravity}
\label{sec:two}

We study the restriction of the Horndeski action \eqref{eqn:action} to a standard kinetic term 
and $G_3 = G_5 = 0$. We also assume: 
\begin{equation}
2G_4 \equiv F(\sigma)=N_\mathrm{pl}^2+\xi\sigma^2 \,,
\end{equation}
where $\xi$ is the coupling to the Ricci scalar which is commonly used in extended quintessence 
\cite{Uzan:1999ch,Perrotta:1999am,Bartolo:1999sq,Amendola:1999qq,Chiba:1999wt}. 
For simplicity we denote by a tilde the quantities normalized to $M_{pl} \equiv 1/\sqrt{8\pi G}$, 
where $G = 6.67 \times 10^{-8}$ cm$^3$ g$^{-1}$ s$^{-2}$ is the gravitational constant measured 
in a Cavendish-like experiment.
We also introduce the notation $\tilde{N}_{pl} \equiv 1 \mp \Delta\tilde{N}_{pl}$ for $\xi \gtrless 0$.

The field equations are obtained by varying the action with respect to the metric:
\be
\label{eqn:EE}
\begin{split}
G_{\mu\nu}&=\frac{1}{F(\sigma)}\left[T_{\mu\nu}+\partial_{\mu}\sigma\partial_{\nu}\sigma
-\frac{1}{2}g_{\mu\nu}\partial^{\rho}\sigma\partial_{\rho}\sigma\right.\\
&\left.\frac{}{}-g_{\mu\nu}V(\sigma)+(\nabla_{\mu}\nabla_{\nu}-g_{\mu\nu}\Box)F(\sigma)\right].
\end{split}
\ee
The Einstein trace equation results:
\be
R=\frac{1}{F}\left[-T+\partial_\mu\sigma\partial^\mu\sigma+4V+3\square F\right] ,
\ee
where $T$ is the trace of the energy-momentum tensor.
The Klein-Gordon (KG) equation can be obtained varying the action with respect to the scalar field:
\be
\label{eqn:KGFR}
-\square\sigma-\frac{1}{2}F_{,\sigma}R+V_{,\sigma} = 0 ,
\ee
and substituting the Einstein trace equation one obtains:
\be
\begin{split}
&-\square \sigma \left(1+\frac{3}{2}\frac{F_{,\sigma}^2}{F}\right)+V_{,\sigma}
-2\frac{V F_{,\sigma}}{F} \\
&+\frac{F_{,\sigma}}{2F}
\left[T-\partial_\mu\sigma\partial^\mu\sigma\left(1+3F_{,\sigma\sigma}\right)\right]=0 \,.
\end{split}
\ee
In this paper, we do not consider a quintessence-like inverse power-law potential (see for 
instance \cite{Uzan:1999ch,Perrotta:1999am,Bartolo:1999sq,Chiba:1999wt}), but we restrict 
ourselves to a potential of the type $V\propto F^2$ in which the scalar field is effectively 
massless. This case generalizes the broken scale invariant case \cite{Wetterich:1987fm,CV,Finelli:2007wb} to NMC
and is a particular case of the class of models with $V\propto F^M$ admitting scaling solutions 
\cite{Amendola:1999qq}.
Note that though for the form of $F(\sigma)$ used in the paper and for large values of $\sigma$, 
this potential looks similar to that in the Higgs inflationary model~\cite{Bezrukov:2007ep}, 
in fact it is crucially different, since it is exactly flat in the Einstein frame\footnote{Although 
our work is based in the original Jordan frame, it is also useful to think about this class 
of theories in the dual Einstein frame where 
$\hat g_{\mu\nu} \propto F g_{\mu\nu} \,, \hat V = {V}/{F^2}$.}
in the absence of other matter and cannot support a metastable inflationary stage in the early Universe. Contrary, this model may be used for description of dark energy in the present Universe.

\subsection{Background cosmology}

\begin{figure}[!h]
\centering
\includegraphics[width=.85\columnwidth]{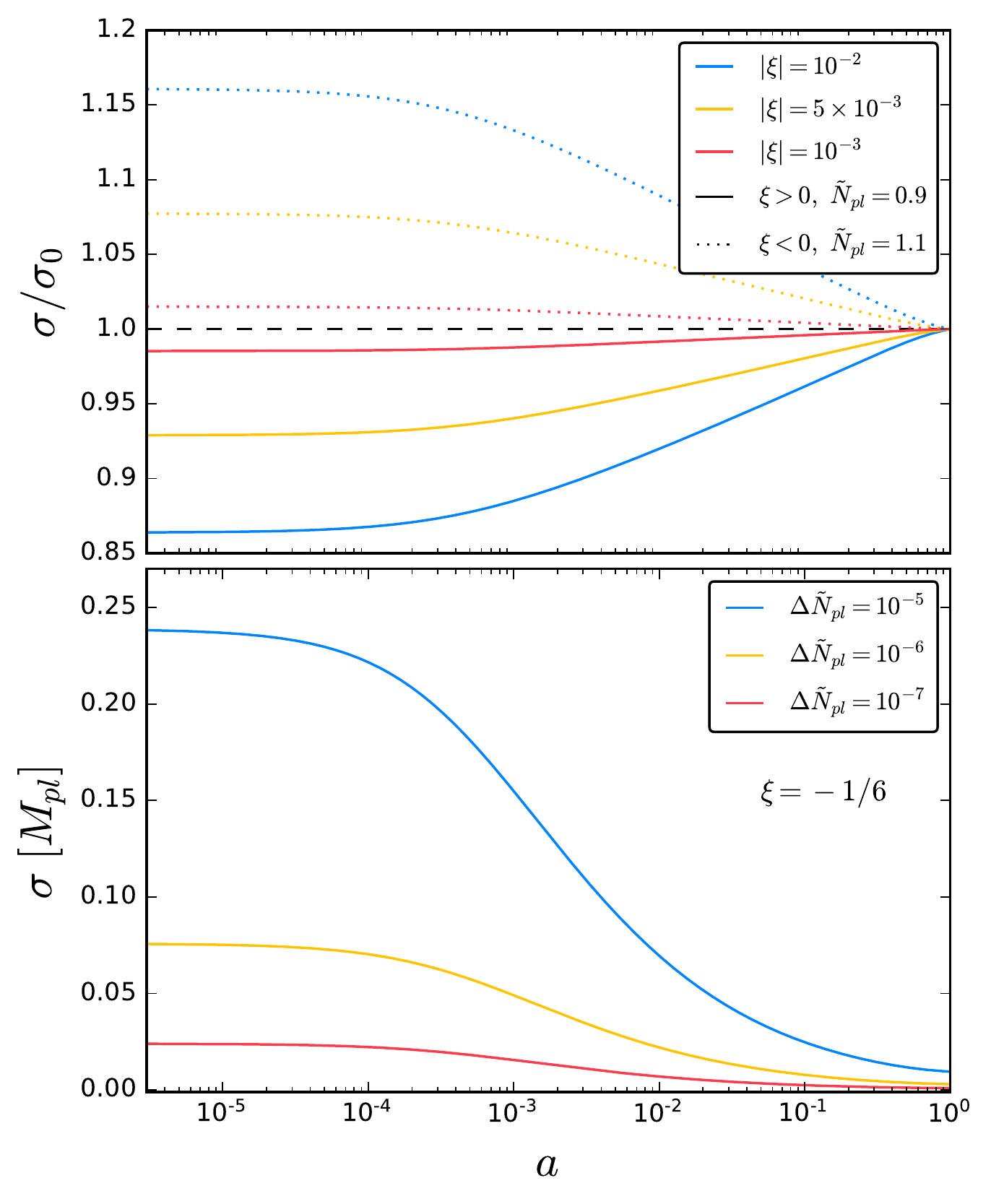}
\caption{Top panel: relative evolution of $\sigma$ for different values of $\xi$.
Bottom panel: evolution of $\sigma$ for different values of $N_{pl}$ for the CC case,
i.e. $\xi=-1/6$.}
\label{fig:phi}
\end{figure}

We consider cosmic time and a flat FLRW metric, for which the unperturbed cosmological 
spacetime metric is given by:
\be
\dd s^2=-\dd t^2+a(t)^2\dd x_i\dd x^i .
\ee
The Friedmann and the KG equations are then given by:
\ba
\label{eqn:NMC_Friedmann}
&3H^{2}F &= \rho+\frac{\dot{\sigma}^{2}}{2}+V(\sigma)-3H\dot{F} \\
& &=\rho+\rho_\sigma, \\
&-2\dot{H}F &= \rho+p+\dot{\sigma}^{2}+\ddot{F}-H\dot{F} \\
& &=\left(\rho+p\right)+\rho_\sigma+p_\sigma.
\ea
\be
\label{eqn:KG}
\begin{split}
\ddot{\sigma} = &-3H\dot{\sigma}+\frac{\xi\sigma}{F+6\xi^{2}\sigma^{2}}\Bigl[\rho_{m}+4V-\frac{F V_{,\sigma}}{\xi\sigma} \\
&-(1+6\xi)\dot{\sigma}^{2}\Bigr].
\end{split}
\ee
In Fig.~\ref{fig:phi} the evolution of the scalar field $\sigma$ is shown for different values 
of $\xi$ for both positive and negative values of the coupling. The natural initial conditions 
for the background displayed in Appendix~\ref{sec:appendix_A} neglect the decaying mode which 
would be rapidly dissipated, but would have destroyed the Universe isotropy at sufficiently 
early times otherwise (see for instance \cite{Finelli:2007wb}). With this natural assumption the scalar 
field is nearly at rest deep in the radiation era whereas it grows (decreases) for positive 
(negative) couplings during the matter era and it reaches a constant value at recent times.
During the matter dominated era in the regime $\xi \sigma^2 \ll N_{\mathrm pl}^2$ 
(which is the only one allowed by observations, see Section V), the evolution of the scalar field 
can be approximated as $\sigma \sim \sigma_i \left[ 1 + 2 \xi (\ln a + 8) \right]$, with 
$\sigma_i$ being the initial value of the scalar field in the radiation era. 
In the bottom panel, we show the evolution of the scalar field for the conformal coupling (CC) 
case with $\xi=-1/6$ for different values of $N_{\mathrm pl}$. In this case the field is always 
sub-Planckian for $\Delta \tilde{N}_{\mathrm pl} \lesssim 0.0005$.

The above equations lead to the straightforward associations: 
\begin{align}
\rho_{\sigma}&=\frac{\dot{\sigma}^{2}}{2}+V(\sigma)-3H\dot{F} \notag\\
&=\frac{\dot{\sigma}^{2}}{2}+V(\sigma)-6H\xi\sigma\dot{\sigma} ,
\end{align}
\be
\begin{split}
p_{\sigma}=&\frac{\dot{\sigma}^{2}}{2}\left[\frac{F(1+4\xi)+2\xi^{2}\sigma^{2}}{F+6\xi^{2}\sigma^{2}}\right]-2H\xi\sigma\dot{\sigma} \\
&+\frac{2\xi^{2}\sigma^{2}}{F+6\xi^{2}\sigma^{2}}\left(\rho_{m}+4V-\frac{FV_{,\sigma}}{\xi\sigma}\right)-V,
\end{split}
\ee
where in the equation for $p_{\sigma}$ we have explicitly substituted the KG equation. 
We can recover an expression for the dark energy (DE) density parameter dividing $\rho_{\sigma}$ 
for the quantity $3H^2 F$ which represents the critical density. 

Alternatively, it is also convenient to define new density parameters in a framework which mimics 
Einstein gravity at present and satisfy the conservation law 
$\dot \rho_{\rm DE}+3H(\rho_{\rm DE}+p_{\rm DE})=0$ \cite{Boisseau:2000pr,Gannouji:2006jm}:
\ba
\label{rhoandp}
\rho_{\rm DE}&=&\frac{F_{0}}{F}\rho_{\sigma}+(\rho_{m}+\rho_{r})\left(\frac{F_{0}}{F}-1\right) \,, \\
p_{\rm DE}&=&\frac{F_{0}}{F}p_{\sigma}+p_r\left(\frac{F_{0}}{F}-1\right) \,.
\ea
The effective parameter of state for DE can be defined as $w_{\rm DE} \equiv p_{\rm DE}/\rho_{\rm DE}$.

\begin{figure}[!h]
\centering
\includegraphics[width=.85\columnwidth]{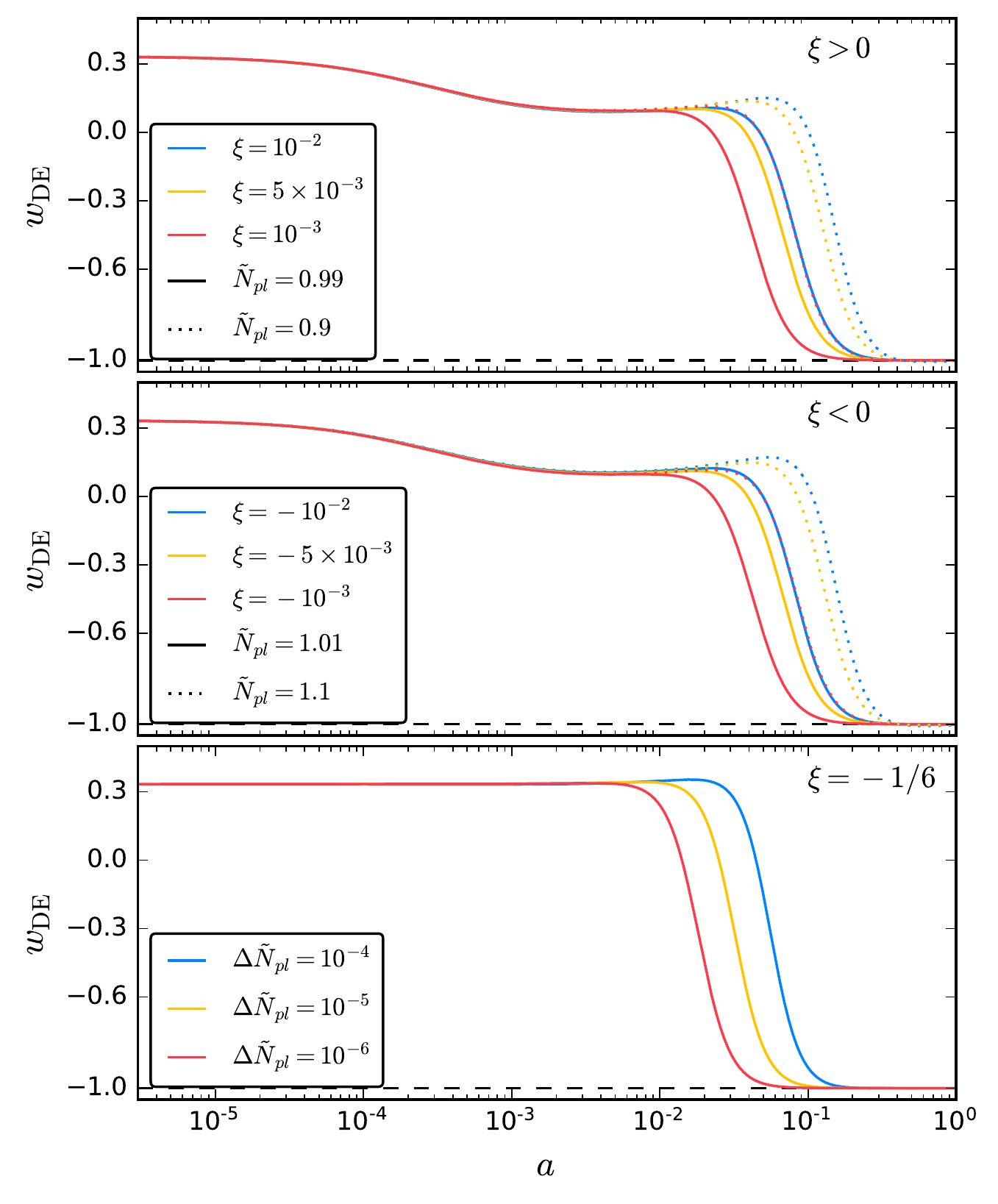}
\caption{Evolution of $w_{\rm DE}$ for different values of $N_{pl}$ and $\xi$.
We plot the effective parameter of state for DE for $\xi>0$ in the upper panel, 
$\xi<0$ in the central panel, and the CC case $\xi=-1/6$ in the bottom panel.}
\label{fig:w}
\end{figure}

In Fig.~\ref{fig:w} the evolution of this effective parameter of state is shown for different 
values of the parameters $N_{pl}$ and $\xi$. In all the cases the parameter of state $w_{\rm DE}$ mimics $1/3$ ($-1$) in the relativistic 
era (at late times): this behaviour can be easily understood from Eq. (\ref{rhoandp}) when $\rho_r$ ($V(\sigma)$) dominates over the energy densities 
of other components. The behavious of $w_{\rm DE}$ at the onset of the matter dominated era is instead model dependent: 
for $\xi \ne -1/6$, we see that $w_{\rm DE} > 0$ from the upper two panels in Fig.~\ref{fig:w}, whereas for $\xi = -1/6$ we obtain 
$w_{\rm DE} \sim 1/3$ when $\sigma_0 \ll \sigma$.
%
%
The absence of an intermediate phase of a matter dominated era for $\xi = -1/6$ is also clear 
in the initial conditions for the scale factor reported in Appendix~\ref{sec:appendix_A}.
It can be seen from Fig.~\ref{fig:w} that there is no phantom behaviour of the effective DE 
component at small redshifts in contrast to more general scalar-tensor DE models studied in 
\cite{Gannouji:2006jm}. 
Indeed a phantom behaviour with $w_{\rm DE} < -1$ \cite{Gannouji:2006jm} is barely visible 
in the transient regime from the tracking value to $w_{\rm DE} \approx -1$ because of the small 
coupling $\xi$ considered in Fig.~\ref{fig:w} and cannot occur in the stable accelerating regime 
for these models with $V(\sigma) \propto F^2 (\sigma)$.

\begin{figure}[!h]
\centering
\includegraphics[width=.85\columnwidth]{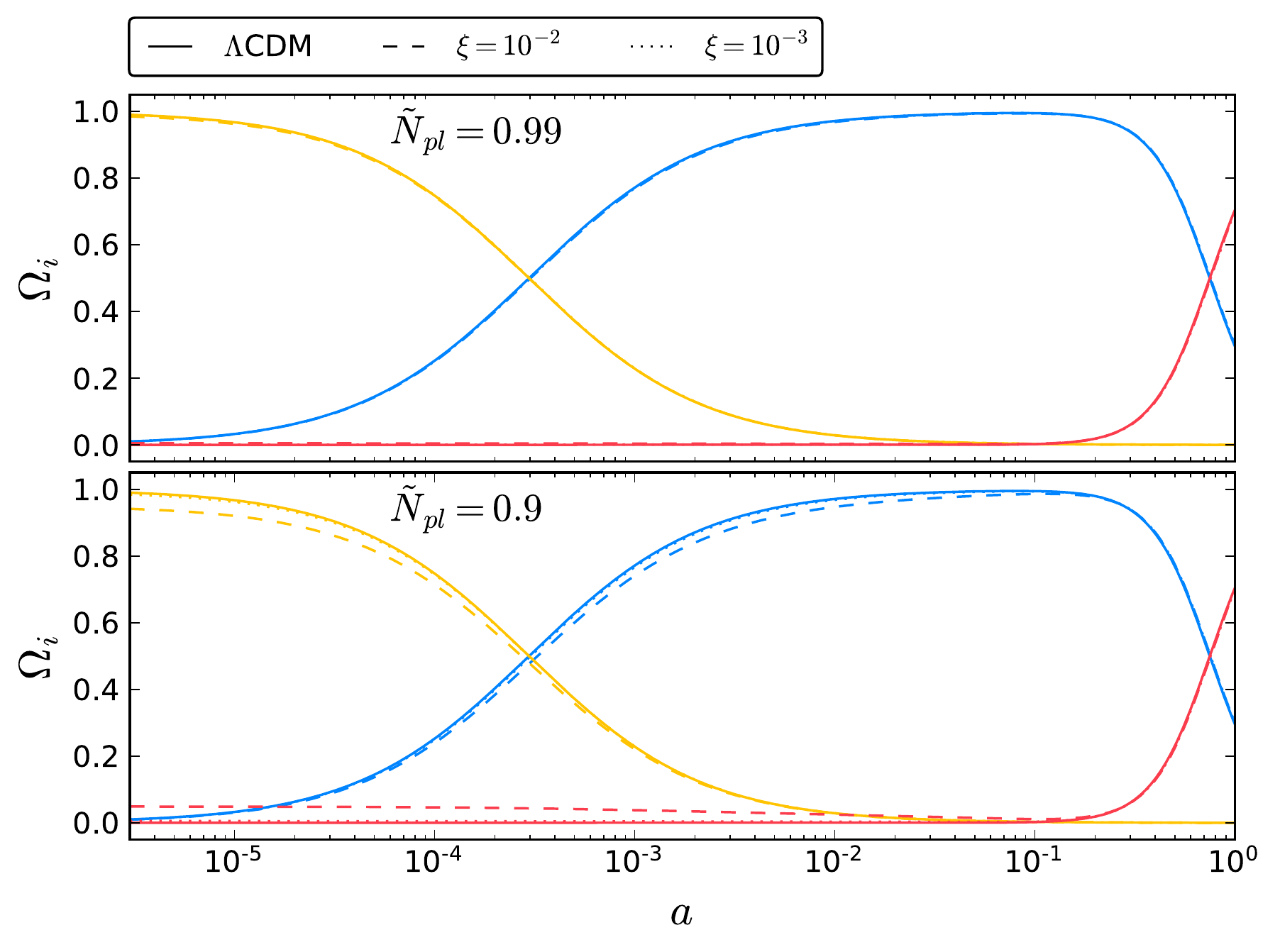}
\caption{Evolution of the density parameters $\Omega_i$: radiation in yellow, 
matter in blue, and effective DE in red.
We plot $\tilde{N}_{pl}=1$ ($\tilde{N}_{pl}=0.9$) for $\xi = 10^{-2},\,10^{-3}$ 
in the top (bottom) panel.}
\label{fig:omega_xp}
\end{figure}

\begin{figure}[!h]
\centering
\includegraphics[width=.85\columnwidth]{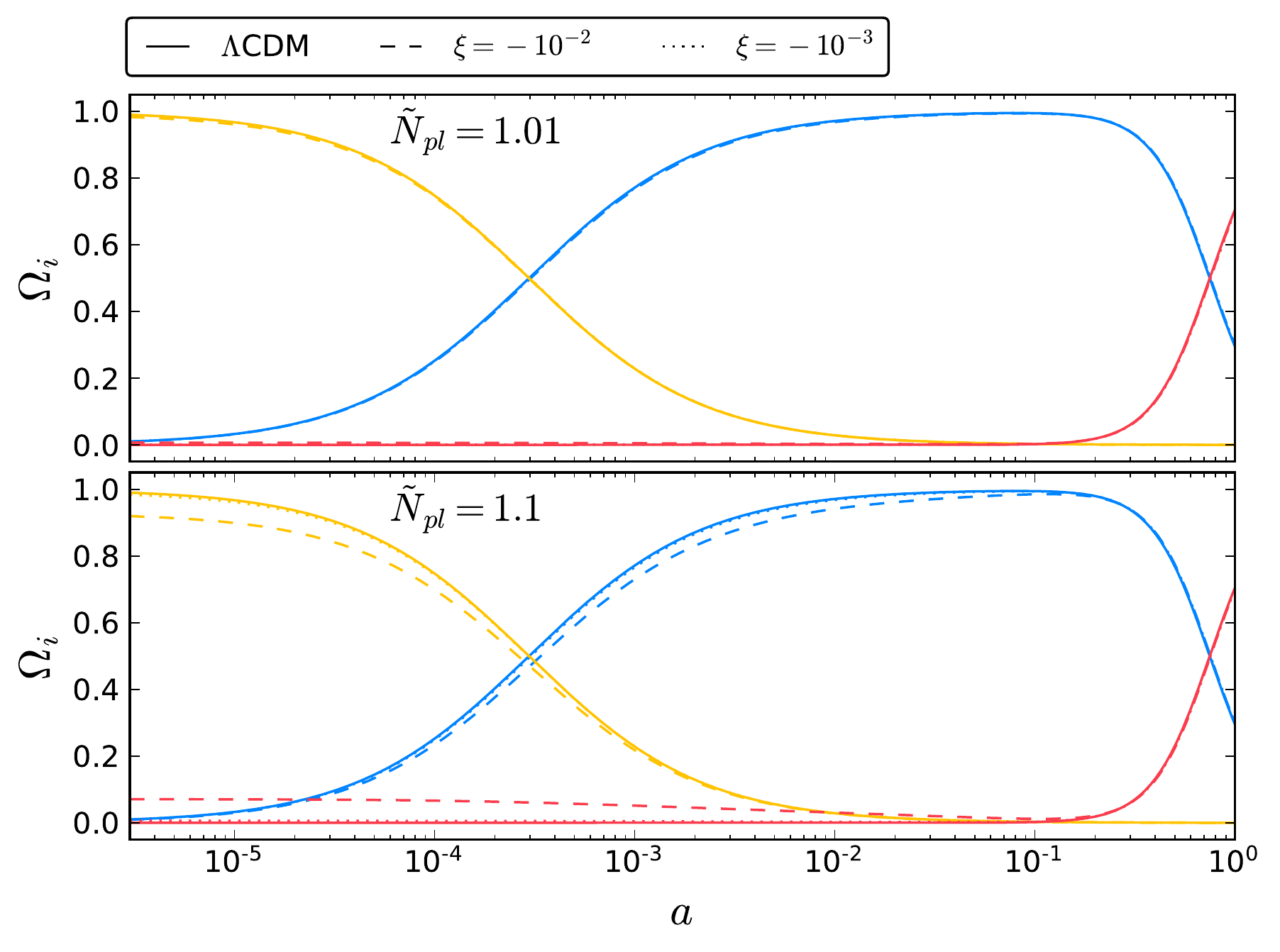}
\caption{Evolution of the density parameters $\Omega_i$: radiation in yellow, 
matter in blue, and effective DE in red.
We plot $\tilde{N}_{pl}=1.01$ ($\tilde{N}_{pl}=1.1$) for $\xi = -10^{-2},\,-10^{-3}$ 
in the top (bottom) panel.}
\label{fig:omega_xn}
\end{figure}

\begin{figure}[!h]
\centering
\includegraphics[width=.85\columnwidth]{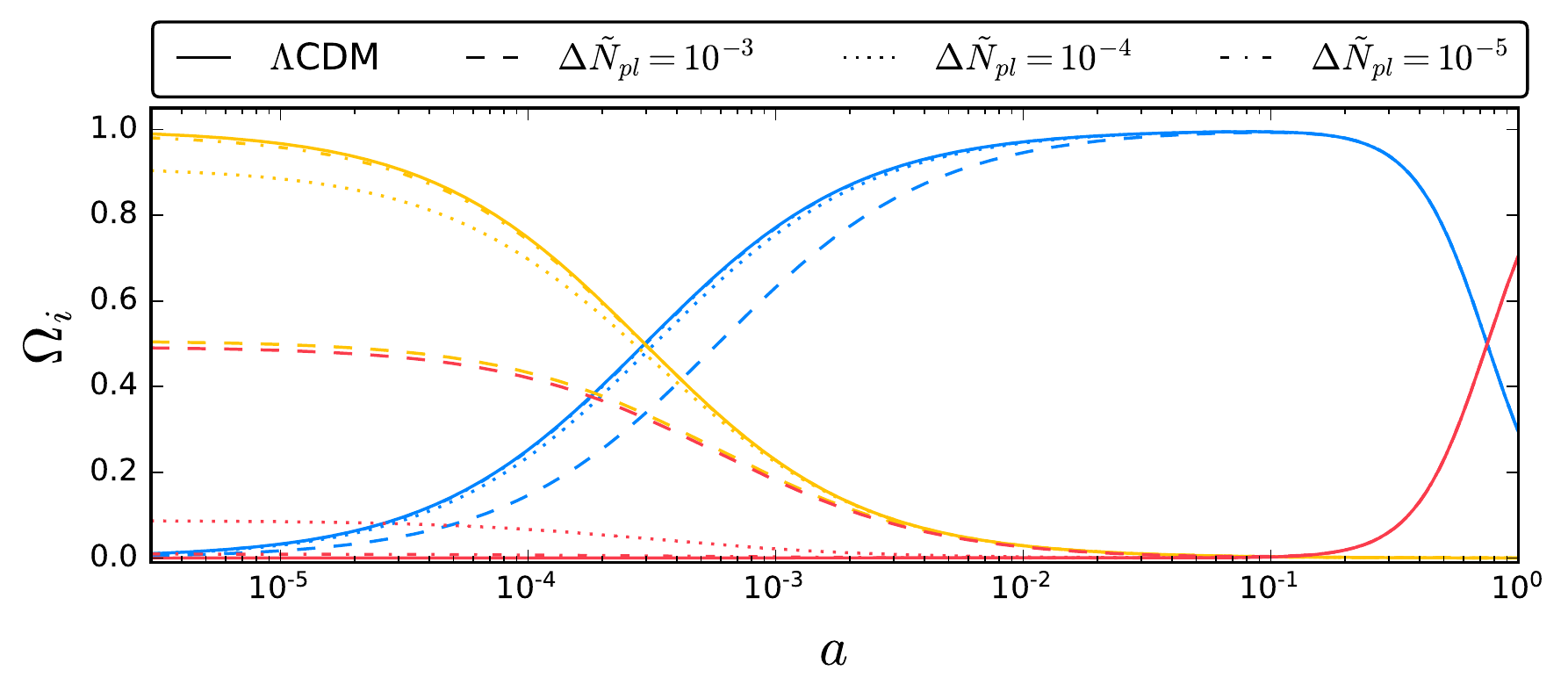}
\caption{Evolution of the density parameters $\Omega_i$: radiation in yellow, 
matter in blue, and effective DE in red.
We plot the CC case $\xi = -1/6$ for 
$\Delta\tilde{N}_{pl}=10^{-3},\,10^{-4},\,10^{-5}$.}
\label{fig:omega_cc}
\end{figure}

In Figs.~\ref{fig:omega_xp}-\ref{fig:omega_xn}-\ref{fig:omega_cc}, we show the 
evolution of the density parameters $\Omega_i$, corresponding to an Einstein 
gravity system with a Newton’s constant given by the current value of the scalar 
field today $G_N=1/(8\pi F_0)$ \cite{Boisseau:2000pr} (also used in \cite{Finelli:2007wb,Umilta:2015cta}) for $\xi >0$, $\xi <0$ and 
$\xi = -1/6$, respectively.

\subsection{Boundary conditions for the the scalar field}

As boundary conditions we impose that the effective Newton's constant at present is compatible 
with the Cavendish-like experiments. The effective gravitational constant for NMC is given 
by \cite{Boisseau:2000pr}:
\be
\label{eqn:Geff}
G_{\mathrm{eff}}=\frac{1}{8\pi F}\left(\frac{2F+4F_{,\sigma}^{2}}{2F+3F_{,\sigma}^{2}}\right) .
\ee  

\begin{figure}[!h]
\centering
\includegraphics[width=.85\columnwidth]{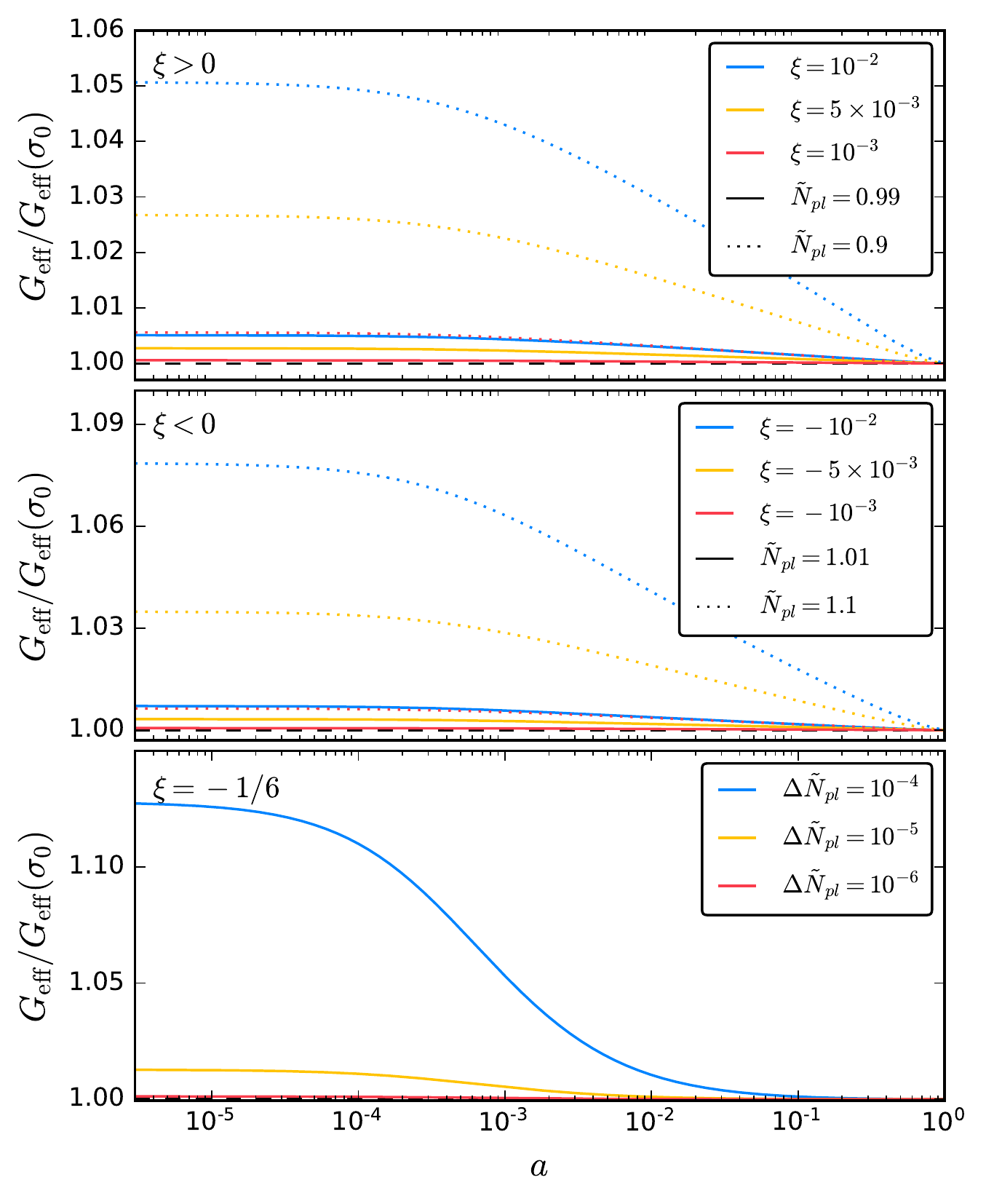}
\caption{Evolution of the effective gravitational constant $G_{\rm eff}$ relative
to its value today for different values of $N_{pl}$ and $\xi$. From top to bottom, the cases
with $\xi > 0$, $\xi < 0$, and $\xi = -1/6$ are displayed, respectively.}
\label{fig:Geff}
\end{figure}

In Fig.~\ref{fig:Geff} is shown the evolution of the relative effective gravitational
constant \eqref{eqn:Geff}. We can see that the effective gravitational constant decreases
in time for all the choices of both $N_{pl}$ and $\xi$.

We can distinguish three different cases beyond GR:
\begin{itemize}
\item $\tilde{N}_{pl}\rightarrow 0$ which is the IG case. This leads to:
\be
\tilde{\sigma}_0^2=\frac{1}{\xi}\frac{1+8\xi}{1+6\xi} ,
\ee
which is the same result as obtained in \cite{Umilta:2015cta};

\item $\xi\rightarrow-1/6$ which is the CC. In this particular case the polynomial equation (\ref{eqn:Geff}) 
in $\sigma_0$ in quadratic and we have:
\be
\tilde{\sigma}_{0}^2=\frac{18\tilde{N}_{pl}^2(\tilde{N}_{pl}^2-1)}{1+3\tilde{N}_{pl}^2} ;
\ee
\item a general NMC case for $\xi \ne -1/6$:
\be
\begin{split}
\tilde{\sigma}_{0}^{2}=&\frac{1-2\tilde{N}_{pl}^{2}+2\xi(4-3\tilde{N}_{pl}^{2})}{2\xi(1+6\xi)} \\
&\pm\frac{\sqrt{1-4\xi(5\tilde{N}_{pl}^{2}-4)+4\xi^{2}(3\tilde{N}_{pl}^{2}-4)^{2}}}{2\xi(1+6\xi)} .
\end{split}
\ee
\end{itemize}
By requiring $\tilde{\sigma}^2\geq0$ and $F\geq 0$, 
we obtain conditions on the two parameters $\tilde{N}_{pl}$ and $\xi$ for the physical  
solution:
\begin{eqnarray}
\tilde{N}_{pl} & < & 1 \, \,\,\text{for}\,\, \, \xi>0 \,, \\
\tilde{N}_{pl} & > & 1 \, \,\,\text{for} \,\,\, \xi<0 \,.
\end{eqnarray}

\subsection{Comparison with general relativity}

The deviations from GR for a theory of gravitation are described by the so called post-Newtonian 
parameters. For NMC only the parameters $\gamma_{\rm PN}$ and $\beta_{\rm PN}$ differ from GR 
predictions, for which they both equal unity. In terms of these parameters the line element can 
be expressed as:
\be
\dd s^2=-(1+2\Phi-2\beta_{\rm PN}\Phi^2)\dd t^2+(1-2\gamma_{\rm PN}\Phi)\dd x_i \dd x^i.
\ee
These parameters are given within NMC by the following equations \cite{Boisseau:2000pr}: 
\ba
\label{eqn:gammaPN}
&\gamma_{\rm PN}&=1-\frac{F_{,\sigma}^{2}}{F+2F_{,\sigma}^{2}},\\
\label{eqn:betaPN}
&\beta_{\rm PN}&=1+\frac{FF_{,\sigma}}{8F+12F_{,\sigma}^{2}}\frac{\dd\gamma_{\rm PN}}{\dd\sigma}.
\ea
We have $\gamma_{\rm PN} \leq 1$ and $\beta_{\rm PN}\leq1$ for $\xi >0$, whereas 
$\gamma_{\rm PN} \leq 1$ and $\beta_{\rm PN}\geq1$ for $\xi <0$. 

\begin{figure}[!h]
	\centering
	\includegraphics[width=.85\columnwidth]{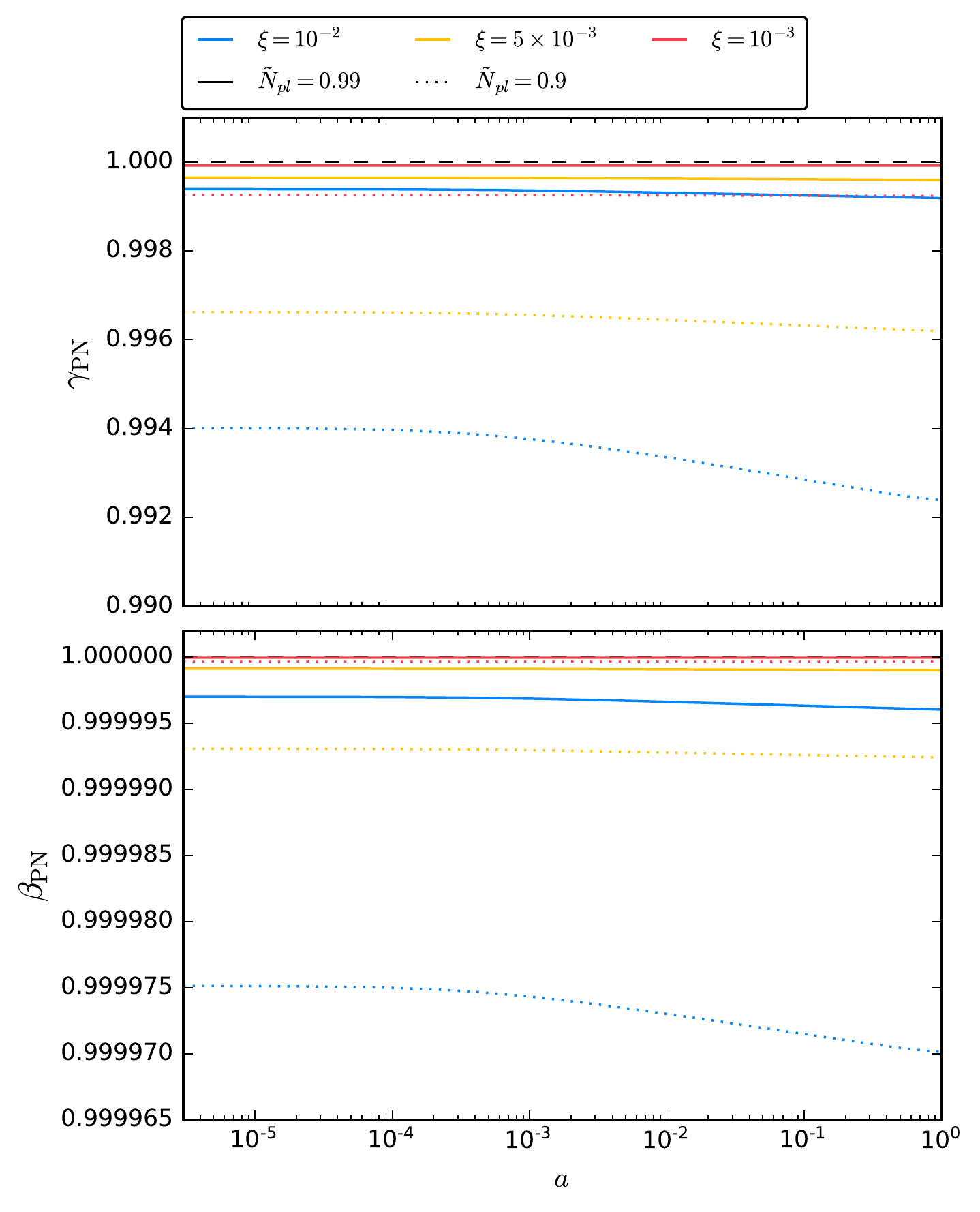} 
	\caption{Evolution of the post-Newtonian parameters $\gamma_{\rm PN}$ and $\beta_{\rm PN}$ 
		for different values of $N_{pl}$ and $\xi$. We show the case with $\xi>0$.}
	\label{fig:ppnPOS}
\end{figure}

\begin{figure}[!h]
	\centering
	\includegraphics[width=.85\columnwidth]{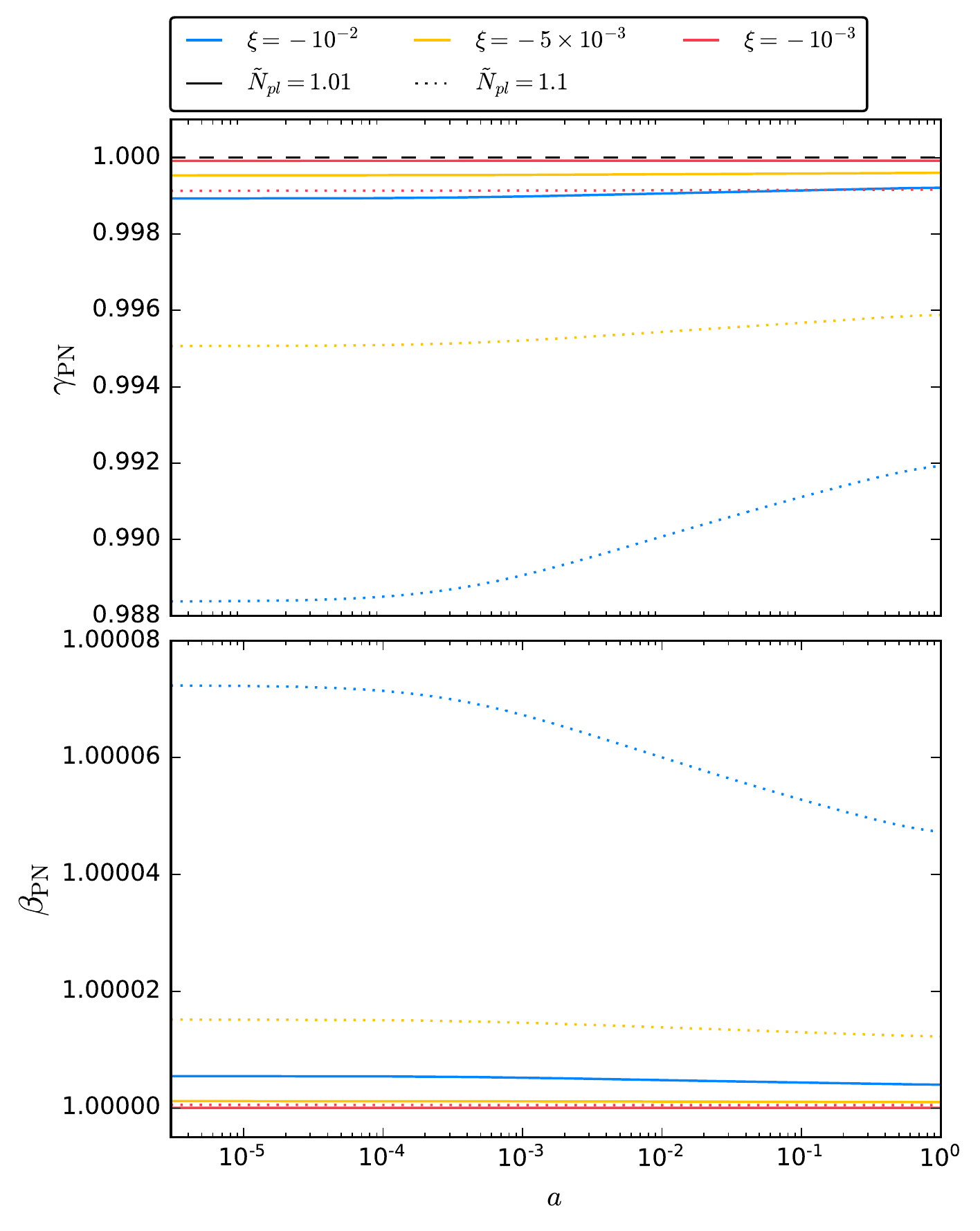}
	\caption{Evolution of the post-Newtonian parameters $\gamma_{\rm PN}$ and $\beta_{\rm PN}$
		for different values of $N_{pl}$ and $\xi$. We show the NMC case with $\xi>0$.}
	\label{fig:ppnNEG}
\end{figure}

\begin{figure}[!h]
	\centering
	\includegraphics[width=.85\columnwidth]{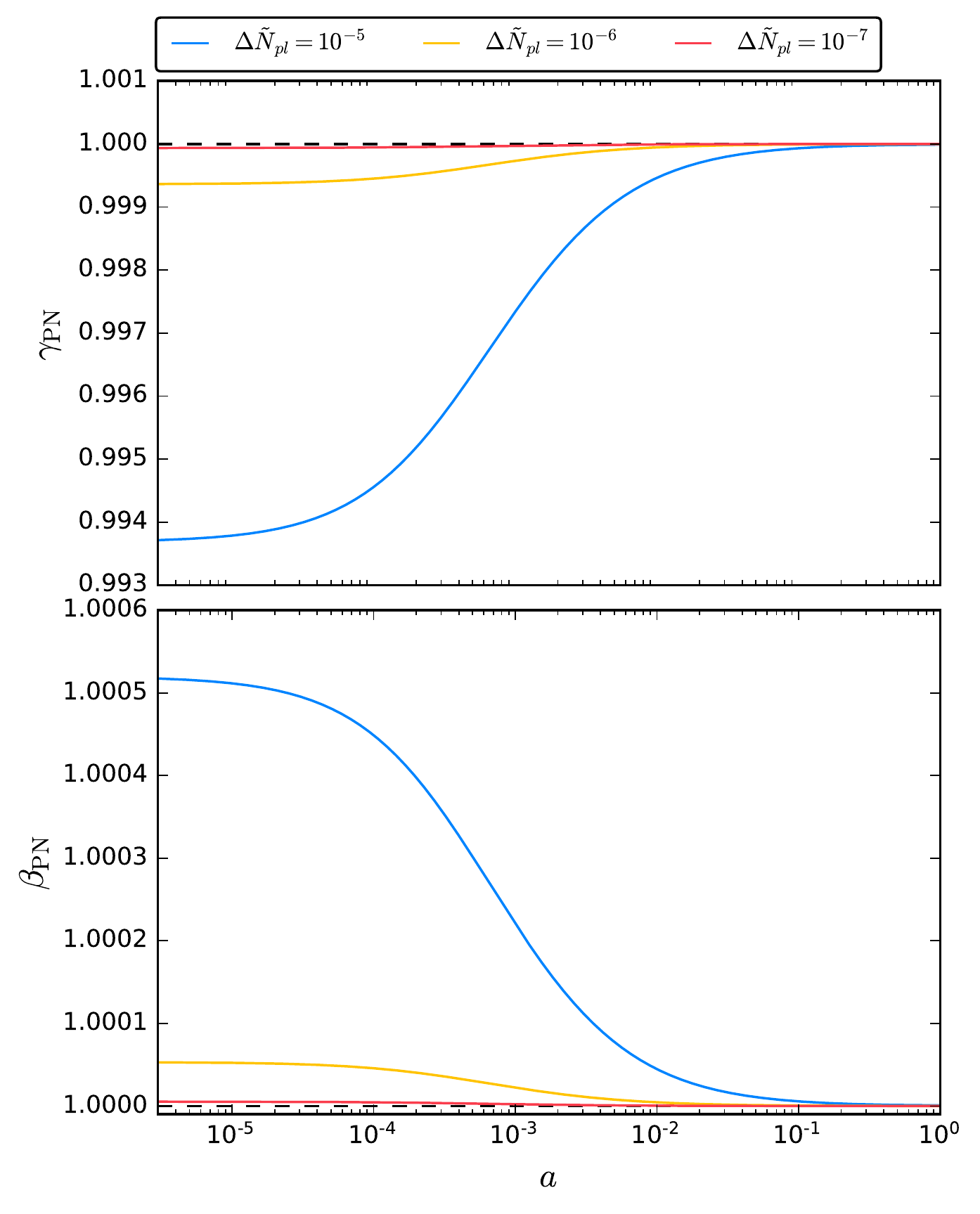}
	\caption{Evolution of the post-Newtonian parameters $\gamma_{\rm PN}$ and $\beta_{\rm PN}$
		for different values of $N_{pl}$. We show the NMC case  $\xi=-1/6$, i.e. the CC case.}
	\label{fig:ppnCC}
\end{figure}

In Figs.~\ref{fig:ppnPOS}, \ref{fig:ppnNEG} and \ref{fig:ppnCC}, we show the evolution of 
these parameters for different values of $N_{pl}$ and $\xi$. It is interesting to note how in the 
CC case $\gamma_{\rm PN}$ and $\beta_{\rm PN}$ approach the GR value more rapidly than for 
$\xi \ne -1/6$.

\section{Linear perturbations}
\label{sec:three}

\begin{figure}[!h]
\centering
\includegraphics[width=.85\columnwidth]{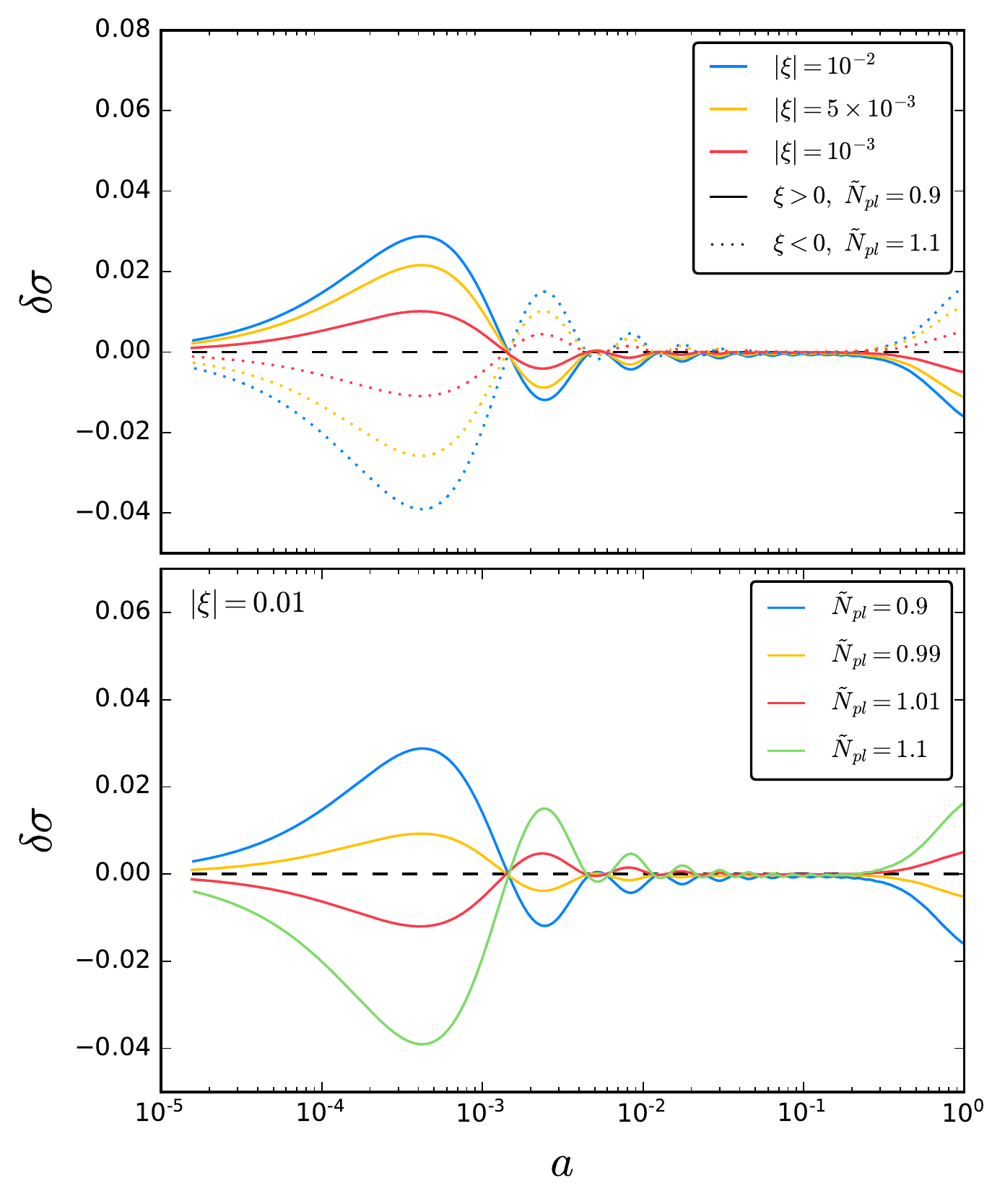}
\caption{Evolution of scalar field perturbations in the synchronous gauge for $k=0.05$ Mpc$^{-1}$.}
\label{fig:fieldfluctuation}
\end{figure}

We study linear fluctuations around the FRW metric in the synchronous gauge:
\be
\dd s^2=a(\tau)^2\left[-\dd\tau^2+(\delta_{ij}+h_{ij})\dd x^i\dd x^j\right],
\ee
where $\tau$ is the conformal time and $h_{ij}$ include both the scalar ($h_{ij}^{S}$) and the 
tensor ($h_{ij}^{T}$) part. We follow the conventions of Ref.~\cite{Ma:1995ey} for scalar metric 
perturbations $h_{ij}$ and scalar field perturbation $\delta \sigma$: 
\begin{align}
h_{ij}^{S}=\int \dd^3k\,e^{i\vec{k}\cdot\vec{x}}\Biggl[&\hat{k}_i\hat{k}_j\,
h(\vec{k},\tau)\notag\\&+\left(\hat{k}_i\hat{k}_j-\frac{1}{3}\delta_{ij}\right)\eta(\vec{k},\tau)\Biggr] ,
\end{align}
\be
\delta \sigma=\int \dd^3k\,e^{i\vec{k}\cdot\vec{x}} \delta \sigma (\vec{k},\tau).
\ee
In Fig.~\ref{fig:fieldfluctuation}, we show the evolution of the scalar field perturbation 
$\delta \sigma$ at $k=0.05$ Mpc$^{-1}$ for different values of $N_{pl}$ and $\xi$.

The modified Einstein equations in Eq.~\eqref{eqn:EE} at first order for scalar perturbations are:
\begin{widetext}
\begin{align}
&\frac{k^2}{a^2}\eta-\frac{1}{2}H\dot{h}=-\frac{1}{2F}\left[\delta\rho+\dot{\sigma}\delta\dot{\sigma}+V_{,\sigma}\delta\sigma-\frac{F_{,\sigma}}{F}\left(\rho+\frac{\dot{\sigma}^2}{2}+V-3H\dot{F}\right)\delta\sigma\right],\\
&\frac{k^2}{a^2}\dot{\eta}=\frac{1}{2F}\left[\sum_i(\rho_i+p_i)\theta_i +k^2\left(\dot{\sigma}\delta\sigma+\delta\dot{F}-H\delta F\right)\right],\\
&\ddot{h}+3H\dot{h}-2\frac{k^2}{a^2}\eta=-\frac{3}{F}\left[p+\dot{\sigma}\delta\dot{\sigma}-V_{,\sigma}\delta\sigma-\frac{F_{,\sigma}}{F}\left(p+\frac{\dot{\sigma}^2}{2}-V+\ddot{F}+2H\dot{F}\right)\delta\sigma\right.\notag \\
&\qquad\qquad\qquad\qquad\left.+\frac{2}{3}\frac{k^2}{a^2}\delta F+\delta\ddot{F}+2H\delta\dot{F}+\frac{1}{3}\dot{h}\dot{F}\right],\\
&\ddot{h}+6\ddot{\eta}+3H(\dot{h}+6\dot{\eta})-2\frac{k^2}{a^2}\eta=-\frac{3}{F}\left[\sum_i(\rho_i+p_i)\sigma_i
+\frac{2}{3}\frac{k^2}{a^2}\delta F+\frac{\dot{F}}{3}(\dot{h}+6\dot{\eta})\right],
\end{align}
\end{widetext}
where all perturbations are considered in the Fourier configuration. The quantities $\theta_i$ 
and $\sigma_i$ represent the velocity potential and the anisotropic stress, respectively.
It can be seen from the last of these equations that the coupling function acts also as a source 
for the anisotropic stress. 

The perturbed Klein-Gordon equation is:
\begin{widetext}
\be
\begin{split}
\delta\ddot{\sigma}=&-\delta\dot{\sigma}\left[3H+\frac{2(1+6\xi)\xi\sigma\dot{\sigma}}{F+6\xi^2\sigma^2}\right] -\delta\sigma\biggl\{\frac{k^2}{a^2}+\frac{FV_{,\sigma,\sigma}}{F+6\xi^2\sigma^2}-\frac{2\xi\sigma V_{,\sigma}}{F+6\xi^2\sigma^2} \biggl[1+\frac{F(1+6\xi)}{F+6\xi^2\sigma^2}\biggr]\\
&+\frac{\xi}{F+6\xi^2\sigma^2} \biggl[1-\frac{2(1+6\xi)\xi\sigma^2}{F+6\xi^2\sigma^2}\biggr]\Big[(1+6\xi)\dot{\sigma}^2-4V +(3p-\rho)\Big]\bigg\}-\frac{\left(3\delta p-\delta\rho\right)\xi\sigma}{F+6\xi^2\sigma^2}-\frac{1}{2}\dot{h}\dot{\sigma}.
\end{split}
\ee
\end{widetext}
As for the homogeneous KG Eq.~\eqref{eqn:KG}, the choice $V\propto F^2$ also leads to an 
effectively massless scalar field fluctuation. 
Both initial conditions for the background and for the linear perturbations at the 
next-to-leading order in $\tau$ are shown in the Appendix~\ref{sec:appendix_A}. 
We consider adiabatic initial condition for the scalar cosmological fluctuations 
\cite{Rossi:2016,Umilta:2015cta}.

\begin{figure}[!h]
\centering
\includegraphics[width=.85\columnwidth]{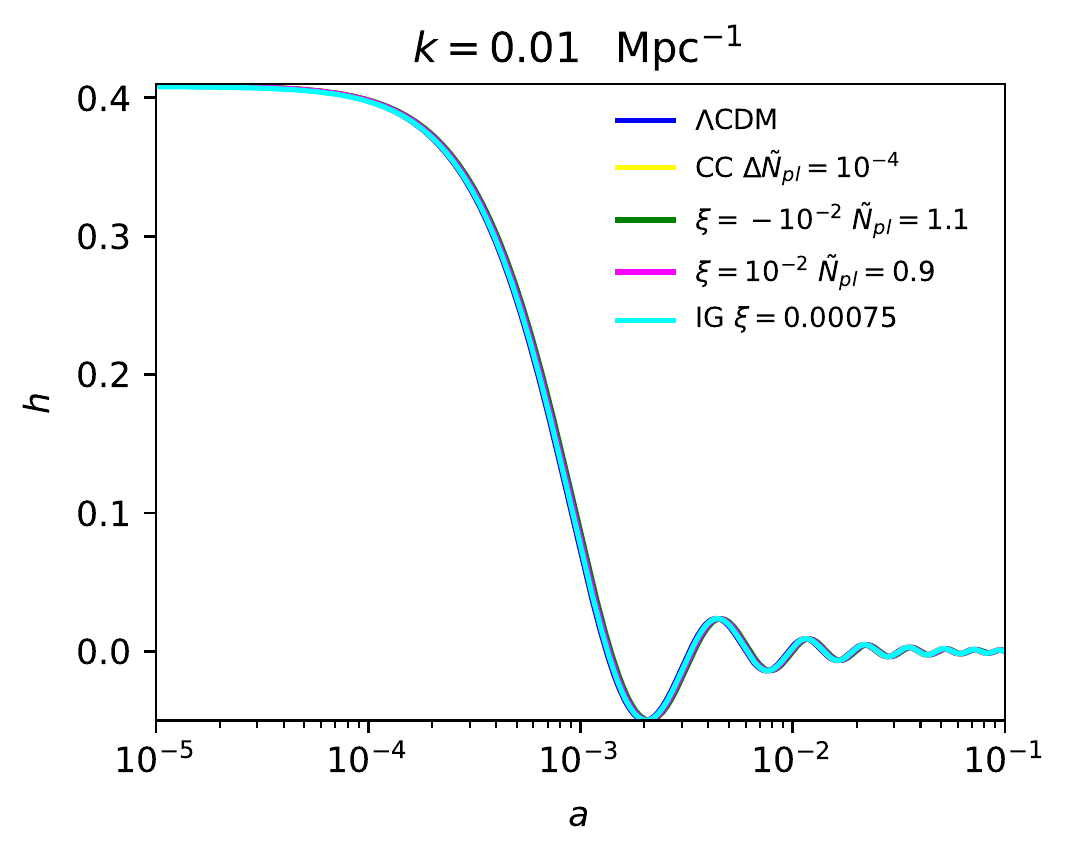}
\caption{Evolution of tensor fluctuations $h^T$ for $k=0.01$ Mpc$^{-1}$.}
\label{fig:gwevolution}
\end{figure}

Analogously, the transverse and traceless part of the metric fluctuation $h_{ij}^{T}$ is expanded as:
\be
h_{i j}^{T} = \int \dd^3k e^{i\vec{k}\cdot\vec{x}}
\left[ h_+ e_{i j}^+ + h_\times e_{i j}^\times \right] \,,
\label{gw}
\ee
where $h_+ \,, h_-$ and $e^+ \,, e^\times$ are the amplitude and normalized tensors of the two independent states to the
direction of propagation of gravitational
waves in Fourier space. The evolution equation for the amplitude is:
\be
\ddot h_{s \,, k} + \left( 3 H + \frac{\dot F}{F} \right) \dot{h}_{s \,, k} + \frac{k^2}{a^2} h_{s \,, k} = \frac{2}{F} \rho_\nu \pi^\nu
\label{eq_motion_gw}
\ee
where $s$ denote the two polarization state of the two independent modes ($s =+,\times$) and the 
right hand side denotes the contribution of the traceless and transverse part of the neutrino 
anisotropic stress.
The importance of the extra-damping term in the evolution equation for gravitational waves has 
been previously stressed \cite{Riazuelo:2000fc,Amendola:2014wma}.
The example of the impact of this term with respect to GR is depicted in Fig.~\ref{fig:gwevolution}. 
Note that the parameters chosen are compatible with the previous figures in this paper and we are 
therefore in a regime in which $\dot F/F \ll 3H$.

\section{CMB anisotropies and matter power spectra}
\label{sec:four}

The footprints of these scalar-tensor theories into the CMB anisotropies angular power spectra 
can be understood as a generalization of the effects in eJBD or equivalently IG theories. 
The redshift of matter-radiation equality is modified in scalar-tensor theories by the motion 
of the scalar field driven by pressureless matter and this results in a shift of the CMB 
acoustic peaks for values $\xi \ne 0$, as for the IG case \cite{Liddle:1998ij,Chen:1999qh}.
In addition, a departure from $\tilde{N}_{pl} = 1$ induces a further change both in the amplitude 
of the peaks and their positions. We note that decreasing the value of $\tilde{N}_{pl}$ is possible 
to suppress the deviations with respect to the $\Lambda$CDM model allowing for larger values of 
the coupling $\xi$ compared to the IG case.

\begin{figure}[!h]
\centering
\includegraphics[width=.85\columnwidth]{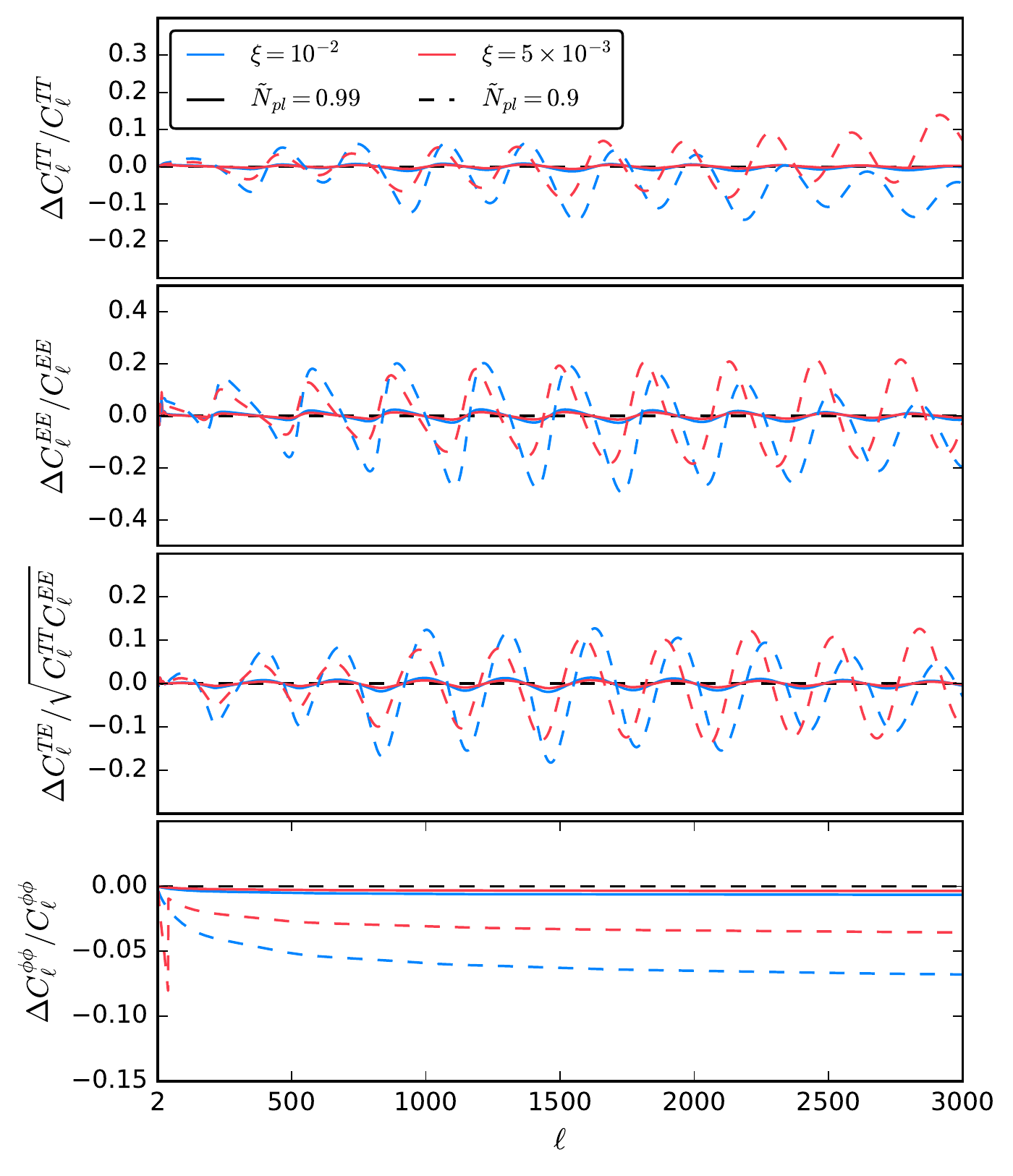}
\caption{From top to bottom: relative differences of the TT-EE-TE-$\phi\phi$ power spectra with respect 
to the $\Lambda$CDM model for $\tilde{N}_{pl}=1,\,0.9$ and different values of $\xi=10^{-2},\,5\times10^{-3}$.}
\label{fig:CMB_xp}
\end{figure}

\begin{figure}[!h]
\centering
\includegraphics[width=.85\columnwidth]{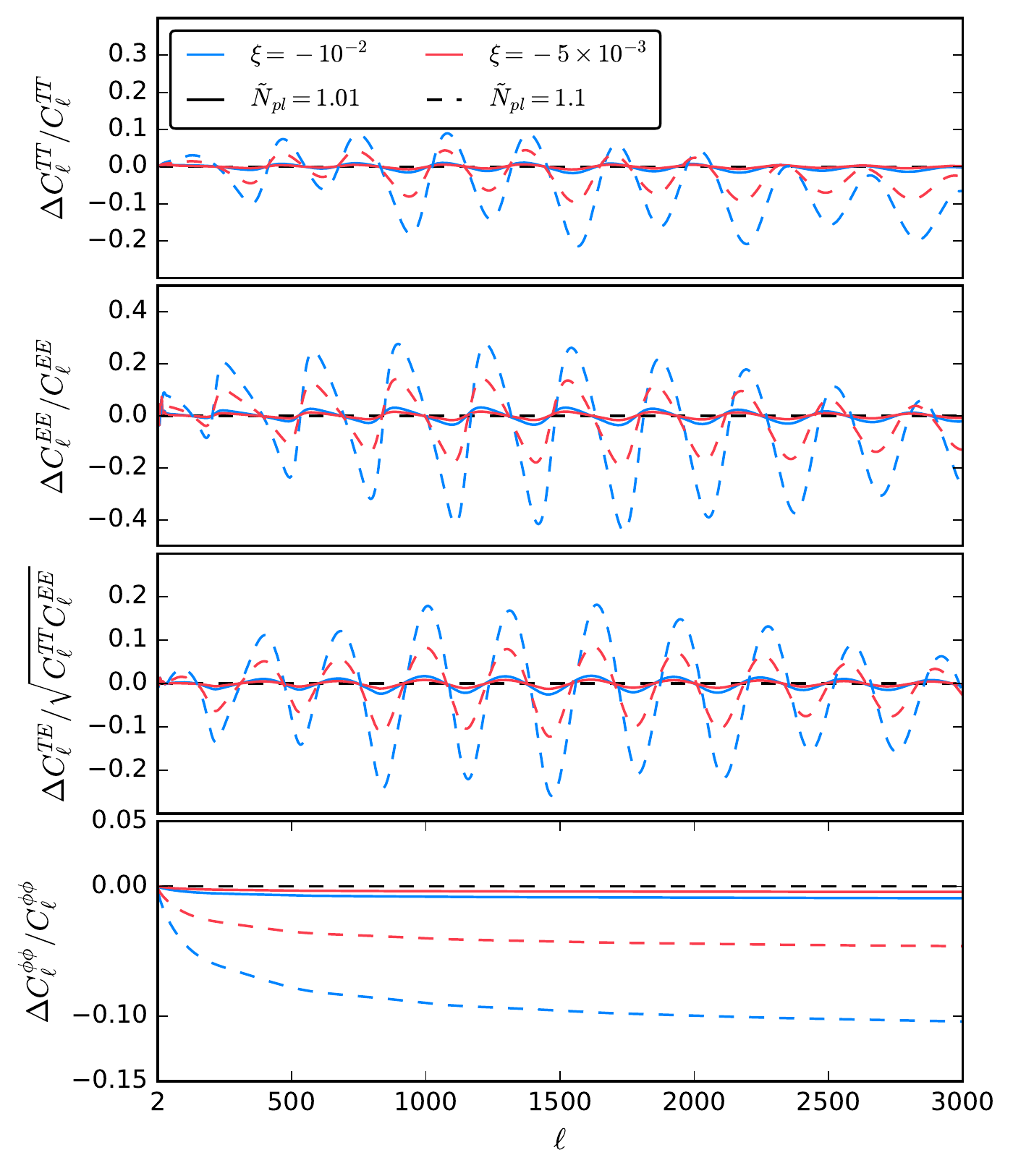}
\caption{From top to bottom: relative differences of the TT-EE-TE-$\phi\phi$ power spectra with respect 
to the $\Lambda$CDM model for $\tilde{N}_{pl}=1.01,\,1.1$ and different values of $\xi=-10^{-2},\,-5\times10^{-3}$.}
\label{fig:CMB_xn}
\end{figure}

\begin{figure}[!h]
\centering
\includegraphics[width=.85\columnwidth]{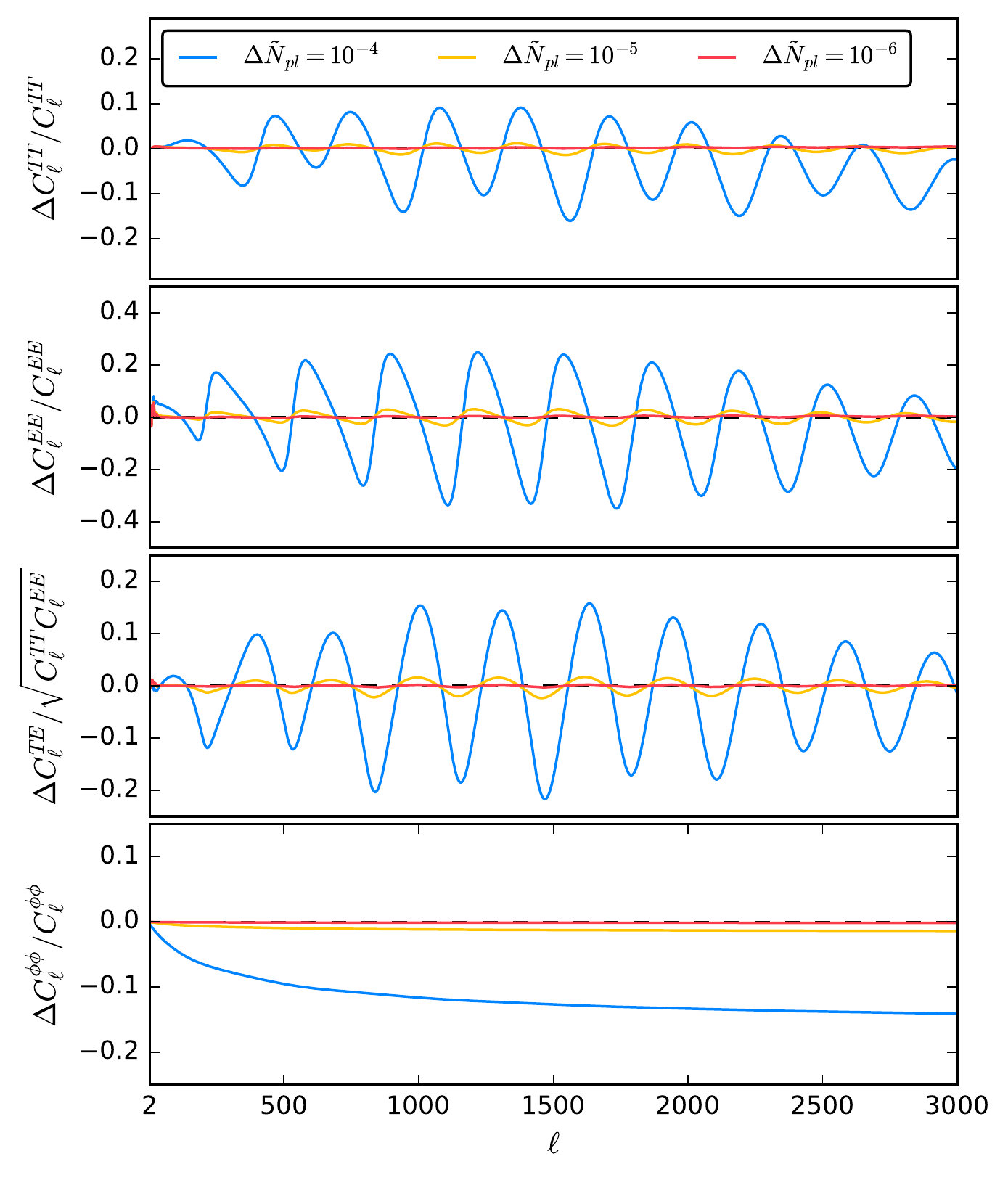}
\caption{From top to bottom: relative differences of the TT-EE-TE-$\phi\phi$ power spectra with respect 
to the $\Lambda$CDM model for the CC case, i.e. $\xi=-1/6$, with different values of 
$\Delta \tilde{N}_{pl}=10^{-4},\,10^{-5},\,10^{-6}$.}
\label{fig:CMB_cc}
\end{figure}

We show the relative differences with respect to the $\Lambda$CDM model for the lensed CMB 
angular power spectra anisotropies in temperature and E-mode polarization, and the CMB lensing 
angular power spectra for different values of $N_{pl}$ for $\xi>0$ in Fig.~\ref{fig:CMB_xp}, 
$\xi<0$ in Fig.~\ref{fig:CMB_xn}, and the CC case $\xi=-1/6$ in Fig.~\ref{fig:CMB_cc}.
We show also the absolute difference of the TE cross-correlation weighted by the square root 
of the product of the two auto-correlators.

\begin{figure}[!h]
\centering
\includegraphics[width=.85\columnwidth]{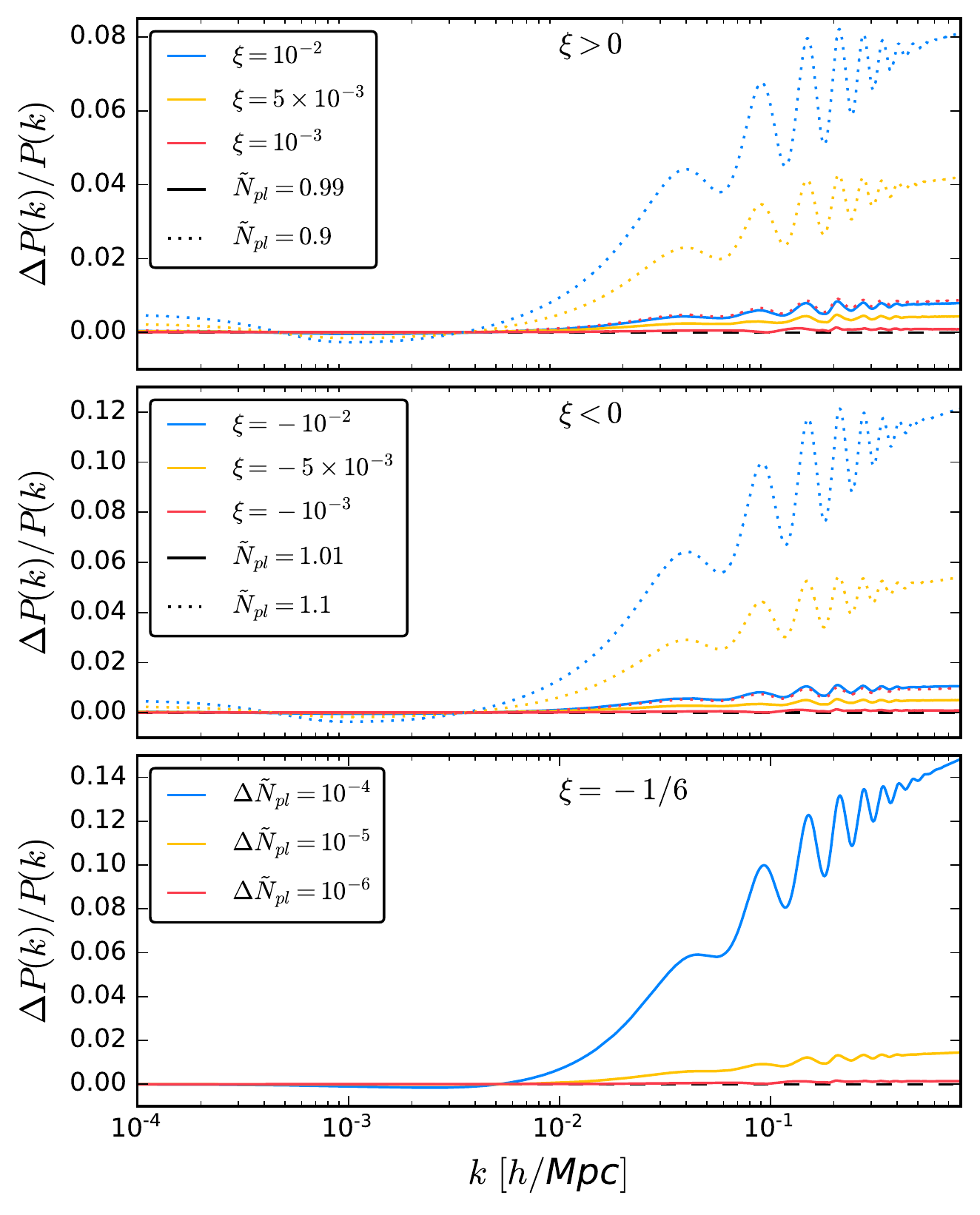}
\caption{From top to bottom: relative differences of the matter power spectra at $z=0$ with respect
to the $\Lambda$CDM model for $\xi>0$, $\xi <0$ and $\xi = -1/6$.}
\label{fig:pk_NMC}
\end{figure}

In Fig.~\ref{fig:pk_NMC} we show the relative differences for the matter power spectrum at $z=0$ 
with respect to the $\Lambda$CDM model for different values of the parameters. 
In all the cases the $P(k)$ is enhanced at small scales, i.e. $k \gtrsim 0.01$ h Mpc$^{-1}$, 
compared to the $\Lambda$CDM model.

\begin{figure}[!h]
\centering
\includegraphics[width=.85\columnwidth]{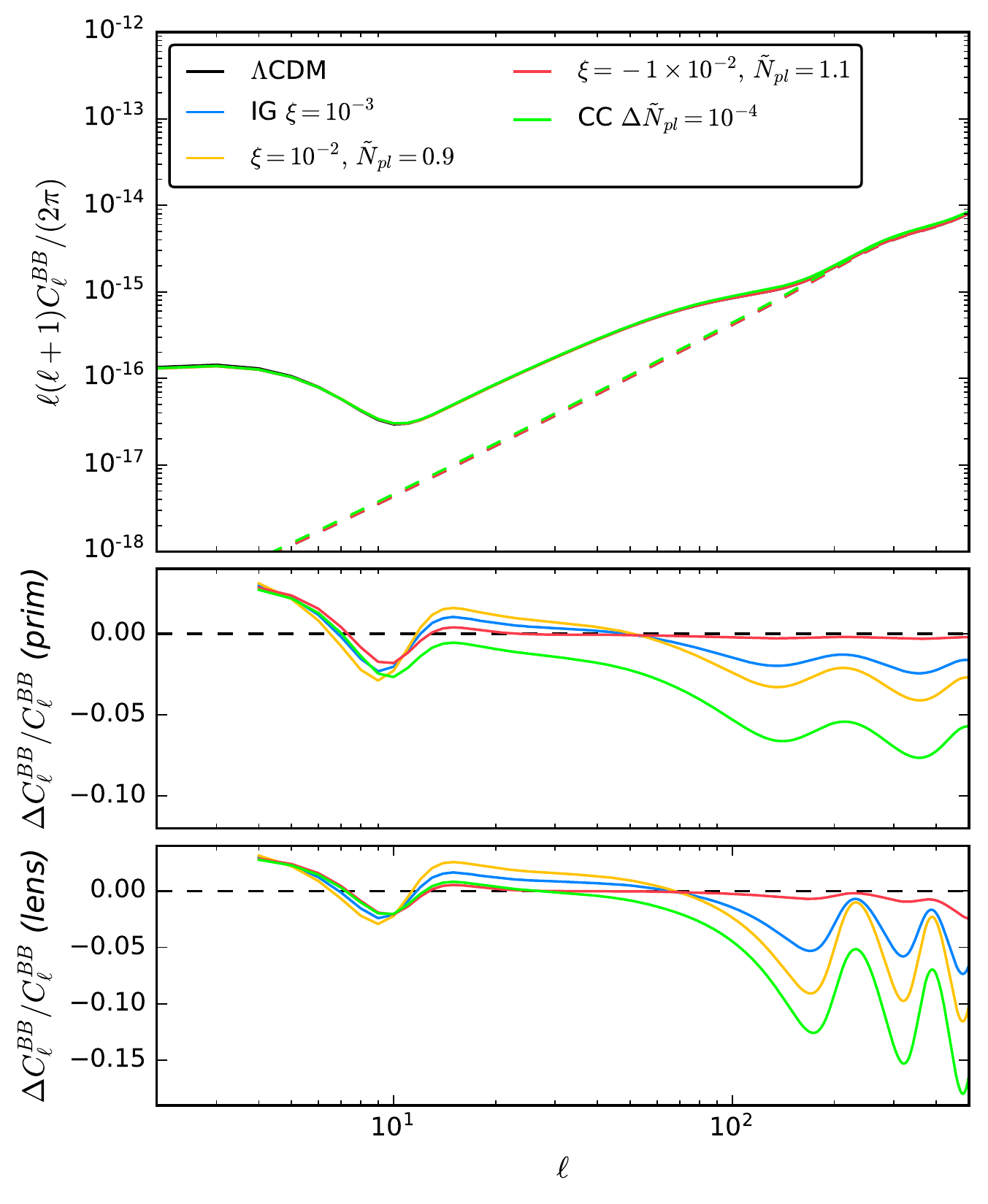}
\caption{From top to bottom: CMB B-mode polarization band power, relative differences of the tensor contribution, 
and relative differences of the lensing contribution with respect
to the $\Lambda$CDM model for $\xi>0$, $\xi <0$, $\xi = -1/6$, and IG. 
Dashed lines refer to the lensing contribution to the B-mode polarization angular power spectrum.}
\label{fig:Bmode}
\end{figure}

We end this section by discussing the B-mode polarization power spectra resulting from the 
evolution of tensor fluctuations in Eq.~\ref{eq_motion_gw}.
Fig.~\ref{fig:Bmode} shows the comparison of the tensor and lensing contributions to B-mode 
polarization in $\Lambda$CDM GR and the scalar-tensor cases of IG ($N_{\mathrm pl}=0$), 
CC ($\xi=-1/6$), positive and negative $\xi$ for a value of a tensor-to-scalar ratio $r=0.05$, 
compatible with the most recent constraints \cite{Akrami:2018odb,Ade:2018gkx}. It is important 
to note that for the values of the couplings chosen in Fig.~\ref{fig:Bmode} the main differences 
in the tensor contribution to B-mode polarization with respect to $\Lambda$CDM GR case is due 
to the different evolution in the Hubble parameter and in the transfer functions in the definition 
of CMB anisotropies.

\section{Constraints from cosmological observations}
\label{sec:five} 

We perform a Monte Carlo Markov Chain analysis by using the publicly available code 
\texttt{MontePython}
\footnote{\href{https://github.com/brinckmann/montepython\_public}{https://github.com/brinckmann/montepython\_public}}
\cite{Audren:2012wb,Brinckmann:2018cvx} connected to our modified version of the code 
\texttt{CLASS}
\footnote{\href{https://github.com/lesgourg/class\_public}{https://github.com/lesgourg/class\_public}}
\cite{Blas:2011rf}, i.e. \texttt{CLASSig} \cite{Umilta:2015cta}.

We use $Planck$ 2015 and BAO likelihoods.
We combine the $Planck$ high-$\ell$ ($\ell > 29$) temperature data with the joint 
temperature-polarization low-$\ell$ ($2 \le \ell \le 29$) likelihood in pixel space at a 
resolution of 3.7 deg, i.e. \texttt{HEALPIX} Nside=16 \cite{Aghanim:2015xee}. The $Planck$ CMB 
lensing likelihood in the conservative multipoles range, i.e. $40 \leq \ell \leq 400$ 
\cite{Ade:2015zua} from the publicly available $Planck$ 2015 release is also combined.
We use BAO data to complement CMB anisotropies at low redshift: we include measurements of 
$D_V/r_s$ at $z_\mathrm{eff}= 0.106$ from 6dFGRS \cite{Beutler:2011hx}, 
at $z_\mathrm{eff} = 0.15$ from SDSS-MGS \cite{Anderson:2013zyy}, 
and from SDSS-DR11 CMASS and LOWZ at $z_\mathrm{eff} = 0.57$ and 
$z_\mathrm{eff} = 0.32$, respectively \cite{Ross:2014qpa}. 

We sample with linear priors the six standard cosmological parameters, 
i.e. $\omega_{\rm b} \equiv \Omega_{\rm b}h^2$, 
$\omega_{\rm c} \equiv \Omega_{\rm c}h^2$, $H_0$, $\tau_\mathrm{re}$, 
$\ln\left(10^{10}A_{\rm s}\right)$, and $n_{\rm s}$, plus the two extra 
parameters for a non-minimally coupled scalar field, i.e. $\Delta \tilde N_{\mathrm pl}$ and $\xi$. 
In the analysis we assume massless neutrinos and marginalize over $Planck$ high-$\ell$ likelihood 
foreground and calibration nuisance parameters \cite{Aghanim:2015xee} which are allowed to vary.

As in \cite{Ballardini:2016cvy}, we take into account the change of the cosmological abundances 
of the light elements during Big Bang Nucleosynthesis (BBN) induced by a different gravitational
constant during the radiation era with respect the theoretical prediction obtained from the 
public code \texttt{PArthENoPE} \cite{Pisanti:2007hk}. We take into account the modified BBN 
consistency condition due to the different value of the effective gravitational constant during 
BBN, by considering this effect as modelled by dark radiation, since the latter effect is 
already tabulated as $Y_\textup{P}^\textup{BBN}(\omega_b,\,N_\textup{eff})$ \cite{Hamann:2007sb} 
in the public version of the \texttt{CLASS} code.
As in \cite{Ballardini:2016cvy}, the posterior probabilities for the primary cosmological 
parameters are hardly affected by the modified BBN consistency condition, and we report a 
small shift for the primordial Helium abundance to higher values.

\subsection{Results}

\begin{figure}[!h]
\centering
\includegraphics[width=.8\columnwidth]{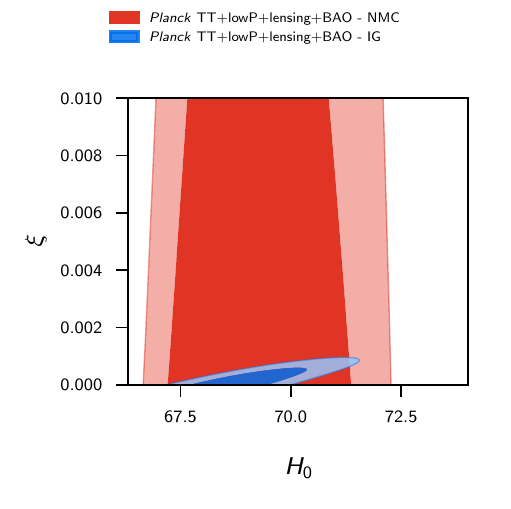}
\caption{2D marginalized confidence levels at 68\% and 95\% for ($H_0$, $\xi$)
for NMC $\xi>0$ (red) and IG (blue) with $Planck$ TT + lowP + lensing + BAO.  }
\label{fig:xp_H0xi}
\end{figure}

\begin{figure}[!h]
\centering
\includegraphics[width=.8\columnwidth]{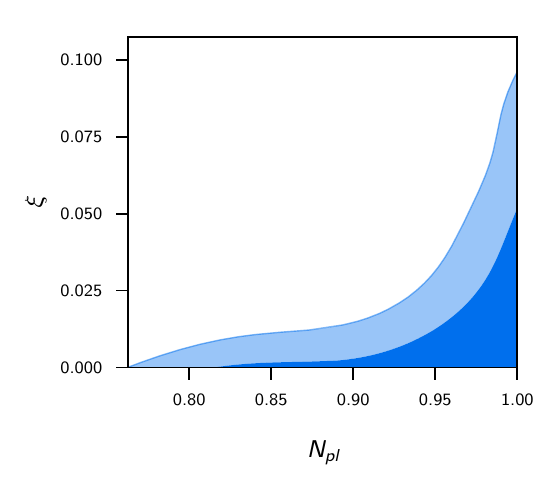}
\caption{2D marginalized confidence levels at 68\% and 95\% for ($N_{pl}$, $\xi$)
for NMC $\xi>0$ with $Planck$ TT + lowP + lensing + BAO.}
\label{fig:xp_Nxi}
\end{figure}

The results from our MCMC exploration are summarized in Table \ref{tab:xi}.
We find for the positive branch of the coupling at 95\% CL:
\begin{align}
&N_{pl} > 0.81\ [\text{M}_{pl}],\\
&\xi < 0.064.
\end{align}
We show in Fig.~\ref{fig:xp_H0xi} a zoom of the 2D parameter space ($H_0$, $\xi$) 
comparing the result of NMC to IG, i.e. $N_{pl}=0$. The constraint on $\xi$ is degradated 
by almost two order of magnitude ($\xi < 0.0075$ at 95\% CL for IG \cite{Ballardini:2016cvy}) 
due to the strong degeneracy between $N_{pl}$ and $\xi$, see Fig.~\ref{fig:xp_Nxi}.

\begin{figure}[!h]
\centering
\includegraphics[width=.8\columnwidth]{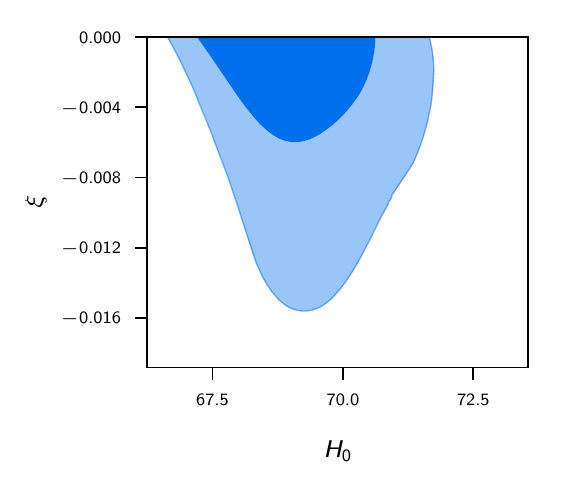}
\caption{2D marginalized confidence levels at 68\% and 95\% for ($H_0$, $\xi$)
for NMC $\xi<0$ with $Planck$ TT + lowP + lensing + BAO.}
\label{fig:xn_H0xi}
\end{figure}

\begin{figure}[!h]
\centering
\includegraphics[width=.8\columnwidth]{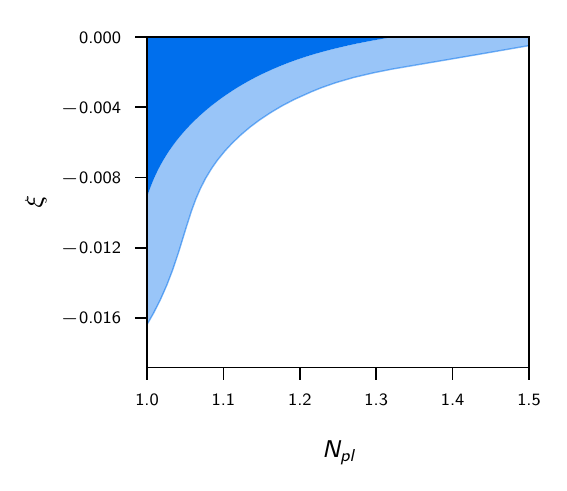}
\caption{2D marginalized confidence levels at 68\% and 95\% for ($N_{pl}$, $\xi$)
for NMC $\xi<0$ with $Planck$ TT + lowP + lensing + BAO.}
\label{fig:xn_Nxi}
\end{figure}

The constraints for the negative branch are (see Figs.~\ref{fig:xn_H0xi}-\ref{fig:xn_Nxi}):
\begin{align}
&N_{pl} < 1.39\ [\text{M}_{pl}],\\
&\xi > -0.11.
\end{align}
at the 95\% CL for $Planck$ TT + lowP + lensing + BAO.

We quote also the derived constraints on the change of the effective Newton's constant 
\eqref{eqn:Geff} evaluated between the radiation era and the present time, and also its 
derivative at present time at 95\% CL: 
\begin{align}
&\frac{\delta G_{\rm eff}}{G} > -0.027\,,\\
&\frac{\dot{G}_{\rm eff}}{G}(z=0) > -1.4 \left[\times10^{-13}\,\text{yr}^{-1}\right]
\,,
\end{align}
for $\xi>0$, and:
\begin{align}
&\frac{\delta G_{\rm eff}}{G} > -0.027\,,\\
&\frac{\dot{G}_{\rm eff}}{G}(z=0) > -0.97 \left[\times10^{-13}\,\text{yr}^{-1}\right]
\,,
\end{align}
for $\xi<0$.

\begin{table*}
\centering{\scriptsize
\begin{tabular}{|l|cccc|}
\hline
                                                      & $Planck$ TT + lowP     & $Planck$ TT + lowP              & $Planck$ TT + lowP        & $Planck$ TT + lowP  \\
                                                      & + lensing + BAO        & + lensing + BAO                 & + lensing + BAO           & + lensing + BAO  \\
                                                      & $\Lambda$CDM           & IG                              & ($\xi > 0$)               & ($\xi < 0$)   \\
\hline
\rule[-1mm]{0mm}{.4cm}
$ \omega_{\rm b}$                                     & $0.02225 \pm 0.00020$  & $0.02224_{-0.00021}^{+0.00020}$ & $0.02226 \pm 0.00019$     & $0.02226 \pm 0.00021$                \\
\rule[-1mm]{0mm}{.4cm}
$ \omega_{\rm c}$                                     & $0.1186 \pm 0.0012$    & $0.1191 \pm {-0.0014}$          & $0.1190 \pm 0.0015$       & $0.1189 \pm 0.0015$                 \\
\rule[-1mm]{0mm}{.4cm}
$H_0$ [km s$^{-1}$ Mpc$^{-1}$]                        & $67.78 \pm 0.57$       & $69.4_{-0.9}^{+0.7}$            & $69.2_{-1.1}^{+0.8}$      & $69.2_{-1.0}^{+0.7}$        \\
\rule[-1mm]{0mm}{.4cm}
$\tau_\mathrm{re}$                                                & $0.066 \pm 0.012$      & $0.063_{-0.014}^{+0.012}$       & $0.068 \pm 0.014$         & $0.069 \pm 0.013$  \\
\rule[-1mm]{0mm}{.4cm}
$\ln \left(10^{10} A_{\rm s} \right)$                 & $3.062 \pm 0.024$      & $3.059^{+0.022}_{-0.026}$       & $3.069_{-0.027}^{+0.023}$ & $3.071 \pm 0.024$    \\
\rule[-1mm]{0mm}{.4cm}
$n_{\rm s}$                                           & $0.9675 \pm 0.0045$    & $0.9669^{+0.0042}_{-0.0047}$    & $0.9674 \pm 0.0046$       & $0.9728 \pm 0.0043$        \\
\rule[-1mm]{0mm}{.4cm}
$\xi$                                                 & $\dots$                & $< 0.00075$ (95\% CL)           & $< 0.064$ (95\% CL)       & $> -0.011$ (95\% CL) 
\rule[-1mm]{0mm}{.4cm}\\
\hline
\rule[-1mm]{0mm}{.4cm}
$N_{pl}\ [\text{M}_{pl}]$                             & $\dots$                & $0$                         & $> 0.81$ (95\% CL)        & $< 1.39$ (95\% CL)        \\
\rule[-1mm]{0mm}{.4cm}
$\gamma_{\rm PN}$                      		      & $1$                    & $> 0.9970$ (95\% CL)            & $> 0.995$ (95\% CL)       & $> 0.997$ (95\% CL)  \\
\rule[-1mm]{0mm}{.4cm}
$\beta_{\rm PN}$  		                      & $1$     	       & $1$                             & $> 0.99987$  (95\% CL)    & $< 1.000011$ (95\% CL)  
\rule[-1mm]{0mm}{.4cm}\\
\hline
\rule[-1mm]{0mm}{.4cm}
$\delta G_{\rm N}/G_{\rm N}$                          & $\dots$                & $-0.009^{+0.003}_{-0.009}$      & $> -0.027$ (95\% CL)       & $> -0.027$ (95\% CL)          \\
\rule[-1mm]{0mm}{.4cm}
$10^{13}\,\dot{G}_{\rm N}(z=0)/G_{\rm N}$ [yr$^{-1}$] & $\dots$                & $-0.37^{+34}_{-12}$             & $> -1.4$ (95\% CL)         & $> -0.97$ (95\% CL) 
\rule[-1mm]{0mm}{.4cm}\\
\hline
\end{tabular}}
\caption{Constraints on main and derived parameters for $Planck$ TT + lowP + lensing + BAO 
(at 68\% CL if not otherwise stated). In the first column we report the results obtained 
for the branch with $\xi>0$ and in the second the branch for $\xi<0$.
In the first column we report the results obtained for the $\Lambda$CDM model with the same dataset 
\cite{Ade:2015xua} and in the second column IG case, i.e. $N_{pl}=0$, for comparison 
\cite{Ballardini:2016cvy}.}
\label{tab:xi}
\end{table*}

For the CC case, i.e. fixing $\xi=-1/6$, results are listed in Tab.~\ref{tab:cc}. 
This model is severely constrained by data leading to tight upper bound on $N_{pl}$ at 95\% CL:
\be
1 < N_{pl} < 1.000038\ [\text{M}_{pl}],
\ee
where $\tilde{N}_{pl}$ can take only values larger than one in this case.

\begin{table}
\centering{\scriptsize
\begin{tabular}{|l|cc|}
\hline
                                       & $Planck$ TT + lowP            & $Planck$ TT + lowP  \\
                                       & + lensing + BAO               & + lensing + BAO  \\
                                       &                               & + HST   \\
\hline
\rule[-1mm]{0mm}{.4cm}
$ \omega_{\rm b}$                      & $0.02223 \pm 0.00021$         & $0.02228 \pm 0.00021$                \\
\rule[-1mm]{0mm}{.4cm}
$ \omega_{\rm c}$                      & $0.1188_{-0.0015}^{+0.0014}$  & $0.1187 \pm 0.0015$                 \\
\rule[-1mm]{0mm}{.4cm}
$H_0$ [km s$^{-1}$ Mpc$^{-1}$]         & $69.19_{-0.93}^{+0.77}$       & $70.20 \pm 0.83$        \\
\rule[-1mm]{0mm}{.4cm}
$\tau_\mathrm{re}$                                 & $0.068_{-0.014}^{+0.012}$     & $0.070_{-0.015}^{+0.013}$  \\
\rule[-1mm]{0mm}{.4cm}
$\ln \left(10^{10} A_{\rm s} \right)$  & $3.070 \pm 0.024$             & $3.074 \pm 0.024$    \\
\rule[-1mm]{0mm}{.4cm}
$n_{\rm s}$                            & $0.9699 \pm 0.0045$           & $0.9728 \pm 0.0043$        
\rule[-1mm]{0mm}{.4cm}\\
\hline
\rule[-1mm]{0mm}{.4cm}
$N_{pl}\ [\text{M}_{pl}]$              & $< 1.000038$ (95\% CL)        & $1.000028^{+0.000012}_{-0.000014}$        \\
\rule[-1mm]{0mm}{.4cm}
$\gamma_{\rm PN}$                      & $> 0.99996$ (95\% CL)         & $1.00003 \pm 0.00001$  \\
\rule[-1mm]{0mm}{.4cm}
$\beta_{\rm PN}$  		       & $< 1.000003$ (95\% CL)        & $0.999998 \pm 0.000001$  
\rule[-1mm]{0mm}{.4cm}\\
\hline
\end{tabular}}
\caption{Constraints on main and derived parameters for $Planck$ TT + lowP + lensing + BAO in the case of the CC model
(at 68\% CL if not otherwise stated).}
\label{tab:cc}
\end{table}

All these models provide a fit to $Planck$ 2015 and BAO data very similar to $\Lambda$CDM: 
we report $\Delta \chi^2 \sim -2.6$ for all the models considered in this paper.
Due to the limited improvement in $\Delta \chi^2$, none of these models is preferred at a statistically significant level with respect 
to $\Lambda$CDM.

\subsection{The Hubble parameter}

\begin{figure}[!h]
\centering
\includegraphics[width=.8\columnwidth]{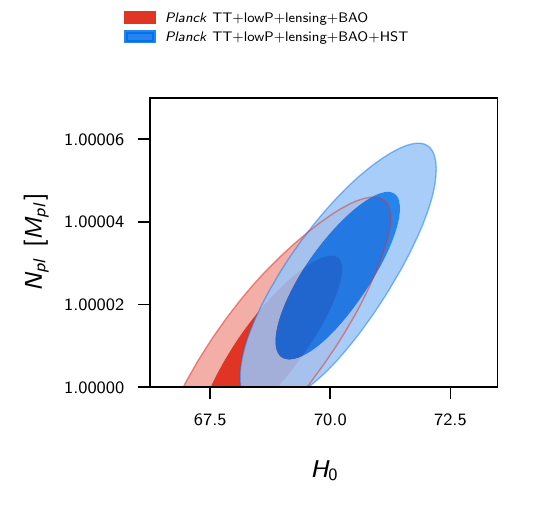}
\caption{2D marginalized confidence levels at 68\% and 95\% for ($H_0$, $N_{pl}$) 
for conformal coupling with $Planck$ TT + lowP + lensing + BAO. 
We include in blue the local estimates of $H_0 = 73.52 \pm 1.62$ [km/s/Mpc] \cite{Riess:2018byc}.}
\label{fig:cc_H0tau}
\end{figure}

We find constraints compatible with the $\Lambda$CDM values for the standard cosmological 
parameters. However, the shifts in $H_0$ deserve a particular mention: as already remarked 
in \cite{Umilta:2015cta,Ballardini:2016cvy} for the IG case, the mean values for $H_0$ are 
larger for all the models studied here. 
Fig.~\ref{fig:cc_H0tau} shows how the 2D marginalized contours for ($H_0$, $N_{pl}$) 
have a degeneracy.
We find:
\be
H_0 = 69.19_{-0.93}^{+0.77}\ [\text{km/s/Mpc}],
\label{H0_CC}
\ee
This value is larger, but compatible at 2$\sigma$ level with the $\Lambda$CDM value 
($H_0 = 67.78 \pm 0.57\ [\text{km/s/Mpc}]$). However, it is still lower than the local 
measurement of the Hubble constant \cite{Riess:2018byc} ($H_0=73.52 \pm 1.62\ [\text{km/s/Mpc}]$) 
obtained by including the new MW parallaxes from HST and Gaia to the rest of the data from 
\cite{Riess:2016jrr}. Therefore the tension between the model dependent estimate of the Hubble 
parameter from $Planck$ 2015 plus BAO data and the local measurement from \cite{Riess:2018byc} 
decreases to 2.3$\sigma$ from the 3.3$\sigma$ of the $\Lambda$CDM model.
For comparison, by varying the number degree of relativistic species $N_{\rm eff}$ in Einstein 
gravity, a lower value for the Hubble parameter, i.e. $H_0=68.00\pm 1.5\ [\text{km/s/Mpc}]$ 
(with $N_{\rm eff}=3.08^{+0.22}_{-0.24}$) for $Planck$ TT + lowP + lensing + BAO at 68\% CL, 
is obtained compared to the CC case reported in Eq. (\ref{H0_CC}).
When the local measurement of the Hubble constant \cite{Riess:2018byc} is included in the fit 
we obtain:
\begin{align}
&H_0 = 70.20 \pm 0.83\ [\text{km/s/Mpc}],\\
&N_{pl} = 1.000028^{+0.000012}_{-0.000014}\ [\text{M}_{pl}] \,.
\end{align}

\begin{figure}[h!]
	\begin{tabular}{c}
		\includegraphics[width=.8\columnwidth]{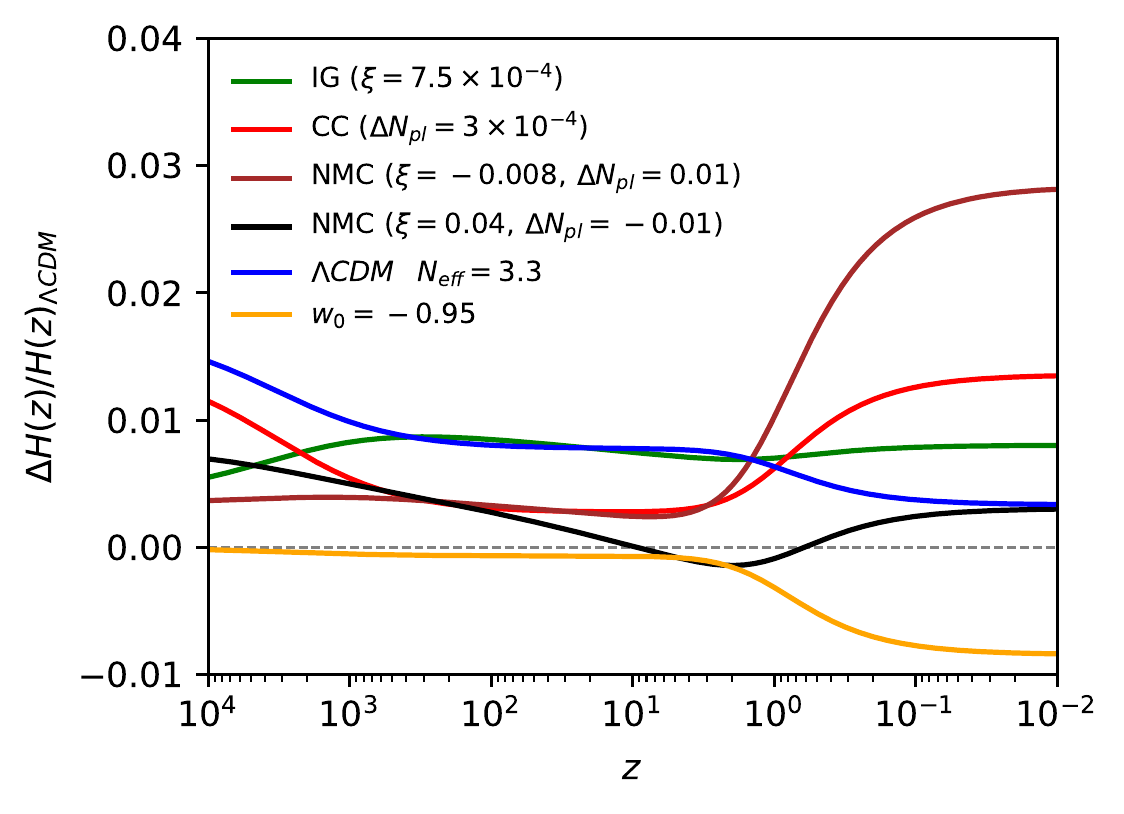}\\
		\includegraphics[width=.8\columnwidth]{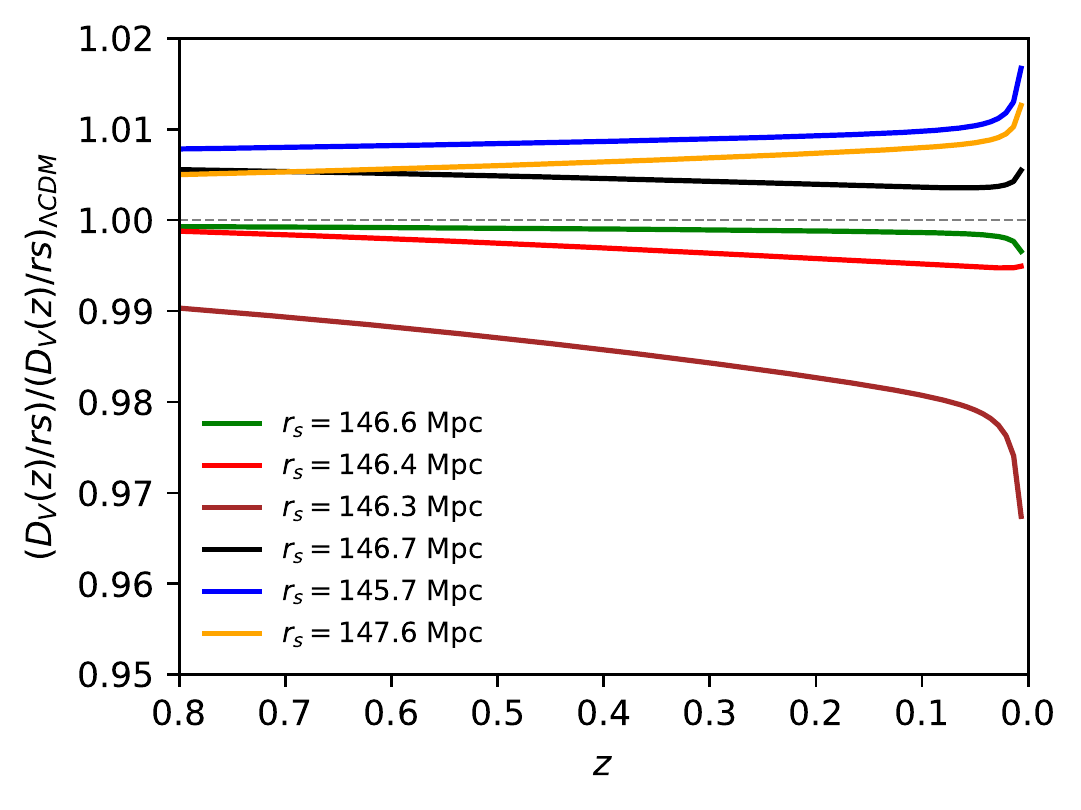}
	\end{tabular}
	\caption{Redshift evolution for relative difference between the Hubble parameter $H(z)$ and its $\Lambda$CDM counterpart  
(upper panel) and the ratio $D_V(z)/r_s$ (lower panel). For $\Lambda$CDM 
quantities we used $Planck$ TT + lowP + lensing + BAO best-fit. The models plotted are IG, 
CC, $\xi<0$, $\xi>0$, $\Lambda CDM$ model with $N_{\textup{eff}}>3.046$ and wCDM with 
$w_0=\textup{const}\neq0$ for green, red, 
brown, black, blue and orange lines respectively.
		\label{H0BAO}}
\end{figure}

Since the marginalized value for $H_0$ in either eJBD and NMC models is larger than in common 
extensions of the $\Lambda$CDM model \cite{Ade:2015xua}, such as $\Lambda$CDM + $N_{\rm eff}$, 
it is useful to understand how the evolution of the Hubble parameter differ at early and late times. 
The differences at early time can be easily understood: since the effective Newton's constant 
can only decrease, if we consider the same $H_0$, this will correspond to an higher $H(z)$ or 
to a larger $N_{\rm eff}$ in the radiation era compared to the $\Lambda$CDM. A second effect 
around recombination is the motion of the scalar field driven by pressureless matter.
At lower redshifts, the differences with respect to $\Lambda$CDM are originated by the onset 
of the acceleration stage by $\sigma$.
The upper panel of Fig.~\ref{H0BAO} shows relative differences of $H(z)$ with respect to the 
$Planck$ TT + lowP + lensing + BAO $\Lambda$CDM best-fit: best-fit (for IG) or NMC models within 
the 1$\sigma$ contours are compared with $\Lambda$CDM + $N_{\rm eff}$ or $w$CDM.
This plot shows how in these scalar-tensor models both early and late time dynamics can contribute 
to a larger value for $H_0$ than in $\Lambda$CDM + $N_{\rm eff}$, for example. 

However, because of this contribution from late time dynamics, the change in $H_0$ cannot be 
interpreted only as a proportional decrement in the comoving sound horizon at the baryon drag 
epoch $r_s$, which is the quantity used to calibrate the BAO standard ruler and is 147.6 Mpc for 
$\Lambda$CDM with the data considered. 
The bottom panel of Fig.~\ref{H0BAO} shows $D_V(z)/r_s\equiv\frac{[c z (1+z)^2 D_A(z)^2 H(z)^{-1}]^{1/3}}{r_s}$, with $D_A$ as the angular diameter distance, normalized to its $\Lambda$CDM value, and 
the value of $r_s$. It is easy to see that both $r_s$ and $H_0$ are lower for $\Lambda$CDM + 
$N_{\rm eff}$ than for the scalar-tensor models studied here and the eJBD model.
These scalar-tensor models therefore differ from those which aim in reducing the tension between 
CMB anisotropies and the local measurements of $H_0$ through a decrement of $r_s$ 
\cite{Cuesta:2014asa,Bernal:2016gxb,Aylor:2018drw}, such as those in which ultralight axion fields 
move slowly around recombination and then dilute away 
\cite{Poulin:2018dzj,Poulin:2018cxd,Agrawal:2019lmo}. In the scalar-tensor models considered here 
the scalar field moves naturally around recombination since is forced by pressureless matter 
and dominates at late time acting as DE.

\subsection{Constraints on the post-Newtonian parameters}

Finally, we quote the derived constraints on the post-Newtonian parameters. 
In this class of models $\gamma_{\rm PN},\,\beta_{\rm PN} \ne 1$ according to 
Eqs.~\eqref{eqn:gammaPN}-\eqref{eqn:betaPN} at 95\% CL:
\begin{align}
&0.995 < \gamma_{\rm PN} < 1,\ (\xi > 0) \\
&0.99987 < \beta_{\rm PN} < 1,\ 
\end{align}
\begin{align}
&0.997 < \gamma_{\rm PN} < 1,\ (\xi < 0) \\
&1 < \beta_{\rm PN} < 1.000011 .
\end{align}
See Fig. 23 for the 2D marginalized constraints in the
$(\gamma_{\rm PN}, \beta_{\rm PN})$ plane.
See Fig. 24 for the 2D marginalized constraints
in the $(H_0, \gamma_{\rm PN})$ plane for $\xi > 0$
compared to the IG case studied in \cite{Ballardini:2016cvy}.

\begin{figure}[!h]
\centering
\includegraphics[width=.8\columnwidth]{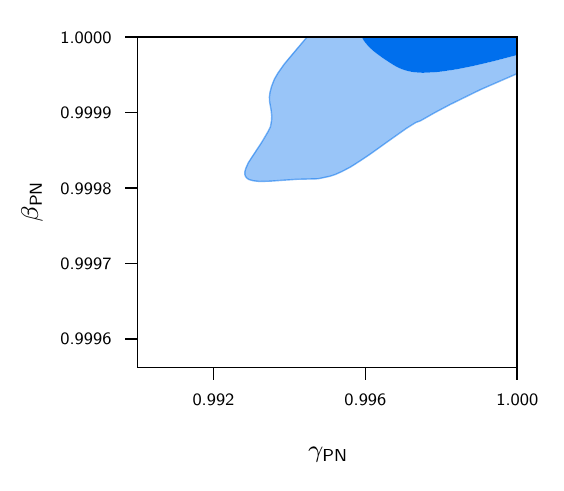}
\includegraphics[width=.8\columnwidth]{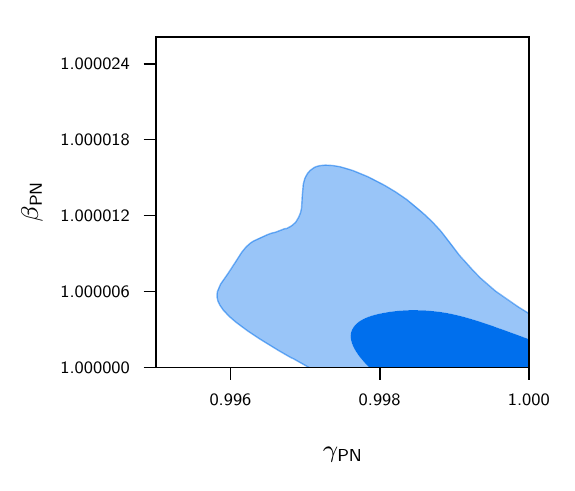}
\caption{2D marginalized confidence levels at 68\% and 95\% for 
($\gamma_{\rm PN}$, $\beta_{\rm PN}$) for NMC $\xi>0$ (left panel) and $\xi<0$ (right panel) 
with $Planck$ TT + lowP + lensing + BAO}
\label{fig:xn}
\end{figure}

The tight constraint on $N_{pl}$ for the CC case correspond at 95\% CL to:
\begin{align}
&0 < 1 - \gamma_{\rm PN} < 4\times 10^{-5},\\
&0 < \beta_{\rm PN}-1 < 3\times 10^{-6},
\end{align}
for $Planck$ TT + lowP + lensing + BAO, where the latter is tighter than the 
constraint from the perihelion shift $\beta_{\rm PN}-1 = (4.1\pm7.8) \times 10^{-5}$ 
\cite{Will:2014kxa} and the former is twice the uncertainty of the Shapiro time 
delay constraint $\gamma_{\rm PN}-1 = (2.1 \pm 2.3) \times 10^{-5}$ \cite{Bertotti:2003rm}.

\begin{figure}[!h]
\centering
\includegraphics[width=.8\columnwidth]{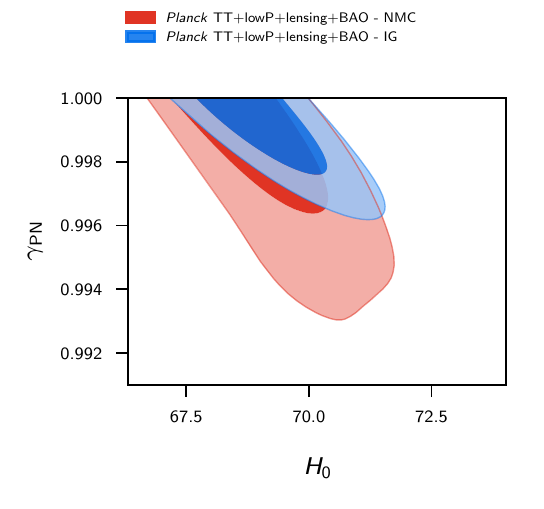}
\caption{2D marginalized confidence levels at 68\% and 95\% for ($H_0$, $\gamma_{\rm PN}$)
for NMC $\xi>0$ (red) and IG (blue) with $Planck$ TT + lowP + lensing + BAO.}    
\label{fig:xp_H0ppn}
\end{figure}

\section{Conclusions}
\label{sec:conclusion}

We have expanded on our previous study of the
observational predictions within the eJBD theory
or, equivalently, IG \cite{Umilta:2015cta,Ballardini:2016cvy},
to the case of a scalar field nonminimally coupled
to the Einstein gravity as in Eq.~\eqref{eqn:action} with $G_3 = G_5 =0$ and $2G_4=F(\sigma)=N_{pl}^2+\xi\sigma^2$.
We have studied this class of model under the assumption that the effective gravitational 
constant in these scalar-tensor theories is compatible with the one measured in a Cavendish-like 
experiment. Whereas in the eJBD theory only the first post-Newtonian parameter $\gamma_{\rm PN}$ 
($=1-\frac{F_{,\sigma}^{2}}{F+2F_{,\sigma}^{2}}$) is not 
vanishing, in this simple extension both the first and second 
post-Newtonian parameter $\beta_{\rm PN}$ ($=1+\frac{F \, F_{,\sigma}}{8F+12F_{,\sigma}^{2}}
\frac{d\gamma_{\rm PN}}{d\sigma}$) are non-zero. The second post-Newtonian 
parameters encodes the sign of the coupling to gravity, i.e. $\beta_\mathrm{PN} > 0$ ($<0$) 
for $\xi > 0$ ($\xi <0$).

For the sake of semplicity, we have restricted ourselves to the class of potential 
$V(\sigma) \propto F^2 (\sigma)$, which makes the field effectively massless 
\cite{Amendola:1999qq} and allows for a direct comparison with the IG model for $N_{pl}=0$ 
\cite{CV,Wetterich:1987fk,Finelli:2007wb,Umilta:2015cta,Ballardini:2016cvy}. 
For this choice of potential $V(\sigma) \propto F^2 (\sigma)$, the scalar field 
is effectively massless. By assuming natural initial conditions in which the decaying mode
is negligible, the scalar field starts at rest deep
in the radiation era and is pushed by pressureless matter
to the final stage in which it drives the Universe
in a nearly de Sitter stage at late times with $\sigma = {\rm const}$. 
In general the effective parameter of state $w_{\rm DE}$ for $\sigma$ defined in 
\cite{Boisseau:2000pr} tracks the one of the dominant matter component before reaching $-1$ 
once the Universe enters in the accelerated stage as for the IG case. We find that the 
conformal case $\xi=-1/6$ is an exception to this general trend: for such a value the 
effective parameter of state $w_{\rm DE}$ interpolates between $1/3$ and $-1$ without 
an intermediate pressureless stage. 
Irrespective of the sign of the coupling $\xi$, $G_N (a) = 1/(8 \pi F)$ 
decrease with time for this class of potential.

As in our previous works in IG, we have considered adiabatic initial conditions for 
fluctuations \cite{Umilta:2015cta,Ballardini:2016cvy,Paoletti:2018xet} which are derived 
in this work for a non-minimally coupled scalar field.
By extending the modification of {\tt CLASSig} \cite{Umilta:2015cta} to a generic coupling 
$F(\sigma)$, we have derived the CMB temperature and polarization anisotropies and the matter 
power spectrum.
Since the effective Newton's constant decrease in time after the relativistic era, we 
observe a shift of the acoustic peaks to higher multipoles and an excess in the matter power 
spectrum at $k \gtrsim 0.01$ Mpc$^{-1}$ proportional to the deviation from GR.

We have used $Planck$ 2015 and BAO data to constrain this class of models. 
As for IG, we obtain a marginalized value for $H_0$ higher than in $\Lambda$CDM for all these 
models, potentially alleviating the tension with the local measurement of the Hubble parameter 
obtained by calibrating with the Cepheids \cite{Riess:2018byc}. The goodness of fit to $Planck$ 
2015 plus BAO data provided by the models studied in this paper is quantitative similar to 
$\Lambda$CDM: since they have one (for the conformal coupled case $\xi=-1/6$) or two (for 
$\xi$ allowed to vary) extra parameters, these models are not preferred with respect to $\Lambda$CDM.
We have derived 95\% CL upper bounds $\xi < 0.064$ ($|\xi| < 0.11$) and 
$0.81 < N_{pl} < 1$ ($1 < N_{pl} < 1.39$) for $\xi >0$ ($\xi <0$). 
It is interesting to note that the bound on $\gamma_{\rm PN}$ and $\delta G_{\rm eff}/G$ have just a small degradation
with respect to eJBD with the same data set ($0.997 < \gamma_{\rm PN} <1$ 
\cite{Ballardini:2016cvy}). Overall, some cosmological constraints do not seem strongly dependent 
on the assumption $\beta_{\rm PN} = 0$ and have a large margin of improvement with future observations \cite{Ballardini:2019tho}.
Although model dependent,
cosmological observations seem more promising than
other independent ways to test scalar-tensor theories
in the strong gravity regime as the search for the presence
of scalar polarization states of gravitational waves \cite{Du:2018txo},
which is also strongly constrained by LIGO/Virgo \cite{Abbott:2018lct}.

The conformal value $\xi = -1/6$ is an interesting and particular case which stands out 
within the general class of non-minimally coupled scalar fields. In addition to what already 
remarked about its effective parameter of state, we find that $Planck$ 2015 + BAO data 
constrain quite tightly the conformal case with $V(\sigma) \propto F^2 (\sigma)$: as 95\% CL 
intervals, we find $1 < 10^5 \Delta \tilde{N}_{\rm Pl} < 3.8$, or equivalently $0.99996 < \gamma_{\rm PN} < 1$, 
$1 < \beta_{\rm PN} < 1.000003$, in terms of the post-Newtonian parameters.
These tight cosmological constraints for the conformal case are comparable to those of obtained within the Solar System bounds \cite{Bertotti:2003rm}.

As from Figs.~\ref{fig:CMB_xp}-\ref{fig:CMB_xn}-\ref{fig:CMB_cc}-\ref{fig:Bmode} CMB polarization anisotropies have a greater  sensitivity to the variation of 
the gravitational strength in these models. It will be therefore 
interesting to see the impact of the latest and more robust measurement 
of CMB polarization anisotropies from Planck 
\cite{Akrami:2018vks,Akrami:2018jnw,Aghanim:2018fcm} and from BICEP2/Keck Array \cite{Akrami:2018vks} 
as well as of the more recent BAO data on the constraints of these models.

\begin{acknowledgments}
We would like to thank Lloyd Knox and Vivian Poulin for discussions.
MBa was supported by the South African Radio Astronomy Observatory, which is a facility of 
the National Research Foundation, an agency of the Department of Science and Technology 
and he was also supported by the Claude Leon Foundation.
MBa, MBr, FF and DP acknowledge financial contribution from the agreement ASI/INAF n. 2018-23-HH.0 
"Attività scientifica per la missione EUCLID – Fase D". 
FF and DP acknowledge also financial support by ASI Grant 2016-24-H.0.
AAS was partly supported by the program KP19-270 ”Questions of the origin
and evolution of the Universe” of the Presidium of the Russian Academy of
Sciences. This research used computational resources of the National Energy Research Scientific Computing Center (NERSC) and 
of INAF OAS Bologna.
\end{acknowledgments}

\setcounter{section}{0}
\appendix
\section{Initial Conditions}
\label{sec:appendix_A}
Here we report the initial conditions adopted in this paper for a non-minimally coupled scalar field, which generalize the 
case of adiabatic initial conditions for IG presented in \cite{Paoletti:2018xet}. 
These quantities reduces to IG and general relativity cases for $N_{pl}=0$ and $(N_{pl}=M_{pl},\xi=0)$, respectively.

\begin{widetext}
For the background cosmology we have as initial conditions:
\be
a(\tau)=\sqrt{\frac{\rho_{r0}}{3F_{i}}}\tau\left[1+\frac{\omega}{4}\tau-\frac{5}{16}\frac{\xi^2\sigma_i^2(1+6\xi)}{F_i+6\xi^2\sigma_i^2}\omega^2\tau^2\right],
\ee
\be
\mathcal{H}(\tau)=\frac{1}{\tau}\left[1+\frac{\omega}{4}\tau-\frac{1}{16}\frac{F_i+4\xi^2\sigma_i^2(4+15\xi)}{F_i+6\xi^2\sigma_i^2}\omega^2\tau^2\right],
\ee
\be
\sigma(\tau)=\sigma_i\left[1+\frac{3}{2}\xi\omega\tau-\frac{2F_i(1-3\xi)+27\xi^2\sigma_i^2(1+2\xi)}{8(F_i+6\xi^2\sigma_i^2)}\omega^2\tau^2\right],
\ee
where $\omega=\frac{\rho_{m0}}{\sqrt{3\rho_{r0}}}\frac{\sqrt{F_i}}{F_i+6\xi^2\sigma_i^2}$.

For cosmological fluctuations in the synchronous gauge we have as adiabatic initial conditions:
\be
\delta_{\gamma}(k,\tau)=\delta_\nu(k,\tau)=\frac{4}{3}\delta_b(k,\tau)=\frac{4}{3}\delta_c(k,\tau)=-\frac{1}{3}k^2\tau^2
\left(1-\frac{\omega}{5}\tau\right),
\ee

\be
\theta_{c}(k,\tau)=0,
\ee

\be
\theta_{\gamma}(k,\tau)=\theta_b(k,\tau)=-\frac{k^4\tau^3}{36}\left[1-\frac{3}{20}\frac{F_i(1-R_\nu+5R_b)+30\xi^2\sigma_i^2}{(1-R_\nu)F_i}\omega\tau\right],
\ee

\be
\theta_{\nu}(k,\tau)=-\frac{k^4\tau^3}{36}\left[\frac{23+4R_\nu}{15+4R_\nu}
-\frac{3(275+50R_\nu+8R_\nu^2)F_i-180(-5+4R_\nu)\xi^2\sigma_i^2}{20(15+2R_\nu)(15+4R_\nu)F_i}\omega\tau\right],
\ee
\be
\sigma_\nu(k,\tau)=\frac{2k^2\tau^2}{3(15+4R_\nu)}\left[1+\frac{(-5+4R_\nu)(F+6\xi^2\sigma_i^2)}{4(15+2R_\nu)F_i}\omega\tau\right],
\ee

\be
\eta(k,\tau)=1-\frac{k^2\tau^2}{12}\left[\frac{5+4R_\nu}{15+4R_\nu}-\frac{150(-5+4R_\nu)\xi^2\sigma_i^2+(325+280R_\nu+16R_\nu^2)F_i}{10(15+4R_\nu)(15+2R_\nu)F_i}\omega\tau\right],
\ee

\be
h(k,\tau)=\frac{k^2\tau^2}{2}\left(1-\frac{\omega}{5}\tau\right),
\ee

\be
\delta\sigma(k,\tau)=-\frac{1}{8}k^2\tau^3\xi\omega\sigma_i\left[1-\frac{2\xi^2\sigma_i^2(24+45\xi)+(4-9\xi)F_i}{10(F_i+6\xi^2\sigma_i^2)}
\omega\tau\right] \,,
\ee
where $R_\nu=\frac{\rho_{\nu0}}{\rho_r0}$ and $R_b=\frac{\rho_{b0}}{\rho_{m0}}$.

\end{widetext}

\bibliography{biblio}

\begin{thebibliography}{67}%
\makeatletter
\providecommand \@ifxundefined [1]{%
 \@ifx{#1\undefined}
}%
\providecommand \@ifnum [1]{%
 \ifnum #1\expandafter \@firstoftwo
 \else \expandafter \@secondoftwo
 \fi
}%
\providecommand \@ifx [1]{%
 \ifx #1\expandafter \@firstoftwo
 \else \expandafter \@secondoftwo
 \fi
}%
\providecommand \natexlab [1]{#1}%
\providecommand \enquote  [1]{``#1''}%
\providecommand \bibnamefont  [1]{#1}%
\providecommand \bibfnamefont [1]{#1}%
\providecommand \citenamefont [1]{#1}%
\providecommand \href@noop [0]{\@secondoftwo}%
\providecommand \href [0]{\begingroup \@sanitize@url \@href}%
\providecommand \@href[1]{\@@startlink{#1}\@@href}%
\providecommand \@@href[1]{\endgroup#1\@@endlink}%
\providecommand \@sanitize@url [0]{\catcode `\\12\catcode `\$12\catcode
  `\&12\catcode `\#12\catcode `\^12\catcode `\_12\catcode `\%12\relax}%
\providecommand \@@startlink[1]{}%
\providecommand \@@endlink[0]{}%
\providecommand \url  [0]{\begingroup\@sanitize@url \@url }%
\providecommand \@url [1]{\endgroup\@href {#1}{\urlprefix }}%
\providecommand \urlprefix  [0]{URL }%
\providecommand \Eprint [0]{\href }%
\providecommand \doibase [0]{http://dx.doi.org/}%
\providecommand \selectlanguage [0]{\@gobble}%
\providecommand \bibinfo  [0]{\@secondoftwo}%
\providecommand \bibfield  [0]{\@secondoftwo}%
\providecommand \translation [1]{[#1]}%
\providecommand \BibitemOpen [0]{}%
\providecommand \bibitemStop [0]{}%
\providecommand \bibitemNoStop [0]{.\EOS\space}%
\providecommand \EOS [0]{\spacefactor3000\relax}%
\providecommand \BibitemShut  [1]{\csname bibitem#1\endcsname}%
\let\auto@bib@innerbib\@empty
\bibitem [{\citenamefont {Uzan}(2011)}]{Uzan:2010pm}%
  \BibitemOpen
  \bibfield  {author} {\bibinfo {author} {\bibfnamefont {Jean-Philippe}\
  \bibnamefont {Uzan}},\ }\bibfield  {title} {\enquote {\bibinfo {title}
  {{Varying Constants, Gravitation and Cosmology}},}\ }\href {\doibase
  10.12942/lrr-2011-2} {\bibfield  {journal} {\bibinfo  {journal} {Living Rev.
  Rel.}\ }\textbf {\bibinfo {volume} {14}},\ \bibinfo {pages} {2} (\bibinfo
  {year} {2011})},\ \Eprint {http://arxiv.org/abs/1009.5514} {arXiv:1009.5514
  [astro-ph.CO]} \BibitemShut {NoStop}%
\bibitem [{\citenamefont {Ade}\ \emph {et~al.}(2014)\citenamefont {Ade} \emph
  {et~al.}}]{Ade:2013zuv}%
  \BibitemOpen
  \bibfield  {author} {\bibinfo {author} {\bibfnamefont {P.~A.~R.}\
  \bibnamefont {Ade}} \emph {et~al.} (\bibinfo {collaboration} {Planck}),\
  }\bibfield  {title} {\enquote {\bibinfo {title} {{Planck 2013 results. XVI.
  Cosmological parameters}},}\ }\href {\doibase 10.1051/0004-6361/201321591}
  {\bibfield  {journal} {\bibinfo  {journal} {Astron. Astrophys.}\ }\textbf
  {\bibinfo {volume} {571}},\ \bibinfo {pages} {A16} (\bibinfo {year}
  {2014})},\ \Eprint {http://arxiv.org/abs/1303.5076} {arXiv:1303.5076
  [astro-ph.CO]} \BibitemShut {NoStop}%
\bibitem [{\citenamefont {Umilt\`a}\ \emph {et~al.}(2015)\citenamefont
  {Umilt\`a}, \citenamefont {Ballardini}, \citenamefont {Finelli},\ and\
  \citenamefont {Paoletti}}]{Umilta:2015cta}%
  \BibitemOpen
  \bibfield  {author} {\bibinfo {author} {\bibfnamefont {C.}~\bibnamefont
  {Umilt\`a}}, \bibinfo {author} {\bibfnamefont {M.}~\bibnamefont
  {Ballardini}}, \bibinfo {author} {\bibfnamefont {F.}~\bibnamefont {Finelli}},
  \ and\ \bibinfo {author} {\bibfnamefont {D.}~\bibnamefont {Paoletti}},\
  }\bibfield  {title} {\enquote {\bibinfo {title} {{CMB and BAO constraints for
  an induced gravity dark energy model with a quartic potential}},}\ }\href
  {\doibase 10.1088/1475-7516/2015/08/017} {\bibfield  {journal} {\bibinfo
  {journal} {JCAP}\ }\textbf {\bibinfo {volume} {1508}},\ \bibinfo {pages}
  {017} (\bibinfo {year} {2015})},\ \Eprint {http://arxiv.org/abs/1507.00718}
  {arXiv:1507.00718 [astro-ph.CO]} \BibitemShut {NoStop}%
\bibitem [{\citenamefont {Ballardini}\ \emph {et~al.}(2016)\citenamefont
  {Ballardini}, \citenamefont {Finelli}, \citenamefont {Umilt\`a},\ and\
  \citenamefont {Paoletti}}]{Ballardini:2016cvy}%
  \BibitemOpen
  \bibfield  {author} {\bibinfo {author} {\bibfnamefont {Mario}\ \bibnamefont
  {Ballardini}}, \bibinfo {author} {\bibfnamefont {Fabio}\ \bibnamefont
  {Finelli}}, \bibinfo {author} {\bibfnamefont {Caterina}\ \bibnamefont
  {Umilt\`a}}, \ and\ \bibinfo {author} {\bibfnamefont {Daniela}\ \bibnamefont
  {Paoletti}},\ }\bibfield  {title} {\enquote {\bibinfo {title} {{Cosmological
  constraints on induced gravity dark energy models}},}\ }\href {\doibase
  10.1088/1475-7516/2016/05/067} {\bibfield  {journal} {\bibinfo  {journal}
  {JCAP}\ }\textbf {\bibinfo {volume} {1605}},\ \bibinfo {pages} {067}
  (\bibinfo {year} {2016})},\ \Eprint {http://arxiv.org/abs/1601.03387}
  {arXiv:1601.03387 [astro-ph.CO]} \BibitemShut {NoStop}%
\bibitem [{\citenamefont {Jordan}(1949)}]{Jordan:1949zz}%
  \BibitemOpen
  \bibfield  {author} {\bibinfo {author} {\bibfnamefont {Pascual}\ \bibnamefont
  {Jordan}},\ }\bibfield  {title} {\enquote {\bibinfo {title} {{Formation of
  the Stars and Development of the Universe}},}\ }\href {\doibase
  10.1038/164637a0} {\bibfield  {journal} {\bibinfo  {journal} {Nature}\
  }\textbf {\bibinfo {volume} {164}},\ \bibinfo {pages} {637--640} (\bibinfo
  {year} {1949})}\BibitemShut {NoStop}%
\bibitem [{\citenamefont {Brans}\ and\ \citenamefont
  {Dicke}(1961)}]{Brans:1961sx}%
  \BibitemOpen
  \bibfield  {author} {\bibinfo {author} {\bibfnamefont {C.}~\bibnamefont
  {Brans}}\ and\ \bibinfo {author} {\bibfnamefont {R.~H.}\ \bibnamefont
  {Dicke}},\ }\bibfield  {title} {\enquote {\bibinfo {title} {{Mach's principle
  and a relativistic theory of gravitation}},}\ }\href {\doibase
  10.1103/PhysRev.124.925} {\bibfield  {journal} {\bibinfo  {journal} {Phys.
  Rev.}\ }\textbf {\bibinfo {volume} {124}},\ \bibinfo {pages} {925--935}
  (\bibinfo {year} {1961})}\BibitemShut {NoStop}%
\bibitem [{\citenamefont {Chen}\ and\ \citenamefont
  {Kamionkowski}(1999)}]{Chen:1999qh}%
  \BibitemOpen
  \bibfield  {author} {\bibinfo {author} {\bibfnamefont {Xue-lei}\ \bibnamefont
  {Chen}}\ and\ \bibinfo {author} {\bibfnamefont {Marc}\ \bibnamefont
  {Kamionkowski}},\ }\bibfield  {title} {\enquote {\bibinfo {title} {{Cosmic
  microwave background temperature and polarization anisotropy in Brans-Dicke
  cosmology}},}\ }\href {\doibase 10.1103/PhysRevD.60.104036} {\bibfield
  {journal} {\bibinfo  {journal} {Phys. Rev.}\ }\textbf {\bibinfo {volume}
  {D60}},\ \bibinfo {pages} {104036} (\bibinfo {year} {1999})},\ \Eprint
  {http://arxiv.org/abs/astro-ph/9905368} {arXiv:astro-ph/9905368 [astro-ph]}
  \BibitemShut {NoStop}%
\bibitem [{\citenamefont {Nagata}\ \emph {et~al.}(2004)\citenamefont {Nagata},
  \citenamefont {Chiba},\ and\ \citenamefont {Sugiyama}}]{Nagata:2003qn}%
  \BibitemOpen
  \bibfield  {author} {\bibinfo {author} {\bibfnamefont {Ryo}\ \bibnamefont
  {Nagata}}, \bibinfo {author} {\bibfnamefont {Takeshi}\ \bibnamefont {Chiba}},
  \ and\ \bibinfo {author} {\bibfnamefont {Naoshi}\ \bibnamefont {Sugiyama}},\
  }\bibfield  {title} {\enquote {\bibinfo {title} {{WMAP constraints on scalar-
  tensor cosmology and the variation of the gravitational constant}},}\ }\href
  {\doibase 10.1103/PhysRevD.69.083512} {\bibfield  {journal} {\bibinfo
  {journal} {Phys. Rev.}\ }\textbf {\bibinfo {volume} {D69}},\ \bibinfo {pages}
  {083512} (\bibinfo {year} {2004})},\ \Eprint
  {http://arxiv.org/abs/astro-ph/0311274} {arXiv:astro-ph/0311274 [astro-ph]}
  \BibitemShut {NoStop}%
\bibitem [{\citenamefont {Acquaviva}\ \emph {et~al.}(2005)\citenamefont
  {Acquaviva}, \citenamefont {Baccigalupi}, \citenamefont {Leach},
  \citenamefont {Liddle},\ and\ \citenamefont {Perrotta}}]{Acquaviva:2004ti}%
  \BibitemOpen
  \bibfield  {author} {\bibinfo {author} {\bibfnamefont {Viviana}\ \bibnamefont
  {Acquaviva}}, \bibinfo {author} {\bibfnamefont {Carlo}\ \bibnamefont
  {Baccigalupi}}, \bibinfo {author} {\bibfnamefont {Samuel~M.}\ \bibnamefont
  {Leach}}, \bibinfo {author} {\bibfnamefont {Andrew~R.}\ \bibnamefont
  {Liddle}}, \ and\ \bibinfo {author} {\bibfnamefont {Francesca}\ \bibnamefont
  {Perrotta}},\ }\bibfield  {title} {\enquote {\bibinfo {title} {{Structure
  formation constraints on the Jordan-Brans-Dicke theory}},}\ }\href {\doibase
  10.1103/PhysRevD.71.104025} {\bibfield  {journal} {\bibinfo  {journal} {Phys.
  Rev.}\ }\textbf {\bibinfo {volume} {D71}},\ \bibinfo {pages} {104025}
  (\bibinfo {year} {2005})},\ \Eprint {http://arxiv.org/abs/astro-ph/0412052}
  {arXiv:astro-ph/0412052 [astro-ph]} \BibitemShut {NoStop}%
\bibitem [{\citenamefont {Avilez}\ and\ \citenamefont
  {Skordis}(2014)}]{Avilez:2013dxa}%
  \BibitemOpen
  \bibfield  {author} {\bibinfo {author} {\bibfnamefont {A.}~\bibnamefont
  {Avilez}}\ and\ \bibinfo {author} {\bibfnamefont {C.}~\bibnamefont
  {Skordis}},\ }\bibfield  {title} {\enquote {\bibinfo {title} {{Cosmological
  constraints on Brans-Dicke theory}},}\ }\href {\doibase
  10.1103/PhysRevLett.113.011101} {\bibfield  {journal} {\bibinfo  {journal}
  {Phys. Rev. Lett.}\ }\textbf {\bibinfo {volume} {113}},\ \bibinfo {pages}
  {011101} (\bibinfo {year} {2014})},\ \Eprint {http://arxiv.org/abs/1303.4330}
  {arXiv:1303.4330 [astro-ph.CO]} \BibitemShut {NoStop}%
\bibitem [{\citenamefont {Li}\ \emph {et~al.}(2013)\citenamefont {Li},
  \citenamefont {Wu},\ and\ \citenamefont {Chen}}]{Li:2013nwa}%
  \BibitemOpen
  \bibfield  {author} {\bibinfo {author} {\bibfnamefont {Yi-Chao}\ \bibnamefont
  {Li}}, \bibinfo {author} {\bibfnamefont {Feng-Quan}\ \bibnamefont {Wu}}, \
  and\ \bibinfo {author} {\bibfnamefont {Xuelei}\ \bibnamefont {Chen}},\
  }\bibfield  {title} {\enquote {\bibinfo {title} {{Constraints on the
  Brans-Dicke gravity theory with the Planck data}},}\ }\href {\doibase
  10.1103/PhysRevD.88.084053} {\bibfield  {journal} {\bibinfo  {journal} {Phys.
  Rev.}\ }\textbf {\bibinfo {volume} {D88}},\ \bibinfo {pages} {084053}
  (\bibinfo {year} {2013})},\ \Eprint {http://arxiv.org/abs/1305.0055}
  {arXiv:1305.0055 [astro-ph.CO]} \BibitemShut {NoStop}%
\bibitem [{\citenamefont {Ooba}\ \emph {et~al.}(2016)\citenamefont {Ooba},
  \citenamefont {Ichiki}, \citenamefont {Chiba},\ and\ \citenamefont
  {Sugiyama}}]{Ooba:2016slp}%
  \BibitemOpen
  \bibfield  {author} {\bibinfo {author} {\bibfnamefont {Junpei}\ \bibnamefont
  {Ooba}}, \bibinfo {author} {\bibfnamefont {Kiyotomo}\ \bibnamefont {Ichiki}},
  \bibinfo {author} {\bibfnamefont {Takeshi}\ \bibnamefont {Chiba}}, \ and\
  \bibinfo {author} {\bibfnamefont {Naoshi}\ \bibnamefont {Sugiyama}},\
  }\bibfield  {title} {\enquote {\bibinfo {title} {{Planck constraints on
  scalar-tensor cosmology and the variation of the gravitational constant}},}\
  }\href {\doibase 10.1103/PhysRevD.93.122002} {\bibfield  {journal} {\bibinfo
  {journal} {Phys. Rev.}\ }\textbf {\bibinfo {volume} {D93}},\ \bibinfo {pages}
  {122002} (\bibinfo {year} {2016})},\ \Eprint
  {http://arxiv.org/abs/1602.00809} {arXiv:1602.00809 [astro-ph.CO]}
  \BibitemShut {NoStop}%
\bibitem [{\citenamefont {Horndeski}(1974)}]{Horndeski:1974wa}%
  \BibitemOpen
  \bibfield  {author} {\bibinfo {author} {\bibfnamefont {Gregory~Walter}\
  \bibnamefont {Horndeski}},\ }\bibfield  {title} {\enquote {\bibinfo {title}
  {{Second-order scalar-tensor field equations in a four-dimensional space}},}\
  }\href {\doibase 10.1007/BF01807638} {\bibfield  {journal} {\bibinfo
  {journal} {Int. J. Theor. Phys.}\ }\textbf {\bibinfo {volume} {10}},\
  \bibinfo {pages} {363--384} (\bibinfo {year} {1974})}\BibitemShut {NoStop}%
\bibitem [{\citenamefont {Ballardini}\ \emph {et~al.}(2019)\citenamefont
  {Ballardini}, \citenamefont {Sapone}, \citenamefont {Umiltà}, \citenamefont
  {Finelli},\ and\ \citenamefont {Paoletti}}]{Ballardini:2019tho}%
  \BibitemOpen
  \bibfield  {author} {\bibinfo {author} {\bibfnamefont {M.}~\bibnamefont
  {Ballardini}}, \bibinfo {author} {\bibfnamefont {D.}~\bibnamefont {Sapone}},
  \bibinfo {author} {\bibfnamefont {C.}~\bibnamefont {Umiltà}}, \bibinfo
  {author} {\bibfnamefont {F.}~\bibnamefont {Finelli}}, \ and\ \bibinfo
  {author} {\bibfnamefont {D.}~\bibnamefont {Paoletti}},\ }\bibfield  {title}
  {\enquote {\bibinfo {title} {{Testing extended Jordan-Brans-Dicke theories
  with future cosmological observations}},}\ }\href@noop {} {\  (\bibinfo
  {year} {2019})},\ \Eprint {http://arxiv.org/abs/1902.01407} {arXiv:1902.01407
  [astro-ph.CO]} \BibitemShut {NoStop}%
\bibitem [{\citenamefont {Alonso}\ \emph {et~al.}(2017)\citenamefont {Alonso},
  \citenamefont {Bellini}, \citenamefont {Ferreira},\ and\ \citenamefont
  {Zumalacárregui}}]{Alonso:2016suf}%
  \BibitemOpen
  \bibfield  {author} {\bibinfo {author} {\bibfnamefont {David}\ \bibnamefont
  {Alonso}}, \bibinfo {author} {\bibfnamefont {Emilio}\ \bibnamefont
  {Bellini}}, \bibinfo {author} {\bibfnamefont {Pedro~G.}\ \bibnamefont
  {Ferreira}}, \ and\ \bibinfo {author} {\bibfnamefont {Miguel}\ \bibnamefont
  {Zumalacárregui}},\ }\bibfield  {title} {\enquote {\bibinfo {title}
  {{Observational future of cosmological scalar-tensor theories}},}\ }\href
  {\doibase 10.1103/PhysRevD.95.063502} {\bibfield  {journal} {\bibinfo
  {journal} {Phys. Rev.}\ }\textbf {\bibinfo {volume} {D95}},\ \bibinfo {pages}
  {063502} (\bibinfo {year} {2017})},\ \Eprint
  {http://arxiv.org/abs/1610.09290} {arXiv:1610.09290 [astro-ph.CO]}
  \BibitemShut {NoStop}%
\bibitem [{\citenamefont {Uzan}(1999)}]{Uzan:1999ch}%
  \BibitemOpen
  \bibfield  {author} {\bibinfo {author} {\bibfnamefont {Jean-Philippe}\
  \bibnamefont {Uzan}},\ }\bibfield  {title} {\enquote {\bibinfo {title}
  {{Cosmological scaling solutions of nonminimally coupled scalar fields}},}\
  }\href {\doibase 10.1103/PhysRevD.59.123510} {\bibfield  {journal} {\bibinfo
  {journal} {Phys. Rev.}\ }\textbf {\bibinfo {volume} {D59}},\ \bibinfo {pages}
  {123510} (\bibinfo {year} {1999})},\ \Eprint
  {http://arxiv.org/abs/gr-qc/9903004} {arXiv:gr-qc/9903004 [gr-qc]}
  \BibitemShut {NoStop}%
\bibitem [{\citenamefont {Perrotta}\ \emph {et~al.}(1999)\citenamefont
  {Perrotta}, \citenamefont {Baccigalupi},\ and\ \citenamefont
  {Matarrese}}]{Perrotta:1999am}%
  \BibitemOpen
  \bibfield  {author} {\bibinfo {author} {\bibfnamefont {Francesca}\
  \bibnamefont {Perrotta}}, \bibinfo {author} {\bibfnamefont {Carlo}\
  \bibnamefont {Baccigalupi}}, \ and\ \bibinfo {author} {\bibfnamefont
  {Sabino}\ \bibnamefont {Matarrese}},\ }\bibfield  {title} {\enquote {\bibinfo
  {title} {{Extended quintessence}},}\ }\href {\doibase
  10.1103/PhysRevD.61.023507} {\bibfield  {journal} {\bibinfo  {journal} {Phys.
  Rev.}\ }\textbf {\bibinfo {volume} {D61}},\ \bibinfo {pages} {023507}
  (\bibinfo {year} {1999})},\ \Eprint {http://arxiv.org/abs/astro-ph/9906066}
  {arXiv:astro-ph/9906066 [astro-ph]} \BibitemShut {NoStop}%
\bibitem [{\citenamefont {Bartolo}\ and\ \citenamefont
  {Pietroni}(2000)}]{Bartolo:1999sq}%
  \BibitemOpen
  \bibfield  {author} {\bibinfo {author} {\bibfnamefont {Nicola}\ \bibnamefont
  {Bartolo}}\ and\ \bibinfo {author} {\bibfnamefont {Massimo}\ \bibnamefont
  {Pietroni}},\ }\bibfield  {title} {\enquote {\bibinfo {title} {{Scalar tensor
  gravity and quintessence}},}\ }\href {\doibase 10.1103/PhysRevD.61.023518}
  {\bibfield  {journal} {\bibinfo  {journal} {Phys. Rev.}\ }\textbf {\bibinfo
  {volume} {D61}},\ \bibinfo {pages} {023518} (\bibinfo {year} {2000})},\
  \Eprint {http://arxiv.org/abs/hep-ph/9908521} {arXiv:hep-ph/9908521 [hep-ph]}
  \BibitemShut {NoStop}%
\bibitem [{\citenamefont {Amendola}(1999)}]{Amendola:1999qq}%
  \BibitemOpen
  \bibfield  {author} {\bibinfo {author} {\bibfnamefont {Luca}\ \bibnamefont
  {Amendola}},\ }\bibfield  {title} {\enquote {\bibinfo {title} {{Scaling
  solutions in general nonminimal coupling theories}},}\ }\href {\doibase
  10.1103/PhysRevD.60.043501} {\bibfield  {journal} {\bibinfo  {journal} {Phys.
  Rev.}\ }\textbf {\bibinfo {volume} {D60}},\ \bibinfo {pages} {043501}
  (\bibinfo {year} {1999})},\ \Eprint {http://arxiv.org/abs/astro-ph/9904120}
  {arXiv:astro-ph/9904120 [astro-ph]} \BibitemShut {NoStop}%
\bibitem [{\citenamefont {Chiba}(1999)}]{Chiba:1999wt}%
  \BibitemOpen
  \bibfield  {author} {\bibinfo {author} {\bibfnamefont {Takeshi}\ \bibnamefont
  {Chiba}},\ }\bibfield  {title} {\enquote {\bibinfo {title} {{Quintessence,
  the gravitational constant, and gravity}},}\ }\href {\doibase
  10.1103/PhysRevD.60.083508} {\bibfield  {journal} {\bibinfo  {journal} {Phys.
  Rev.}\ }\textbf {\bibinfo {volume} {D60}},\ \bibinfo {pages} {083508}
  (\bibinfo {year} {1999})},\ \Eprint {http://arxiv.org/abs/gr-qc/9903094}
  {arXiv:gr-qc/9903094 [gr-qc]} \BibitemShut {NoStop}%
\bibitem [{\citenamefont {Baker}\ \emph {et~al.}(2017)\citenamefont {Baker},
  \citenamefont {Bellini}, \citenamefont {Ferreira}, \citenamefont {Lagos},
  \citenamefont {Noller},\ and\ \citenamefont {Sawicki}}]{Baker:2017hug}%
  \BibitemOpen
  \bibfield  {author} {\bibinfo {author} {\bibfnamefont {T.}~\bibnamefont
  {Baker}}, \bibinfo {author} {\bibfnamefont {E.}~\bibnamefont {Bellini}},
  \bibinfo {author} {\bibfnamefont {P.~G.}\ \bibnamefont {Ferreira}}, \bibinfo
  {author} {\bibfnamefont {M.}~\bibnamefont {Lagos}}, \bibinfo {author}
  {\bibfnamefont {J.}~\bibnamefont {Noller}}, \ and\ \bibinfo {author}
  {\bibfnamefont {I.}~\bibnamefont {Sawicki}},\ }\bibfield  {title} {\enquote
  {\bibinfo {title} {{Strong constraints on cosmological gravity from GW170817
  and GRB 170817A}},}\ }\href {\doibase 10.1103/PhysRevLett.119.251301}
  {\bibfield  {journal} {\bibinfo  {journal} {Phys. Rev. Lett.}\ }\textbf
  {\bibinfo {volume} {119}},\ \bibinfo {pages} {251301} (\bibinfo {year}
  {2017})},\ \Eprint {http://arxiv.org/abs/1710.06394} {arXiv:1710.06394
  [astro-ph.CO]} \BibitemShut {NoStop}%
\bibitem [{\citenamefont {Creminelli}\ and\ \citenamefont
  {Vernizzi}(2017)}]{Creminelli:2017sry}%
  \BibitemOpen
  \bibfield  {author} {\bibinfo {author} {\bibfnamefont {Paolo}\ \bibnamefont
  {Creminelli}}\ and\ \bibinfo {author} {\bibfnamefont {Filippo}\ \bibnamefont
  {Vernizzi}},\ }\bibfield  {title} {\enquote {\bibinfo {title} {{Dark Energy
  after GW170817 and GRB170817A}},}\ }\href {\doibase
  10.1103/PhysRevLett.119.251302} {\bibfield  {journal} {\bibinfo  {journal}
  {Phys. Rev. Lett.}\ }\textbf {\bibinfo {volume} {119}},\ \bibinfo {pages}
  {251302} (\bibinfo {year} {2017})},\ \Eprint
  {http://arxiv.org/abs/1710.05877} {arXiv:1710.05877 [astro-ph.CO]}
  \BibitemShut {NoStop}%
\bibitem [{\citenamefont {Ezquiaga}\ and\ \citenamefont
  {Zumalacárregui}(2017)}]{Ezquiaga:2017ekz}%
  \BibitemOpen
  \bibfield  {author} {\bibinfo {author} {\bibfnamefont {Jose~María}\
  \bibnamefont {Ezquiaga}}\ and\ \bibinfo {author} {\bibfnamefont {Miguel}\
  \bibnamefont {Zumalacárregui}},\ }\bibfield  {title} {\enquote {\bibinfo
  {title} {{Dark Energy After GW170817: Dead Ends and the Road Ahead}},}\
  }\href {\doibase 10.1103/PhysRevLett.119.251304} {\bibfield  {journal}
  {\bibinfo  {journal} {Phys. Rev. Lett.}\ }\textbf {\bibinfo {volume} {119}},\
  \bibinfo {pages} {251304} (\bibinfo {year} {2017})},\ \Eprint
  {http://arxiv.org/abs/1710.05901} {arXiv:1710.05901 [astro-ph.CO]}
  \BibitemShut {NoStop}%
\bibitem [{\citenamefont {Abbott}\ \emph {et~al.}(2017)\citenamefont {Abbott}
  \emph {et~al.}}]{ligobinary}%
  \BibitemOpen
  \bibfield  {author} {\bibinfo {author} {\bibfnamefont {B.~P.}\ \bibnamefont
  {Abbott}} \emph {et~al.} (\bibinfo {collaboration} {LIGO scientific,
  Virgo}),\ }\bibfield  {title} {\enquote {\bibinfo {title} {{GW170817:
  Observation of Gravitational Waves from a Binary Neutron Star Inspiral}},}\
  }\href {\doibase 10.1103/PhysRevLett.119.161101} {\bibfield  {journal}
  {\bibinfo  {journal} {Phys. Rev. Lett.}\ }\textbf {\bibinfo {volume} {119}},\
  \bibinfo {pages} {161101} (\bibinfo {year} {2017})},\ \Eprint
  {http://arxiv.org/abs/1710.05832} {arXiv:1710.05832 [gr-qc]} \BibitemShut
  {NoStop}%
\bibitem [{\citenamefont {Lombriser}\ and\ \citenamefont
  {Taylor}(2016)}]{Lombriser:2015sxa}%
  \BibitemOpen
  \bibfield  {author} {\bibinfo {author} {\bibfnamefont {Lucas}\ \bibnamefont
  {Lombriser}}\ and\ \bibinfo {author} {\bibfnamefont {Andy}\ \bibnamefont
  {Taylor}},\ }\bibfield  {title} {\enquote {\bibinfo {title} {{Breaking a Dark
  Degeneracy with Gravitational Waves}},}\ }\href {\doibase
  10.1088/1475-7516/2016/03/031} {\bibfield  {journal} {\bibinfo  {journal}
  {JCAP}\ }\textbf {\bibinfo {volume} {1603}},\ \bibinfo {pages} {031}
  (\bibinfo {year} {2016})},\ \Eprint {http://arxiv.org/abs/1509.08458}
  {arXiv:1509.08458 [astro-ph.CO]} \BibitemShut {NoStop}%
\bibitem [{\citenamefont {Lombriser}\ and\ \citenamefont
  {Lima}(2017)}]{Lombriser:2016yzn}%
  \BibitemOpen
  \bibfield  {author} {\bibinfo {author} {\bibfnamefont {Lucas}\ \bibnamefont
  {Lombriser}}\ and\ \bibinfo {author} {\bibfnamefont {Nelson~A.}\ \bibnamefont
  {Lima}},\ }\bibfield  {title} {\enquote {\bibinfo {title} {{Challenges to
  Self-Acceleration in Modified Gravity from Gravitational Waves and
  Large-Scale Structure}},}\ }\href {\doibase 10.1016/j.physletb.2016.12.048}
  {\bibfield  {journal} {\bibinfo  {journal} {Phys. Lett.}\ }\textbf {\bibinfo
  {volume} {B765}},\ \bibinfo {pages} {382--385} (\bibinfo {year} {2017})},\
  \Eprint {http://arxiv.org/abs/1602.07670} {arXiv:1602.07670 [astro-ph.CO]}
  \BibitemShut {NoStop}%
\bibitem [{\citenamefont {Wetterich}(1988{\natexlab{a}})}]{Wetterich:1987fm}%
  \BibitemOpen
  \bibfield  {author} {\bibinfo {author} {\bibfnamefont {C.}~\bibnamefont
  {Wetterich}},\ }\bibfield  {title} {\enquote {\bibinfo {title} {{Cosmology
  and the Fate of Dilatation Symmetry}},}\ }\href {\doibase
  10.1016/0550-3213(88)90193-9} {\bibfield  {journal} {\bibinfo  {journal}
  {Nucl. Phys.}\ }\textbf {\bibinfo {volume} {B302}},\ \bibinfo {pages}
  {668--696} (\bibinfo {year} {1988}{\natexlab{a}})}\BibitemShut {NoStop}%
\bibitem [{\citenamefont {Cooper}\ and\ \citenamefont {Venturi}(1981)}]{CV}%
  \BibitemOpen
  \bibfield  {author} {\bibinfo {author} {\bibfnamefont {Fred}\ \bibnamefont
  {Cooper}}\ and\ \bibinfo {author} {\bibfnamefont {Giovanni}\ \bibnamefont
  {Venturi}},\ }\bibfield  {title} {\enquote {\bibinfo {title} {{Cosmology and
  Broken Scale Invariance}},}\ }\href {\doibase 10.1103/PhysRevD.24.3338}
  {\bibfield  {journal} {\bibinfo  {journal} {Phys. Rev.}\ }\textbf {\bibinfo
  {volume} {D24}},\ \bibinfo {pages} {3338} (\bibinfo {year}
  {1981})}\BibitemShut {NoStop}%
\bibitem [{\citenamefont {Finelli}\ \emph {et~al.}(2008)\citenamefont
  {Finelli}, \citenamefont {Tronconi},\ and\ \citenamefont
  {Venturi}}]{Finelli:2007wb}%
  \BibitemOpen
  \bibfield  {author} {\bibinfo {author} {\bibfnamefont {F.}~\bibnamefont
  {Finelli}}, \bibinfo {author} {\bibfnamefont {A.}~\bibnamefont {Tronconi}}, \
  and\ \bibinfo {author} {\bibfnamefont {Giovanni}\ \bibnamefont {Venturi}},\
  }\bibfield  {title} {\enquote {\bibinfo {title} {{Dark Energy, Induced
  Gravity and Broken Scale Invariance}},}\ }\href {\doibase
  10.1016/j.physletb.2007.11.053} {\bibfield  {journal} {\bibinfo  {journal}
  {Phys. Lett.}\ }\textbf {\bibinfo {volume} {B659}},\ \bibinfo {pages}
  {466--470} (\bibinfo {year} {2008})},\ \Eprint
  {http://arxiv.org/abs/0710.2741} {arXiv:0710.2741 [astro-ph]} \BibitemShut
  {NoStop}%
\bibitem [{\citenamefont {Bezrukov}\ and\ \citenamefont
  {Shaposhnikov}(2008)}]{Bezrukov:2007ep}%
  \BibitemOpen
  \bibfield  {author} {\bibinfo {author} {\bibfnamefont {Fedor~L.}\
  \bibnamefont {Bezrukov}}\ and\ \bibinfo {author} {\bibfnamefont {Mikhail}\
  \bibnamefont {Shaposhnikov}},\ }\bibfield  {title} {\enquote {\bibinfo
  {title} {{The Standard Model Higgs boson as the inflaton}},}\ }\href
  {\doibase 10.1016/j.physletb.2007.11.072} {\bibfield  {journal} {\bibinfo
  {journal} {Phys. Lett.}\ }\textbf {\bibinfo {volume} {B659}},\ \bibinfo
  {pages} {703--706} (\bibinfo {year} {2008})},\ \Eprint
  {http://arxiv.org/abs/0710.3755} {arXiv:0710.3755 [hep-th]} \BibitemShut
  {NoStop}%
\bibitem [{\citenamefont {Boisseau}\ \emph {et~al.}(2000)\citenamefont
  {Boisseau}, \citenamefont {Esposito-Farese}, \citenamefont {Polarski},\ and\
  \citenamefont {Starobinsky}}]{Boisseau:2000pr}%
  \BibitemOpen
  \bibfield  {author} {\bibinfo {author} {\bibfnamefont {B.}~\bibnamefont
  {Boisseau}}, \bibinfo {author} {\bibfnamefont {Gilles}\ \bibnamefont
  {Esposito-Farese}}, \bibinfo {author} {\bibfnamefont {D.}~\bibnamefont
  {Polarski}}, \ and\ \bibinfo {author} {\bibfnamefont {Alexei~A.}\
  \bibnamefont {Starobinsky}},\ }\bibfield  {title} {\enquote {\bibinfo {title}
  {{Reconstruction of a scalar tensor theory of gravity in an accelerating
  universe}},}\ }\href {\doibase 10.1103/PhysRevLett.85.2236} {\bibfield
  {journal} {\bibinfo  {journal} {Phys. Rev. Lett.}\ }\textbf {\bibinfo
  {volume} {85}},\ \bibinfo {pages} {2236} (\bibinfo {year} {2000})},\ \Eprint
  {http://arxiv.org/abs/gr-qc/0001066} {arXiv:gr-qc/0001066 [gr-qc]}
  \BibitemShut {NoStop}%
\bibitem [{\citenamefont {Gannouji}\ \emph {et~al.}(2006)\citenamefont
  {Gannouji}, \citenamefont {Polarski}, \citenamefont {Ranquet},\ and\
  \citenamefont {Starobinsky}}]{Gannouji:2006jm}%
  \BibitemOpen
  \bibfield  {author} {\bibinfo {author} {\bibfnamefont {Radouane}\
  \bibnamefont {Gannouji}}, \bibinfo {author} {\bibfnamefont {David}\
  \bibnamefont {Polarski}}, \bibinfo {author} {\bibfnamefont {Andre}\
  \bibnamefont {Ranquet}}, \ and\ \bibinfo {author} {\bibfnamefont {Alexei~A.}\
  \bibnamefont {Starobinsky}},\ }\bibfield  {title} {\enquote {\bibinfo {title}
  {{Scalar-Tensor Models of Normal and Phantom Dark Energy}},}\ }\href
  {\doibase 10.1088/1475-7516/2006/09/016} {\bibfield  {journal} {\bibinfo
  {journal} {JCAP}\ }\textbf {\bibinfo {volume} {0609}},\ \bibinfo {pages}
  {016} (\bibinfo {year} {2006})},\ \Eprint
  {http://arxiv.org/abs/astro-ph/0606287} {arXiv:astro-ph/0606287 [astro-ph]}
  \BibitemShut {NoStop}%
\bibitem [{\citenamefont {Ma}\ and\ \citenamefont
  {Bertschinger}(1995)}]{Ma:1995ey}%
  \BibitemOpen
  \bibfield  {author} {\bibinfo {author} {\bibfnamefont {Chung-Pei}\
  \bibnamefont {Ma}}\ and\ \bibinfo {author} {\bibfnamefont {Edmund}\
  \bibnamefont {Bertschinger}},\ }\bibfield  {title} {\enquote {\bibinfo
  {title} {{Cosmological perturbation theory in the synchronous and conformal
  Newtonian gauges}},}\ }\href {\doibase 10.1086/176550} {\bibfield  {journal}
  {\bibinfo  {journal} {Astrophys. J.}\ }\textbf {\bibinfo {volume} {455}},\
  \bibinfo {pages} {7--25} (\bibinfo {year} {1995})},\ \Eprint
  {http://arxiv.org/abs/astro-ph/9506072} {arXiv:astro-ph/9506072 [astro-ph]}
  \BibitemShut {NoStop}%
\bibitem [{\citenamefont {Rossi}(2016)}]{Rossi:2016}%
  \BibitemOpen
  \bibfield  {author} {\bibinfo {author} {\bibfnamefont {Massimo}\ \bibnamefont
  {Rossi}},\ }\bibfield  {title} {\enquote {\bibinfo {title} {{Dark Energy as a
  scalar field non-minimally coupled to gravity}},}\ }\href@noop {} {\
  (\bibinfo {year} {2016})}\BibitemShut {NoStop}%
\bibitem [{\citenamefont {Riazuelo}\ and\ \citenamefont
  {Uzan}(2000)}]{Riazuelo:2000fc}%
  \BibitemOpen
  \bibfield  {author} {\bibinfo {author} {\bibfnamefont {Alain}\ \bibnamefont
  {Riazuelo}}\ and\ \bibinfo {author} {\bibfnamefont {Jean-Philippe}\
  \bibnamefont {Uzan}},\ }\bibfield  {title} {\enquote {\bibinfo {title}
  {{Quintessence and gravitational waves}},}\ }\href {\doibase
  10.1103/PhysRevD.62.083506} {\bibfield  {journal} {\bibinfo  {journal} {Phys.
  Rev.}\ }\textbf {\bibinfo {volume} {D62}},\ \bibinfo {pages} {083506}
  (\bibinfo {year} {2000})},\ \Eprint {http://arxiv.org/abs/astro-ph/0004156}
  {arXiv:astro-ph/0004156 [astro-ph]} \BibitemShut {NoStop}%
\bibitem [{\citenamefont {Amendola}\ \emph {et~al.}(2014)\citenamefont
  {Amendola}, \citenamefont {Ballesteros},\ and\ \citenamefont
  {Pettorino}}]{Amendola:2014wma}%
  \BibitemOpen
  \bibfield  {author} {\bibinfo {author} {\bibfnamefont {Luca}\ \bibnamefont
  {Amendola}}, \bibinfo {author} {\bibfnamefont {Guillermo}\ \bibnamefont
  {Ballesteros}}, \ and\ \bibinfo {author} {\bibfnamefont {Valeria}\
  \bibnamefont {Pettorino}},\ }\bibfield  {title} {\enquote {\bibinfo {title}
  {{Effects of modified gravity on B-mode polarization}},}\ }\href {\doibase
  10.1103/PhysRevD.90.043009} {\bibfield  {journal} {\bibinfo  {journal} {Phys.
  Rev.}\ }\textbf {\bibinfo {volume} {D90}},\ \bibinfo {pages} {043009}
  (\bibinfo {year} {2014})},\ \Eprint {http://arxiv.org/abs/1405.7004}
  {arXiv:1405.7004 [astro-ph.CO]} \BibitemShut {NoStop}%
\bibitem [{\citenamefont {Liddle}\ \emph {et~al.}(1998)\citenamefont {Liddle},
  \citenamefont {Mazumdar},\ and\ \citenamefont {Barrow}}]{Liddle:1998ij}%
  \BibitemOpen
  \bibfield  {author} {\bibinfo {author} {\bibfnamefont {Andrew~R.}\
  \bibnamefont {Liddle}}, \bibinfo {author} {\bibfnamefont {Anupam}\
  \bibnamefont {Mazumdar}}, \ and\ \bibinfo {author} {\bibfnamefont {John~D.}\
  \bibnamefont {Barrow}},\ }\bibfield  {title} {\enquote {\bibinfo {title}
  {{Radiation matter transition in Jordan-Brans-Dicke theory}},}\ }\href
  {\doibase 10.1103/PhysRevD.58.027302} {\bibfield  {journal} {\bibinfo
  {journal} {Phys. Rev.}\ }\textbf {\bibinfo {volume} {D58}},\ \bibinfo {pages}
  {027302} (\bibinfo {year} {1998})},\ \Eprint
  {http://arxiv.org/abs/astro-ph/9802133} {arXiv:astro-ph/9802133 [astro-ph]}
  \BibitemShut {NoStop}%
\bibitem [{\citenamefont {Akrami}\ \emph
  {et~al.}(2018{\natexlab{a}})\citenamefont {Akrami} \emph
  {et~al.}}]{Akrami:2018odb}%
  \BibitemOpen
  \bibfield  {author} {\bibinfo {author} {\bibfnamefont {Y.}~\bibnamefont
  {Akrami}} \emph {et~al.} (\bibinfo {collaboration} {Planck}),\ }\bibfield
  {title} {\enquote {\bibinfo {title} {{Planck 2018 results. X. Constraints on
  inflation}},}\ }\href@noop {} {\  (\bibinfo {year} {2018}{\natexlab{a}})},\
  \Eprint {http://arxiv.org/abs/1807.06211} {arXiv:1807.06211 [astro-ph.CO]}
  \BibitemShut {NoStop}%
\bibitem [{\citenamefont {Ade}\ \emph {et~al.}(2018)\citenamefont {Ade} \emph
  {et~al.}}]{Ade:2018gkx}%
  \BibitemOpen
  \bibfield  {author} {\bibinfo {author} {\bibfnamefont {P.~A.~R.}\
  \bibnamefont {Ade}} \emph {et~al.} (\bibinfo {collaboration} {BICEP2, Keck
  Array}),\ }\bibfield  {title} {\enquote {\bibinfo {title} {{BICEP2 / Keck
  Array x: Constraints on Primordial Gravitational Waves using Planck, WMAP,
  and New BICEP2/Keck Observations through the 2015 Season}},}\ }\href
  {\doibase 10.1103/PhysRevLett.121.221301} {\bibfield  {journal} {\bibinfo
  {journal} {Phys. Rev. Lett.}\ }\textbf {\bibinfo {volume} {121}},\ \bibinfo
  {pages} {221301} (\bibinfo {year} {2018})},\ \Eprint
  {http://arxiv.org/abs/1810.05216} {arXiv:1810.05216 [astro-ph.CO]}
  \BibitemShut {NoStop}%
\bibitem [{\citenamefont {Audren}\ \emph {et~al.}(2013)\citenamefont {Audren},
  \citenamefont {Lesgourgues}, \citenamefont {Benabed},\ and\ \citenamefont
  {Prunet}}]{Audren:2012wb}%
  \BibitemOpen
  \bibfield  {author} {\bibinfo {author} {\bibfnamefont {Benjamin}\
  \bibnamefont {Audren}}, \bibinfo {author} {\bibfnamefont {Julien}\
  \bibnamefont {Lesgourgues}}, \bibinfo {author} {\bibfnamefont {Karim}\
  \bibnamefont {Benabed}}, \ and\ \bibinfo {author} {\bibfnamefont {Simon}\
  \bibnamefont {Prunet}},\ }\bibfield  {title} {\enquote {\bibinfo {title}
  {{Conservative Constraints on Early Cosmology: an illustration of the Monte
  Python cosmological parameter inference code}},}\ }\href {\doibase
  10.1088/1475-7516/2013/02/001} {\bibfield  {journal} {\bibinfo  {journal}
  {JCAP}\ }\textbf {\bibinfo {volume} {1302}},\ \bibinfo {pages} {001}
  (\bibinfo {year} {2013})},\ \Eprint {http://arxiv.org/abs/1210.7183}
  {arXiv:1210.7183 [astro-ph.CO]} \BibitemShut {NoStop}%
\bibitem [{\citenamefont {Brinckmann}\ and\ \citenamefont
  {Lesgourgues}(2018)}]{Brinckmann:2018cvx}%
  \BibitemOpen
  \bibfield  {author} {\bibinfo {author} {\bibfnamefont {Thejs}\ \bibnamefont
  {Brinckmann}}\ and\ \bibinfo {author} {\bibfnamefont {Julien}\ \bibnamefont
  {Lesgourgues}},\ }\bibfield  {title} {\enquote {\bibinfo {title}
  {{MontePython 3: boosted MCMC sampler and other features}},}\ }\href@noop {}
  {\  (\bibinfo {year} {2018})},\ \Eprint {http://arxiv.org/abs/1804.07261}
  {arXiv:1804.07261 [astro-ph.CO]} \BibitemShut {NoStop}%
\bibitem [{\citenamefont {Blas}\ \emph {et~al.}(2011)\citenamefont {Blas},
  \citenamefont {Lesgourgues},\ and\ \citenamefont {Tram}}]{Blas:2011rf}%
  \BibitemOpen
  \bibfield  {author} {\bibinfo {author} {\bibfnamefont {Diego}\ \bibnamefont
  {Blas}}, \bibinfo {author} {\bibfnamefont {Julien}\ \bibnamefont
  {Lesgourgues}}, \ and\ \bibinfo {author} {\bibfnamefont {Thomas}\
  \bibnamefont {Tram}},\ }\bibfield  {title} {\enquote {\bibinfo {title} {{The
  Cosmic Linear Anisotropy Solving System (CLASS) II: Approximation
  schemes}},}\ }\href {\doibase 10.1088/1475-7516/2011/07/034} {\bibfield
  {journal} {\bibinfo  {journal} {JCAP}\ }\textbf {\bibinfo {volume} {1107}},\
  \bibinfo {pages} {034} (\bibinfo {year} {2011})},\ \Eprint
  {http://arxiv.org/abs/1104.2933} {arXiv:1104.2933 [astro-ph.CO]} \BibitemShut
  {NoStop}%
\bibitem [{\citenamefont {Aghanim}\ \emph {et~al.}(2016)\citenamefont {Aghanim}
  \emph {et~al.}}]{Aghanim:2015xee}%
  \BibitemOpen
  \bibfield  {author} {\bibinfo {author} {\bibfnamefont {N.}~\bibnamefont
  {Aghanim}} \emph {et~al.} (\bibinfo {collaboration} {Planck}),\ }\bibfield
  {title} {\enquote {\bibinfo {title} {{Planck 2015 results. XI. CMB power
  spectra, likelihoods, and robustness of parameters}},}\ }\href {\doibase
  10.1051/0004-6361/201526926} {\bibfield  {journal} {\bibinfo  {journal}
  {Astron. Astrophys.}\ }\textbf {\bibinfo {volume} {594}},\ \bibinfo {pages}
  {A11} (\bibinfo {year} {2016})},\ \Eprint {http://arxiv.org/abs/1507.02704}
  {arXiv:1507.02704 [astro-ph.CO]} \BibitemShut {NoStop}%
\bibitem [{\citenamefont {Ade}\ \emph {et~al.}(2016{\natexlab{a}})\citenamefont
  {Ade} \emph {et~al.}}]{Ade:2015zua}%
  \BibitemOpen
  \bibfield  {author} {\bibinfo {author} {\bibfnamefont {P.~A.~R.}\
  \bibnamefont {Ade}} \emph {et~al.} (\bibinfo {collaboration} {Planck}),\
  }\bibfield  {title} {\enquote {\bibinfo {title} {{Planck 2015 results. XV.
  Gravitational lensing}},}\ }\href {\doibase 10.1051/0004-6361/201525941}
  {\bibfield  {journal} {\bibinfo  {journal} {Astron. Astrophys.}\ }\textbf
  {\bibinfo {volume} {594}},\ \bibinfo {pages} {A15} (\bibinfo {year}
  {2016}{\natexlab{a}})},\ \Eprint {http://arxiv.org/abs/1502.01591}
  {arXiv:1502.01591 [astro-ph.CO]} \BibitemShut {NoStop}%
\bibitem [{\citenamefont {Beutler}\ \emph {et~al.}(2011)\citenamefont
  {Beutler}, \citenamefont {Blake}, \citenamefont {Colless}, \citenamefont
  {Jones}, \citenamefont {Staveley-Smith}, \citenamefont {Campbell},
  \citenamefont {Parker}, \citenamefont {Saunders},\ and\ \citenamefont
  {Watson}}]{Beutler:2011hx}%
  \BibitemOpen
  \bibfield  {author} {\bibinfo {author} {\bibfnamefont {Florian}\ \bibnamefont
  {Beutler}}, \bibinfo {author} {\bibfnamefont {Chris}\ \bibnamefont {Blake}},
  \bibinfo {author} {\bibfnamefont {Matthew}\ \bibnamefont {Colless}}, \bibinfo
  {author} {\bibfnamefont {D.~Heath}\ \bibnamefont {Jones}}, \bibinfo {author}
  {\bibfnamefont {Lister}\ \bibnamefont {Staveley-Smith}}, \bibinfo {author}
  {\bibfnamefont {Lachlan}\ \bibnamefont {Campbell}}, \bibinfo {author}
  {\bibfnamefont {Quentin}\ \bibnamefont {Parker}}, \bibinfo {author}
  {\bibfnamefont {Will}\ \bibnamefont {Saunders}}, \ and\ \bibinfo {author}
  {\bibfnamefont {Fred}\ \bibnamefont {Watson}},\ }\bibfield  {title} {\enquote
  {\bibinfo {title} {{The 6dF Galaxy Survey: Baryon Acoustic Oscillations and
  the Local Hubble Constant}},}\ }\href {\doibase
  10.1111/j.1365-2966.2011.19250.x} {\bibfield  {journal} {\bibinfo  {journal}
  {Mon. Not. Roy. Astron. Soc.}\ }\textbf {\bibinfo {volume} {416}},\ \bibinfo
  {pages} {3017--3032} (\bibinfo {year} {2011})},\ \Eprint
  {http://arxiv.org/abs/1106.3366} {arXiv:1106.3366 [astro-ph.CO]} \BibitemShut
  {NoStop}%
\bibitem [{\citenamefont {Anderson}\ \emph {et~al.}(2014)\citenamefont
  {Anderson} \emph {et~al.}}]{Anderson:2013zyy}%
  \BibitemOpen
  \bibfield  {author} {\bibinfo {author} {\bibfnamefont {Lauren}\ \bibnamefont
  {Anderson}} \emph {et~al.} (\bibinfo {collaboration} {BOSS}),\ }\bibfield
  {title} {\enquote {\bibinfo {title} {{The clustering of galaxies in the
  SDSS-III Baryon Oscillation Spectroscopic Survey: baryon acoustic
  oscillations in the Data Releases 10 and 11 Galaxy samples}},}\ }\href
  {\doibase 10.1093/mnras/stu523} {\bibfield  {journal} {\bibinfo  {journal}
  {Mon. Not. Roy. Astron. Soc.}\ }\textbf {\bibinfo {volume} {441}},\ \bibinfo
  {pages} {24--62} (\bibinfo {year} {2014})},\ \Eprint
  {http://arxiv.org/abs/1312.4877} {arXiv:1312.4877 [astro-ph.CO]} \BibitemShut
  {NoStop}%
\bibitem [{\citenamefont {Ross}\ \emph {et~al.}(2015)\citenamefont {Ross},
  \citenamefont {Samushia}, \citenamefont {Howlett}, \citenamefont {Percival},
  \citenamefont {Burden},\ and\ \citenamefont {Manera}}]{Ross:2014qpa}%
  \BibitemOpen
  \bibfield  {author} {\bibinfo {author} {\bibfnamefont {Ashley~J.}\
  \bibnamefont {Ross}}, \bibinfo {author} {\bibfnamefont {Lado}\ \bibnamefont
  {Samushia}}, \bibinfo {author} {\bibfnamefont {Cullan}\ \bibnamefont
  {Howlett}}, \bibinfo {author} {\bibfnamefont {Will~J.}\ \bibnamefont
  {Percival}}, \bibinfo {author} {\bibfnamefont {Angela}\ \bibnamefont
  {Burden}}, \ and\ \bibinfo {author} {\bibfnamefont {Marc}\ \bibnamefont
  {Manera}},\ }\bibfield  {title} {\enquote {\bibinfo {title} {{The clustering
  of the SDSS DR7 main Galaxy sample – I. A 4 per cent distance measure at $z
  = 0.15$}},}\ }\href {\doibase 10.1093/mnras/stv154} {\bibfield  {journal}
  {\bibinfo  {journal} {Mon. Not. Roy. Astron. Soc.}\ }\textbf {\bibinfo
  {volume} {449}},\ \bibinfo {pages} {835--847} (\bibinfo {year} {2015})},\
  \Eprint {http://arxiv.org/abs/1409.3242} {arXiv:1409.3242 [astro-ph.CO]}
  \BibitemShut {NoStop}%
\bibitem [{\citenamefont {Pisanti}\ \emph {et~al.}(2008)\citenamefont
  {Pisanti}, \citenamefont {Cirillo}, \citenamefont {Esposito}, \citenamefont
  {Iocco}, \citenamefont {Mangano}, \citenamefont {Miele},\ and\ \citenamefont
  {Serpico}}]{Pisanti:2007hk}%
  \BibitemOpen
  \bibfield  {author} {\bibinfo {author} {\bibfnamefont {O.}~\bibnamefont
  {Pisanti}}, \bibinfo {author} {\bibfnamefont {A.}~\bibnamefont {Cirillo}},
  \bibinfo {author} {\bibfnamefont {S.}~\bibnamefont {Esposito}}, \bibinfo
  {author} {\bibfnamefont {F.}~\bibnamefont {Iocco}}, \bibinfo {author}
  {\bibfnamefont {G.}~\bibnamefont {Mangano}}, \bibinfo {author} {\bibfnamefont
  {G.}~\bibnamefont {Miele}}, \ and\ \bibinfo {author} {\bibfnamefont {P.~D.}\
  \bibnamefont {Serpico}},\ }\bibfield  {title} {\enquote {\bibinfo {title}
  {{PArthENoPE: Public Algorithm Evaluating the Nucleosynthesis of Primordial
  Elements}},}\ }\href {\doibase 10.1016/j.cpc.2008.02.015} {\bibfield
  {journal} {\bibinfo  {journal} {Comput. Phys. Commun.}\ }\textbf {\bibinfo
  {volume} {178}},\ \bibinfo {pages} {956--971} (\bibinfo {year} {2008})},\
  \Eprint {http://arxiv.org/abs/0705.0290} {arXiv:0705.0290 [astro-ph]}
  \BibitemShut {NoStop}%
\bibitem [{\citenamefont {Hamann}\ \emph {et~al.}(2008)\citenamefont {Hamann},
  \citenamefont {Lesgourgues},\ and\ \citenamefont {Mangano}}]{Hamann:2007sb}%
  \BibitemOpen
  \bibfield  {author} {\bibinfo {author} {\bibfnamefont {Jan}\ \bibnamefont
  {Hamann}}, \bibinfo {author} {\bibfnamefont {Julien}\ \bibnamefont
  {Lesgourgues}}, \ and\ \bibinfo {author} {\bibfnamefont {Gianpiero}\
  \bibnamefont {Mangano}},\ }\bibfield  {title} {\enquote {\bibinfo {title}
  {{Using BBN in cosmological parameter extraction from CMB: A Forecast for
  PLANCK}},}\ }\href {\doibase 10.1088/1475-7516/2008/03/004} {\bibfield
  {journal} {\bibinfo  {journal} {JCAP}\ }\textbf {\bibinfo {volume} {0803}},\
  \bibinfo {pages} {004} (\bibinfo {year} {2008})},\ \Eprint
  {http://arxiv.org/abs/0712.2826} {arXiv:0712.2826 [astro-ph]} \BibitemShut
  {NoStop}%
\bibitem [{\citenamefont {Ade}\ \emph {et~al.}(2016{\natexlab{b}})\citenamefont
  {Ade} \emph {et~al.}}]{Ade:2015xua}%
  \BibitemOpen
  \bibfield  {author} {\bibinfo {author} {\bibfnamefont {P.~A.~R.}\
  \bibnamefont {Ade}} \emph {et~al.} (\bibinfo {collaboration} {Planck}),\
  }\bibfield  {title} {\enquote {\bibinfo {title} {{Planck 2015 results. XIII.
  Cosmological parameters}},}\ }\href {\doibase 10.1051/0004-6361/201525830}
  {\bibfield  {journal} {\bibinfo  {journal} {Astron. Astrophys.}\ }\textbf
  {\bibinfo {volume} {594}},\ \bibinfo {pages} {A13} (\bibinfo {year}
  {2016}{\natexlab{b}})},\ \Eprint {http://arxiv.org/abs/1502.01589}
  {arXiv:1502.01589 [astro-ph.CO]} \BibitemShut {NoStop}%
\bibitem [{\citenamefont {Riess}\ \emph {et~al.}(2018)\citenamefont {Riess}
  \emph {et~al.}}]{Riess:2018byc}%
  \BibitemOpen
  \bibfield  {author} {\bibinfo {author} {\bibfnamefont {Adam~G.}\ \bibnamefont
  {Riess}} \emph {et~al.},\ }\bibfield  {title} {\enquote {\bibinfo {title}
  {{Milky Way Cepheid Standards for Measuring Cosmic Distances and Application
  to Gaia DR2: Implications for the Hubble Constant}},}\ }\href {\doibase
  10.3847/1538-4357/aac82e} {\bibfield  {journal} {\bibinfo  {journal}
  {Astrophys. J.}\ }\textbf {\bibinfo {volume} {861}},\ \bibinfo {pages} {126}
  (\bibinfo {year} {2018})},\ \Eprint {http://arxiv.org/abs/1804.10655}
  {arXiv:1804.10655 [astro-ph.CO]} \BibitemShut {NoStop}%
\bibitem [{\citenamefont {Riess}\ \emph {et~al.}(2016)\citenamefont {Riess}
  \emph {et~al.}}]{Riess:2016jrr}%
  \BibitemOpen
  \bibfield  {author} {\bibinfo {author} {\bibfnamefont {Adam~G.}\ \bibnamefont
  {Riess}} \emph {et~al.},\ }\bibfield  {title} {\enquote {\bibinfo {title} {{A
  2.4\% Determination of the Local Value of the Hubble Constant}},}\ }\href
  {\doibase 10.3847/0004-637X/826/1/56} {\bibfield  {journal} {\bibinfo
  {journal} {Astrophys. J.}\ }\textbf {\bibinfo {volume} {826}},\ \bibinfo
  {pages} {56} (\bibinfo {year} {2016})},\ \Eprint
  {http://arxiv.org/abs/1604.01424} {arXiv:1604.01424 [astro-ph.CO]}
  \BibitemShut {NoStop}%
\bibitem [{\citenamefont {Cuesta}\ \emph {et~al.}(2015)\citenamefont {Cuesta},
  \citenamefont {Verde}, \citenamefont {Riess},\ and\ \citenamefont
  {Jimenez}}]{Cuesta:2014asa}%
  \BibitemOpen
  \bibfield  {author} {\bibinfo {author} {\bibfnamefont {Antonio~J.}\
  \bibnamefont {Cuesta}}, \bibinfo {author} {\bibfnamefont {Licia}\
  \bibnamefont {Verde}}, \bibinfo {author} {\bibfnamefont {Adam}\ \bibnamefont
  {Riess}}, \ and\ \bibinfo {author} {\bibfnamefont {Raul}\ \bibnamefont
  {Jimenez}},\ }\bibfield  {title} {\enquote {\bibinfo {title} {{Calibrating
  the cosmic distance scale ladder: the role of the sound horizon scale and the
  local expansion rate as distance anchors}},}\ }\href {\doibase
  10.1093/mnras/stv261} {\bibfield  {journal} {\bibinfo  {journal} {Mon. Not.
  Roy. Astron. Soc.}\ }\textbf {\bibinfo {volume} {448}},\ \bibinfo {pages}
  {3463--3471} (\bibinfo {year} {2015})},\ \Eprint
  {http://arxiv.org/abs/1411.1094} {arXiv:1411.1094 [astro-ph.CO]} \BibitemShut
  {NoStop}%
\bibitem [{\citenamefont {Bernal}\ \emph {et~al.}(2016)\citenamefont {Bernal},
  \citenamefont {Verde},\ and\ \citenamefont {Riess}}]{Bernal:2016gxb}%
  \BibitemOpen
  \bibfield  {author} {\bibinfo {author} {\bibfnamefont {Jose~Luis}\
  \bibnamefont {Bernal}}, \bibinfo {author} {\bibfnamefont {Licia}\
  \bibnamefont {Verde}}, \ and\ \bibinfo {author} {\bibfnamefont {Adam~G.}\
  \bibnamefont {Riess}},\ }\bibfield  {title} {\enquote {\bibinfo {title} {{The
  trouble with $H_0$}},}\ }\href {\doibase 10.1088/1475-7516/2016/10/019}
  {\bibfield  {journal} {\bibinfo  {journal} {JCAP}\ }\textbf {\bibinfo
  {volume} {1610}},\ \bibinfo {pages} {019} (\bibinfo {year} {2016})},\ \Eprint
  {http://arxiv.org/abs/1607.05617} {arXiv:1607.05617 [astro-ph.CO]}
  \BibitemShut {NoStop}%
\bibitem [{\citenamefont {Aylor}\ \emph {et~al.}(2019)\citenamefont {Aylor},
  \citenamefont {Joy}, \citenamefont {Knox}, \citenamefont {Millea},
  \citenamefont {Raghunathan},\ and\ \citenamefont {Wu}}]{Aylor:2018drw}%
  \BibitemOpen
  \bibfield  {author} {\bibinfo {author} {\bibfnamefont {Kevin}\ \bibnamefont
  {Aylor}}, \bibinfo {author} {\bibfnamefont {MacKenzie}\ \bibnamefont {Joy}},
  \bibinfo {author} {\bibfnamefont {Lloyd}\ \bibnamefont {Knox}}, \bibinfo
  {author} {\bibfnamefont {Marius}\ \bibnamefont {Millea}}, \bibinfo {author}
  {\bibfnamefont {Srinivasan}\ \bibnamefont {Raghunathan}}, \ and\ \bibinfo
  {author} {\bibfnamefont {W.~L.~Kimmy}\ \bibnamefont {Wu}},\ }\bibfield
  {title} {\enquote {\bibinfo {title} {{Sounds Discordant: Classical Distance
  Ladder \& $\Lambda$CDM -based Determinations of the Cosmological Sound
  Horizon}},}\ }\href {\doibase 10.3847/1538-4357/ab0898} {\bibfield  {journal}
  {\bibinfo  {journal} {Astrophys. J.}\ }\textbf {\bibinfo {volume} {874}},\
  \bibinfo {pages} {4} (\bibinfo {year} {2019})},\ \Eprint
  {http://arxiv.org/abs/1811.00537} {arXiv:1811.00537 [astro-ph.CO]}
  \BibitemShut {NoStop}%
\bibitem [{\citenamefont {Poulin}\ \emph
  {et~al.}(2018{\natexlab{a}})\citenamefont {Poulin}, \citenamefont {Smith},
  \citenamefont {Grin}, \citenamefont {Karwal},\ and\ \citenamefont
  {Kamionkowski}}]{Poulin:2018dzj}%
  \BibitemOpen
  \bibfield  {author} {\bibinfo {author} {\bibfnamefont {Vivian}\ \bibnamefont
  {Poulin}}, \bibinfo {author} {\bibfnamefont {Tristan~L.}\ \bibnamefont
  {Smith}}, \bibinfo {author} {\bibfnamefont {Daniel}\ \bibnamefont {Grin}},
  \bibinfo {author} {\bibfnamefont {Tanvi}\ \bibnamefont {Karwal}}, \ and\
  \bibinfo {author} {\bibfnamefont {Marc}\ \bibnamefont {Kamionkowski}},\
  }\bibfield  {title} {\enquote {\bibinfo {title} {{Cosmological implications
  of ultralight axionlike fields}},}\ }\href {\doibase
  10.1103/PhysRevD.98.083525} {\bibfield  {journal} {\bibinfo  {journal} {Phys.
  Rev.}\ }\textbf {\bibinfo {volume} {D98}},\ \bibinfo {pages} {083525}
  (\bibinfo {year} {2018}{\natexlab{a}})},\ \Eprint
  {http://arxiv.org/abs/1806.10608} {arXiv:1806.10608 [astro-ph.CO]}
  \BibitemShut {NoStop}%
\bibitem [{\citenamefont {Poulin}\ \emph
  {et~al.}(2018{\natexlab{b}})\citenamefont {Poulin}, \citenamefont {Smith},
  \citenamefont {Karwal},\ and\ \citenamefont {Kamionkowski}}]{Poulin:2018cxd}%
  \BibitemOpen
  \bibfield  {author} {\bibinfo {author} {\bibfnamefont {Vivian}\ \bibnamefont
  {Poulin}}, \bibinfo {author} {\bibfnamefont {Tristan~L.}\ \bibnamefont
  {Smith}}, \bibinfo {author} {\bibfnamefont {Tanvi}\ \bibnamefont {Karwal}}, \
  and\ \bibinfo {author} {\bibfnamefont {Marc}\ \bibnamefont {Kamionkowski}},\
  }\bibfield  {title} {\enquote {\bibinfo {title} {{Early Dark Energy Can
  Resolve The Hubble Tension}},}\ }\href@noop {} {\  (\bibinfo {year}
  {2018}{\natexlab{b}})},\ \Eprint {http://arxiv.org/abs/1811.04083}
  {arXiv:1811.04083 [astro-ph.CO]} \BibitemShut {NoStop}%
\bibitem [{\citenamefont {Agrawal}\ \emph {et~al.}(2019)\citenamefont
  {Agrawal}, \citenamefont {Cyr-Racine}, \citenamefont {Pinner},\ and\
  \citenamefont {Randall}}]{Agrawal:2019lmo}%
  \BibitemOpen
  \bibfield  {author} {\bibinfo {author} {\bibfnamefont {Prateek}\ \bibnamefont
  {Agrawal}}, \bibinfo {author} {\bibfnamefont {Francis-Yan}\ \bibnamefont
  {Cyr-Racine}}, \bibinfo {author} {\bibfnamefont {David}\ \bibnamefont
  {Pinner}}, \ and\ \bibinfo {author} {\bibfnamefont {Lisa}\ \bibnamefont
  {Randall}},\ }\bibfield  {title} {\enquote {\bibinfo {title} {{Rock 'n' Roll
  Solutions to the Hubble Tension}},}\ }\href@noop {} {\  (\bibinfo {year}
  {2019})},\ \Eprint {http://arxiv.org/abs/1904.01016} {arXiv:1904.01016
  [astro-ph.CO]} \BibitemShut {NoStop}%
\bibitem [{\citenamefont {Will}(2014)}]{Will:2014kxa}%
  \BibitemOpen
  \bibfield  {author} {\bibinfo {author} {\bibfnamefont {Clifford~M.}\
  \bibnamefont {Will}},\ }\bibfield  {title} {\enquote {\bibinfo {title} {{The
  Confrontation between General Relativity and Experiment}},}\ }\href {\doibase
  10.12942/lrr-2014-4} {\bibfield  {journal} {\bibinfo  {journal} {Living Rev.
  Rel.}\ }\textbf {\bibinfo {volume} {17}},\ \bibinfo {pages} {4} (\bibinfo
  {year} {2014})},\ \Eprint {http://arxiv.org/abs/1403.7377} {arXiv:1403.7377
  [gr-qc]} \BibitemShut {NoStop}%
\bibitem [{\citenamefont {Bertotti}\ \emph {et~al.}(2003)\citenamefont
  {Bertotti}, \citenamefont {Iess},\ and\ \citenamefont
  {Tortora}}]{Bertotti:2003rm}%
  \BibitemOpen
  \bibfield  {author} {\bibinfo {author} {\bibfnamefont {B.}~\bibnamefont
  {Bertotti}}, \bibinfo {author} {\bibfnamefont {L.}~\bibnamefont {Iess}}, \
  and\ \bibinfo {author} {\bibfnamefont {P.}~\bibnamefont {Tortora}},\
  }\bibfield  {title} {\enquote {\bibinfo {title} {{A test of general
  relativity using radio links with the Cassini spacecraft}},}\ }\href
  {\doibase 10.1038/nature01997} {\bibfield  {journal} {\bibinfo  {journal}
  {Nature}\ }\textbf {\bibinfo {volume} {425}},\ \bibinfo {pages} {374--376}
  (\bibinfo {year} {2003})}\BibitemShut {NoStop}%
\bibitem [{\citenamefont {Wetterich}(1988{\natexlab{b}})}]{Wetterich:1987fk}%
  \BibitemOpen
  \bibfield  {author} {\bibinfo {author} {\bibfnamefont {C.}~\bibnamefont
  {Wetterich}},\ }\bibfield  {title} {\enquote {\bibinfo {title} {{Cosmologies
  With Variable Newton's 'Constant'}},}\ }\href {\doibase
  10.1016/0550-3213(88)90192-7} {\bibfield  {journal} {\bibinfo  {journal}
  {Nucl. Phys.}\ }\textbf {\bibinfo {volume} {B302}},\ \bibinfo {pages}
  {645--667} (\bibinfo {year} {1988}{\natexlab{b}})}\BibitemShut {NoStop}%
\bibitem [{\citenamefont {Paoletti}\ \emph {et~al.}(2018)\citenamefont
  {Paoletti}, \citenamefont {Braglia}, \citenamefont {Finelli}, \citenamefont
  {Ballardini},\ and\ \citenamefont {Umilt\`a}}]{Paoletti:2018xet}%
  \BibitemOpen
  \bibfield  {author} {\bibinfo {author} {\bibfnamefont {D.}~\bibnamefont
  {Paoletti}}, \bibinfo {author} {\bibfnamefont {M.}~\bibnamefont {Braglia}},
  \bibinfo {author} {\bibfnamefont {F.}~\bibnamefont {Finelli}}, \bibinfo
  {author} {\bibfnamefont {M.}~\bibnamefont {Ballardini}}, \ and\ \bibinfo
  {author} {\bibfnamefont {C.}~\bibnamefont {Umilt\`a}},\ }\bibfield  {title}
  {\enquote {\bibinfo {title} {{Isocurvature fluctuations in the effective
  Newton's constant}},}\ }\href@noop {} {\  (\bibinfo {year} {2018})},\ \Eprint
  {http://arxiv.org/abs/1809.03201} {arXiv:1809.03201 [astro-ph.CO]}
  \BibitemShut {NoStop}%
\bibitem [{\citenamefont {Du}(2019)}]{Du:2018txo}%
  \BibitemOpen
  \bibfield  {author} {\bibinfo {author} {\bibfnamefont {Song~Ming}\
  \bibnamefont {Du}},\ }\bibfield  {title} {\enquote {\bibinfo {title} {{Scalar
  Stochastic Gravitational-Wave Background in Brans-Dicke Theory of
  Gravity}},}\ }\href {\doibase 10.1103/PhysRevD.99.044057} {\bibfield
  {journal} {\bibinfo  {journal} {Phys. Rev.}\ }\textbf {\bibinfo {volume}
  {D99}},\ \bibinfo {pages} {044057} (\bibinfo {year} {2019})},\ \Eprint
  {http://arxiv.org/abs/1812.06068} {arXiv:1812.06068 [gr-qc]} \BibitemShut
  {NoStop}%
\bibitem [{\citenamefont {Abbott}\ \emph {et~al.}(2018)\citenamefont {Abbott}
  \emph {et~al.}}]{Abbott:2018lct}%
  \BibitemOpen
  \bibfield  {author} {\bibinfo {author} {\bibfnamefont {B.~P.}\ \bibnamefont
  {Abbott}} \emph {et~al.} (\bibinfo {collaboration} {LIGO Scientific,
  Virgo}),\ }\bibfield  {title} {\enquote {\bibinfo {title} {{Tests of General
  Relativity with GW170817}},}\ }\href@noop {} {\  (\bibinfo {year} {2018})},\
  \Eprint {http://arxiv.org/abs/1811.00364} {arXiv:1811.00364 [gr-qc]}
  \BibitemShut {NoStop}%
\bibitem [{\citenamefont {Akrami}\ \emph
  {et~al.}(2018{\natexlab{b}})\citenamefont {Akrami} \emph
  {et~al.}}]{Akrami:2018vks}%
  \BibitemOpen
  \bibfield  {author} {\bibinfo {author} {\bibfnamefont {Y.}~\bibnamefont
  {Akrami}} \emph {et~al.} (\bibinfo {collaboration} {Planck}),\ }\bibfield
  {title} {\enquote {\bibinfo {title} {{Planck 2018 results. I. Overview and
  the cosmological legacy of Planck}},}\ }\href@noop {} {\  (\bibinfo {year}
  {2018}{\natexlab{b}})},\ \Eprint {http://arxiv.org/abs/1807.06205}
  {arXiv:1807.06205 [astro-ph.CO]} \BibitemShut {NoStop}%
\bibitem [{\citenamefont {Akrami}\ \emph
  {et~al.}(2018{\natexlab{c}})\citenamefont {Akrami} \emph
  {et~al.}}]{Akrami:2018jnw}%
  \BibitemOpen
  \bibfield  {author} {\bibinfo {author} {\bibfnamefont {Y.}~\bibnamefont
  {Akrami}} \emph {et~al.} (\bibinfo {collaboration} {Planck}),\ }\bibfield
  {title} {\enquote {\bibinfo {title} {{Planck 2018 results. II. Low Frequency
  Instrument data processing}},}\ }\href@noop {} {\  (\bibinfo {year}
  {2018}{\natexlab{c}})},\ \Eprint {http://arxiv.org/abs/1807.06206}
  {arXiv:1807.06206 [astro-ph.CO]} \BibitemShut {NoStop}%
\bibitem [{\citenamefont {Aghanim}\ \emph {et~al.}(2018)\citenamefont {Aghanim}
  \emph {et~al.}}]{Aghanim:2018fcm}%
  \BibitemOpen
  \bibfield  {author} {\bibinfo {author} {\bibfnamefont {N.}~\bibnamefont
  {Aghanim}} \emph {et~al.} (\bibinfo {collaboration} {Planck}),\ }\bibfield
  {title} {\enquote {\bibinfo {title} {{Planck 2018 results. III. High
  Frequency Instrument data processing and frequency maps}},}\ }\href@noop {}
  {\  (\bibinfo {year} {2018})},\ \Eprint {http://arxiv.org/abs/1807.06207}
  {arXiv:1807.06207 [astro-ph.CO]} \BibitemShut {NoStop}%
\end{thebibliography}%

\end{document}